\documentclass[leqno,oneside]{article}
\usepackage{amsfonts}
\usepackage{amssymb}
\usepackage{amsmath}
\newcommand{\field}[1]{\mathbb{#1}}
\newcommand{\bn}{\mbox{\boldmath$\na$}}

\newcommand{\ben}{\begin{displaymath}}
\newcommand{\een}{\end{displaymath}}

\newcommand{\hK}{\hat{K}}

\newcommand{\n}{\noindent}
\newcommand{\beq}{\begin{equation}}
\newcommand{\eeq}{\end{equation}}
\newcommand{\bc}{\begin{center}}
\newcommand{\ec}{\end{center}}
\newcommand{\na}{\nabla}
\newcommand{\dt}{{\bf \Pi}}
\newcommand{\W}{{\bf W}}
\newcommand{\La}{\Lambda}
\newcommand{\atlas}{{\mathcal{A}}}
\newcommand{\dod}{\,\dot{}}
\newcommand{\doo}{\,\dot{}\ \dot{}}
\newcommand{\dooo}{\,\dot{}\ \dot{}\ \dot{}}
\newcommand{\doooo}{\,\dot{}\ \dot{}\ \dot{}\ \dot{}}
\newcommand{\ep}{\hspace{\stretch{1}}$\Box$

\vspace{0.2cm}}
\usepackage{epsfig}
\usepackage{latexsym}
\usepackage{mathrsfs}

\newtheorem{Def}{{\bf Definition}}
\newtheorem{Prop}{{\bf Proposition}}
\newtheorem{T}{{\bf Theorem}}

\newtheorem{Lem}{{\bf Lemma}}

\newtheorem{Remark}{\bf Remark}
\newtheorem{Fact}{\it Fact}

\begin{document}

\begin{center}
{\Large\bf The Constant Mean Curvature Einstein flow and the Bel-Robinson energy}

\vspace{0.4cm}

{\large Martin Reiris}\footnote{e-mail: reiris@math.mit.edu.}\\

\vspace{0.1cm}

\textsc{Math. Dep. Massachusetts Institute of Technology}\\

\end{center}

\vspace{0.3cm}
\begin{center}
\begin{minipage}[c]{11cm}
\linespread{.9}%
\selectfont
{\small We give an extensive treatment of the Constant Mean Curvature (CMC) Einstein flow
from the point of view of the Bel-Robinson energies. The article, in particular, stresses on estimates 
showing how the Bel-Robinson energies and the volume of the evolving states control intrinsically 
the flow along evolution. The treatment is for flows over compact three-manifolds of arbitrary topological type,
although the form of the estimates may vary depending on the Yamabe invariant of the manifold. We end
up showing well posedness of the CMC Einstein flow with $H^{3}\times H^{2}$ regularity, and 
proving a criteria for a flow to be a long-time flow on manifolds with non-positive Yamabe invariant
in terms only of the first order Bel-Robinson energy.}
\end{minipage}
\end{center}

\vspace{.2cm}
\begin{center}
\begin{minipage}[c]{10cm}{\small
{\center\tableofcontents}}
\end{minipage}
\end{center}

\newpage
{\center \section{Introduction.}}

The aim of this article is to provide an intrinsic treatment of the Constant Mean Curvature (or simply CMC) gauge,
entirely in terms of the space-time Bel-Robinson energy and the space-like volume. Roughly speaking, the CMC gauge
foliates the space-time in such a way that, at every leaf, the local rate of (normal) volume expansion is constant 
(i.e. independent of the point). The space-like volume, in turn, is linked to the space-like scalar curvature and 
(though it) to the Yamabe invariant of the three-manifold. As it turns out, the link between the space-like volume 
and the spatial geometry is strengthened if the Bel-Robinson energy associated with the Riemann tensor of the space-time 
is taken into the picture. We will display 
precise estimates showing how the volume and the Bel-Robinson energies control the geometry of the states. 
This, in particular, makes the Bel-Robinson energies and the space-like volume a set of appealing variables to control 
the CMC flow along evolution. We will give a detailed discussion of intrinsic as well as elliptic estimates. 

The CMC gauge and Bel-Robinson energies have been used together several times in the past. In \cite{CK}, for instance, 
Christodoulou and 
Klainerman approached the stability of the Minkowski space-time using a maximal foliation and a elaborated control of the Bel-Robinson 
energies of appropriate Weyl fields. In their work, Weyl fields were used in particular as the fundamental variables from which to 
reconstruct the space-time metric. On the other hand, in \cite{AM}, Andersson and Moncrief gave a proof of the stability of flat cones 
following essentially the same argumental lines. This case, which can be considered as a compact version of \cite{CK}, is however greatly 
simplified thanks to the expansion in volume (and the compactness of the CMC Cauchy slices). In the context of the initial value formulation 
of the Einstein theory, Weyl fields were used by Friedrich in \cite{HF}. Friedrich included the conformal Weyl tensor of the space-time as a 
variable, and by doing so he obtained different hyperbolic reductions of the Einstein equations from which to launch initial value 
formulations. We will take, in spirit, several elements from these works. Namely, we will study how to control the space-time metric from the 
Bel-Robinson energies (of suitable Weyl fields) and use that knowledge to give a treatment of the Cauchy problem in General Relativity 
entirely inside the framework of Weyl fields.    
  
To step further 
in the description of the contents, let us introduce some
terminology first and then state, to exemplify, some of the main estimates that will proved. The reader can consult the background section for a 
detailed account on notions such as Bel-Robinson energy, or harmonic radius that we will mention below. Consider a cosmological solution $({\bf M},{\bf g})$ and a Cauchy
slice $\Sigma$. Denote by $g$ the spatial three-metric and by $K$ the second fundamental form of $\Sigma$. As is well known
$K=-\frac{1}{2}g\dod$ where $g\dod$ is the time derivative of $g$ in the normal direction to $\Sigma$. Thus it will be natural 
to call the pair $(g,K)=(g,-\frac{1}{2}g\dod)$ {\it a state}. A state is in particular a {\it CMC state} if the 
{\it constant mean curvature} $k=tr_{g}K$ is constant on $\Sigma$. Consider the space-time Riemann tensor ${\bf Rm}$. If the space-time
solution ${\bf g}$ is a vacuum solution the curvature ${\bf Rm}$ is a Weyl field as are its covariant derivatives $\bn_{T}^{j}{\bf Rm}$ 
in the normal direction $T$ to $\Sigma$. Consider their associated Bel-Robinson energies and denote them by $Q_{j}$. One of the main
results will be to show 

\vspace{0.2cm}
\begin{Lem} {\rm (Sobolev norms vs. Bel-Robinson norms)}\label{Intr1} Say $\bar{I}\geq 0$ and say $\Sigma$ is a compact
three-manifold. Then the functional (defined over states $(g,K)$ with $k=-3$)
\ben
\|(g,K)\|_{BR}=\frac{1}{\nu}+{\mathcal{V}}+\sum_{j=0}^{j=\bar{I}}Q_{j}=\frac{1}{\nu}+{\mathcal{V}}+{\mathcal{E}}_{\bar{I}},
\een

\n controls the $H^{\bar{I}+2}$-harmonic radius $r_{\bar{I}+2}$ and the 
$H^{\bar{I}+1}_{g}$-norm of $K$.
\end{Lem}

\n Above, ${\mathcal{V}}=(\frac{-k}{3})^{3}Vol_{g}(\Sigma)$ is the so called {\it reduced volume} (see later), 
which is essentially the usual volume if $k$ was fixed and equal to $-3$. $\nu$ is the so called 
{\it volume radius} which is a measure of the local ``flatness" (see later). $H^{\bar{I}+1}_{g}$ is the 
$\bar{I}+1$-Sobolev space and the subindex $g$ indicates that the Sobolev-norm is the natural constructed out of the metric $g$ 
(see the background section). The proof of this Lemma emerges basically as a corollary of a series of partial 
results which have, however, independent value. Consider for instance the case $\bar{I}=0$ in Lemma \ref{Intr1}. We show in 
Proposition \ref{hatK4} that the $H^{1}_{g}$-norm of $K$, namely,
\ben
\|K\|_{H^{1}_{g}}^{2}=\int_{\Sigma}|\nabla K|^{2}+|K|^{2}dv_{g},
\een

\n is controlled only by ${\mathcal{V}}$ and $Q_{0}$. This will follow from the explicit intrinsic estimate
\ben
\int_{\Sigma}2|\na \hat{K}|^{2}+|\hat{K}|^{4}dv_{g}\leq C(|k|{\mathcal{V}}+Q_{0}),
\een

\n in the case the Yamabe invariant $Y(\Sigma)$ is $Y(\Sigma)>0$, and 
\ben
\int_{\Sigma}2|\na \hat{K}|^{2}+|\hat{K}|^{4}dv_{g}\leq C(|k|({\mathcal{V}}-{\mathcal{V}}_{inf})+Q_{0}),
\een

\n if $Y(\Sigma)\leq 0$, where $\hat{K}$ is the traceless part of $K$, $C$ is a numeric constant and ${\mathcal{V}}_{inf}$ is the
infimum of the reduced volume ${\mathcal{V}}$ among all CMC states $(g,K)$. It turns out that 
${\mathcal{V}}_{inf}=(-\frac{1}{6}Y(\Sigma))^{\frac{3}{2}}$ (see the background section).
Being formal, an intrinsic estimate is an inequality (or equality) involving intrinsic Sobolev
norms $H^{\star}_{g}$ but not requiring the volume radius $\nu$. The intrinsic estimates on $K$ above,
imply, as we prove in Proposition \ref{Ric}, an intrinsic estimate on the $L^{2}_{g}$-norm of the Ricci curvature of the three-metric $g$. 
Namely we show
\ben
\|Ric\|^{2}_{L^{2}_{g}}\leq C(|k|{\mathcal{V}}+Q_{0}).
\een

\n It follows from the fundamental theorem of convergence of Riemannian manifolds (see the background section) that a priori control on 
${\mathcal{V}}$ and $Q_{0}$, and in addition a priori control on $\nu$, implies control on the $H^{2}$-harmonic radius $r_{2}$. This is the way 
Lemma \ref{Intr1} is proved when $\bar{I}=0$. The situation when $\bar{I}>0$ requires the use of elliptic estimates and therefore 
a priori control on $\nu$. For instance when $\bar{I}=1$, it is true that ${\mathcal{V}}$, $\nu$ and ${\mathcal{E}}_{1}$ control $\|K\|_{H^{2}_{g}}$
but through an inequality that involves implicitly the use of $\nu$, and which arises through the use of suitable elliptic estimates.   
To handle elliptic estimates when the background metric $g$ is one of the variables, we use Sobolev norms defined with respect to 
atlas composed of harmonic coordinates. Being more precise,
we define $H^{i}$-{\it canonic harmonic atlas}, and measure Sobolev norms with respect to them. A chart $\{x_{\alpha}\}$ in a $H^{i}$-canonic
harmonic atlas, has the property that if we scale $g$ as $\tilde{g}=\frac{1}{r_{i}^{2}}g$ and $\{x_{\alpha}\}$ as $\tilde{x}_{\alpha}^{k}=
\frac{1}{r_{i}}x_{\alpha}^{k}$ then $\tilde{x}_{\alpha}$ gets defined over a ball of radius one in the metric $\tilde{g}$ and moreover 
the usual $H^{i}_{\tilde{x}_{\alpha}}$-norm of $\tilde{g}$ is bounded above by a fixed (but arbitrary) constant. Supplied with some 
additional technical requirements, $H^{i}$-canonic harmonic atlas provide us with a more or less standardized way to define Sobolev norms 
with respect to atlas made of harmonic coordinates. We will denote by $H^{\star}_{\atlas}$ Sobolev spaces defined with respect to 
an atlas $\atlas$ and in particular with respect to a canonic harmonic atlas $\atlas$. Once an 
estimate, or a certain inequality is proved using the Sobolev spaces 
$H^{\star}_{\atlas}$, one can guarantee that a similar inequality or estimate holds with respect to the intrinsic norms $H^{\star}_{g}$, 
but the given estimate or inequality involves necessarily $\nu$. For instance, in Proposition \ref{1/N} we prove that 
${\nu}$, ${\mathcal{V}}$ and $Q_{0}$ control the $H^{2}_{g}$-norm of $1/N$ where $N$ is the
lapse function in the CMC gauge when one take $k$ as a choice of time. This (not self-evident) result, which in particular implies that
$N$ is never zero (even for states $(g,K)$ with $H^{2}\times H^{1}$ regularity) and controlled from below by $\nu$, ${\mathcal{V}}$ and 
$Q_{0}$, requires the intermediate use of the norms $H^{\star}_{\atlas}$, where $\atlas$ is a $H^{2}$-canonic harmonic atlas.    

Lemma \ref{Intr1} naturally points towards a proof of well-posedness of the CMC Einstein flow, entirely in terms of the 
intrinsic quantities $\nu$, ${\mathcal{V}}$ and ${\mathcal{E}}_{\bar{I}}$. In this respect we will prove the following version of 
well-posedness of the Einstein CMC-flow.

\begin{T}\label{Intr2} Let $(\Sigma,\atlas)$ be a $H^{4}$-three-manifold and say $(g_{0},K_{0})$ 
is an initial state in $H^{3}\times H^{2}$ with $k_{0}<0$. Then

\begin{enumerate}

\item There is a unique $H^{3}$-flow solution over an interval $I=(k_{-1},k_{1})$ with $-\infty\leq k_{-1}<k_{0}<k_{1}\leq 0$. 
Moreover the size $inf\{|k_{-1}-k_{0}|,|k_{0}-k_{1}|\}$ of the time interval on which the solution is guaranteed to exist is 
controlled from below by 
$1/\nu(k_{0})$, $\ln |k_{0}|$, ${\mathcal{V}}(g_{0},K_{0})$ and ${\mathcal{E}}_{1}(k_{0})$.

\item There is continuity with respect to the initial conditions if we measure the space of initial conditions with
the $H^{3}_{\atlas}\times H^{2}_{\atlas}$ norm and the space of flow solutions with the BR-norm.

\item Because of {\it item 1} above, we have the following continuity principle: a flow solution $((g,K),(N,X))(k)$ is defined 
until past of $k^{*}<0$ (or before $k_{*}<0$ if the flow is running in the past direction) iff 
$lim sup_{k\rightarrow k^{*}} 1/\nu+{\mathcal{V}}(k)+{\mathcal{E}}_{1}(k)<\infty$.

\end{enumerate} 

\end{T}

\n The BR-norm (Bel-Robinson norm) of a flow in {\it item 2} of the Theorem above is defined by 
\beq\label{BRR}
\|{\bf g}\|_{BR}=\|g\|_{C^{0}(I,2)(H_{{\mathcal{A}}})}+\|K\|_{C^{0}(I,1)(H_{{\mathcal{A}}})}+
\sum_{k=0}^{k=1}\|(E_{k},B_{k})\|_{C^{0}(I,0)(H_{\atlas})},
\eeq

\n where $(E_{0},B_{0})$ and $(E_{1},B_{1})$ are the electric and magnetic components of the Weyl tensors $\W_{0}={\bf Rm}$ and $\W_{1}$
respectively. The atlas $\atlas$ used in (\ref{BRR}) is the atlas that is given from the manifold $(\Sigma,\atlas)$. The space 
$C(I,j)(H_{\atlas})$ is defined as usual as the space of continuous functions over an interval $I$ with values 
in the $H^{j}_{\atlas}$-Sobolev space of tensors of a fixed rank. 

The initial value formulation of General Relativity in the CMC gauge has been considered in the literature some times in the past. 
In particular in \cite{CHY}\footnote{We would like to thank the referee for pointing out this reference to us.}, Choquet-Bruhat 
and York gave an exposition of the Cauchy problem applicable to the CMC gauge and for initial states with $H^{3}\times H^{2}$ regularity. 
In \cite{AM1}, Andersson and 
Moncrief considered the CMC gauge over hyperbolic-three manifolds, in the spatially harmonic gauge and for initial states with 
$H^{2.5+\epsilon}\times H^{1.5+\epsilon}$ regularity. Theorem \ref{Intr2} partially (but not entirely) 
overlaps the results in \cite{CHY} and \cite{AM1}.  

Note from (\ref{BRR}) that the flow of three-metrics $g(k)$, is measured in 
$H^{2}_{\atlas}$ and not in $H^{3}_{\atlas}$. In our approach we cannot guarantee that the metrics $g(k)$ will lie in 
$H^{3}_{\atlas}$ (see however the claims in \cite{CHY}). To circumvent this problem 
we included the terms $(E_{0},B_{0})$ and $(E_{1},B_{1})$ in (\ref{BRR}).  
Indeed, by Lemma \ref{Intr1}, $\frac{1}{\nu}+{\mathcal{V}}+Q_{0}+Q_{1}$ controls the $r_{3}$-harmonic radius and therefore 
(see the background section) the $H^{3}_{\atlas_{h}}$-norm
of $g$ where $\atlas_{h}$ is a $H^{3}$-canonic harmonic atlas. As $\|g\|_{H^{2}_{\atlas}}$ controls $\frac{1}{\nu}$ and ${\mathcal{V}}$ 
(when $|k|$ is bounded)
it follows that the ``norm'' $\|g\|_{H^{2}_{\atlas}}+\|(E_{0},B_{0})\|_{L^{2}_{\atlas}}+\|(E_{1},B_{1})\|_{L^{2}_{\atlas}}$ controls the norm
$\|g\|_{H^{3}_{\atlas_{h}}}$. To transform this fact into a technical tool we introduce in Section \ref{3.4} 
a dynamical $H^{3}$-harmonic atlas $\atlas(k)$. The BR-norm acquires then special relevance when $\atlas(k)$ is taken into account. 
The dynamical atlas will be a fundamental piece in all our treatment of the Einstein CMC flow. 

Consider a $C^{\infty}$ CMC initial state $(g_{0},K_{0})$ with $k_{0}\neq 0$ over a three-manifold that we label as 
$\Sigma_{0}$. It is well known the state $(g,k)$ gives 
rise to a $C^{\infty}$ space-time $({\bf M}, {\bf g})$. By the method of barriers it is shown\footnote{This fact is well known (see the 
discussion in \cite{Re}).} in the background section that there is a unique 
$C^{\infty}$ CMC foliation of ${\bf M}$ around the slice $\Sigma_{0}$ where $k$ varies strictly monotonically. Thus, for 
$C^{\infty}$ space times arising from initial $C^{\infty}$ CMC states, the CMC gauge is well defined at least on a neighborhood of the initial slice. The strategy we will follow to prove 
Theorem \ref{Intr2} is to show that the completion of the space of $C^{\infty}$ CMC flow solutions under the BR-norm (\ref{BRR}) is 
precisely the space of $H^{3}$-CMC flow solutions (see Definition \ref{g} in Section \ref{3.4}). Note that this is not the traditional 
approach to prove well-posedness of a PDE but it illustrates the applicability of the estimates in Section \ref{3}. 

There is an underlying reason of why
to describe the evolution in terms of the intrinsic quantities ${\mathcal{V}}$, $\nu$ and ${\mathcal{E}}_{1}$ and this has to do with
the continuity principle as in {\it item 3} of Theorem \ref{Intr2}. A common strategy to analyze the long-time evolution 
of flows is to design a suitable continuity principle in terms of appropriate ``continuity" variables that could ``describe"\footnote{Note
that ``describe"  in GR means ``describe the space-time" an not only the flow. This fact makes GR a particularly difficult theory to 
treat only from the point of view of dynamical flows.}
the flow when at least one of the variables breaks down. In other words, and being slightly ambiguous, if one defines the notion of
{\it singularity} as equivalent to the blow up of the ``continuity" variables, then no extra information other than the one provided by 
the variables themselves, would be required to analyze the structure of the singularities. In still vague but heuristic terms, one would like to have 
a ``complete set of continuity variables". The right choice of the ``continuity" variables is thus of 
central importance. Indeed this is one of the major obstacles to understand the long-time evolution of the Einstein flow. Although it is
unlikely that the quantities $\nu$, ${\mathcal{V}}$ and 
${\mathcal{E}}_{\bar{I}}$ are best adapted for this purpose, we believe they conform a interesting set, 
deserving to be considered with care and 
in depth. A related problem of considerably technical as well as conceptual difficulty, 
is whether ${\mathcal{E}}_{0}=Q_{0}$ can be used instead of ${\mathcal{E}}_{1}=Q_{0}+Q_{1}$ in Theorem \ref{Intr2}. This problem is known 
(with variations in the formulation) as the $H^{2}$-conjecture. Its difficulty lies however well beyond the scope and the techniques developed in 
this article. Still some of the estimates here developed may be of tangential interest in this goal. 

We require, as part of the definition of {\it flow solution} (Definition \ref{g} in Section \ref{3.4}), 
that a ($H^{3}$) flow $(g,N,X)$ (of three-metrics, lapse and shift) is in fact the flow induced by a CMC foliation on a ($H^{3}$) space-time 
$({\bf M},{\bf g})$ (see Definition \ref{STsol} in Section \ref{3.4}). It follows from Theorem \ref{Intr2} that an initial data $(g_{0},K_{0})$ in 
$H^{3}\times H^{2}$ gives rise to a unique flow solution in a given shift $X$ or, equivalently, 
a unique space-time $({\bf M},{\bf g})$ up to space-time diffeomorphisms. We note that the equivalence between flow and space-time solutions 
is not always imposed as a requirement in some treatments of the initial value formulation in General Relativity.     

As the variables $\nu$, ${\mathcal{V}}$ and $Q_{0}$ are intrinsic to the space-time, we can see that Theorem \ref{Intr2} 
is in fact independent of which shift $X$ we take. Indeed, 
our approach to the Einstein flow is independent of the shift. For this reason we have introduced a general notion of {\it admissible shift},
made up (in some sense) out of the minimum requirements that a function $X(g,K)$ must satisfy to be the shift of an Einstein CMC flow. 
We may well take $X=0$ (which is an admissible shift) all though the article.

Let us give now an overview of the contents of the article. In the background section we provide a detailed account on the notions of: 
Einstein flows (Section \ref{2.1A}), the CMC gauge (Section \ref{2.2}), Weyl fields and Bel-Robinson energies (Section \ref{2.5}) and
scaling (Section \ref{2.6}). Although most of the article is based on standard functional analysis, we introduce some new terminology 
(as {\it canonic atlas}) that needs to be presented with care. This is done in Section \ref{2.4}. Section \ref{3} contains the core of the article.
In Sections \ref{3.1} and \ref{3.2} we discuss the notion of {\it reduced volume} and give a cosmological interpretation of its monotonicity.
The fact that the reduced volume is monotonically decreasing in the expanding direction has important consequences 
for flows over three-manifolds $\Sigma$ with non-positive Yamabe invariant. One can read an easy implication of the monotonicity 
in {\it item 3} of Theorem \ref{Intr2}. In fact, because ${\mathcal{V}}$ is decreasing, it remains bounded to the future, and therefore
we can dispense with it in the continuity criteria. In Section \ref{3.3} we show how the variables $\nu$, ${\mathcal{V}}$ and ${\mathcal{E}}_{I}$
control CMC states $(g,K)$. To clarify the difference between intrinsic estimates and estimates of elliptic type, 
we have divided the section into two subsections, Sections \ref{3.3.1} and Section \ref{3.3.2}.
In Section \ref{3.3.1} we introduce the intrinsic estimates. These estimates are in terms of ${\mathcal{V}}$ and $Q_{0}$ only. In Section
\ref{3.3.2} we introduce the estimates of elliptic type. This necessitates of a laborious treatment of the currents ${\bf J}(\W_{i})$
associated to the Weyl fields $\W_{i}$. It ends up with a proof of Lemma \ref{Intr1}. Section \ref{3.4} is essentially an extrapolation of the
analysis done in Section \ref{3.3} for single states $(g,K)$ but now for flows $(g,K)(k)$. We introduce a BR-norm
and show how it controls the flow and the space-time. In Section \ref{3.4} too, we introduce the notion of a dynamical harmonic atlas. 
We spend much of Section \ref{3.4} analyzing this concept. The dynamical harmonic atlas is the necessary ingredient to 
show that flow solutions are actually space-time solutions. Section \ref{3.5} is devoted to the initial value formulation. 
All the treatment is
entirely in terms of Bel-Robinson energies and Weyl fields. Finally, we present in Section \ref{3.6} some partial results on {\it long time}
CMC flows over manifolds with non-positive Yamabe invariant. A long time CMC flow is one for which the range of $k$ contains an
interval of the form $(k_{0},0)$\footnote{It is conjectured that any flow over a manifold $\Sigma$ with non-positive Yamabe invariant is in effect
a long time flow \cite{Re} (see also \cite{AL} and refferences therein).}. What we prove is the following
\begin{T} Say $Y(\Sigma)\leq 0$ and $({\bf M}\sim \Sigma\times \field{R},{\bf g})$ a smooth $C^{\infty}$ maximally globally hyperbolic 
space-time. Say $(g,K)(k)$ with $k\in [a,b)$
is a CMC flow where $b$ is the lim sup of the range of $k$ and say ${\mathcal{E}}_{1}\leq \Lambda$. Then
if $({\bf M},{\bf g})$ is future geodesically complete the CMC flow is a long time flow i.e. $b=0$.
\end{T}
 
Except in a few exceptions, all the article deals with the {\it vacuum} Einstein theory. In those cases where matter is considered, we will do so assuming that the energy-momentum tensor satisfies the dominant energy condition.     

{\center \section{Background.}\label{2}}

{\center \subsection{The (vacuum) Einstein flow.}\label{2.1A}}

All manifolds and tensors in this
section and the next (Section \ref{2.2}) are considered to be smooth 
i.e. $C^{\infty}$. We will
denote space-times (i.e. formally a four manifold ${\bf M}$ and a Lorentz
$(-+++)$ metric ${\bf g}$) by ${\bf (M,g)}$. All through, 
space-time tensors will be boldfaced. 
We will think the Einstein equations in vacuum
\ben
{\bf Ric} -\frac{1}{2}{\bf Rg}=0, (\ {\rm or}\ {\bf Ric=0}),
\een

\noindent not as global equations on (a given) ${\bf M}$ but rather as
evolution equations on manifolds of the form
${\bf M}\sim\Sigma\times {\field R}$, where $\Sigma$ is an orientable and
compact three-manifold\footnote{Solutions in this kind of topology are called {\it cosmological
solutions}. As usual, hypersurfaces diffeomorphic to $\Sigma$
having Riemannian inherited metrics $g$ are called {\it Cauchy hypersurfaces}}.
 
Say ${\bf M}$ is 3+1-{\it split} by the diffeomorphism $\phi:
\Sigma\times {\field R}\rightarrow {\bf M}$, in a way that
$\phi^{*}({\bf g})|_{\Sigma\times \{t\}}=g(t)$ is Riemannian for every $t$
($t:{\bf M}=\Sigma\times {\field R}\rightarrow {\field R}$ is the {\it time
function} coming from projecting into the second factor). Writing
$\partial_{t}=NT+X$ with $T$ a normal to the time foliation and $X$
tangential to it, the space-time metric splits into
$\Sigma$ and $\partial_{t}$ components as
\begin{equation}\label{metric}
\phi^{*}{\bf g}=-(N^{2}-|X|^{2})dt^{2}+dt\otimes X^{*}+X^{*}\otimes dt + g.
\end{equation}

\noindent In the formula above $g$ was extended to ${\bf M}$ to be zero along
$\partial_{t}$ (i.e. zero when one of the entrances is
$\partial_{t}$) and $X^{*}_{a}=X^{b}g_{ab}$. $N$ is called the {\it lapse}
and $X$ the {\it shift vector}. As seen in a 3+1-splitting, a space-time metric
${\bf g}$ is characterized uniquely by a {\it flow} (a path) of three-metrics
$g(t)$ and lapse-shift $(N,X)(t)$ on a fixed manifold $\Sigma$. We will denote
${\bf g}_{\phi}(t)=((g),(N,X))(t)$ and call it the {\it Einstein flow} in the {\it $\phi$-splitting}\footnote{From 
a Lagrangian point of view one studies the flow ${\bf
g}(t)=((g,v),(N,X))(t)$ of position $g$,
velocity $v=\dot{g}={\mathcal{L}}_{\partial_{t}}g=-2NK+{\mathcal{L}}_{X}g$ ($K$
is the
second fundamental form of the time foliation, note the sign convention) and
lapse-shift $(N,X)$ as a solution to
the Euler-Lagrange equations of the Einstein-Hilbert action (see \cite{W}) 
${\mathcal{S}}=\int_{\Sigma\times I} {\bf R}dv_{{\bf g}}+2\int_{\partial (\Sigma\times
I)}kdv_{g}$. $tr_{g}K=k$ is the {\it mean (extrinsic)
curvature}. Alternatively,
from a Hamiltonian point of view one studies
the evolution of the flow ${\bf g}_{\phi}=((g,P),(N,X))$,
of position $g$, momentum $P=(K-kg)dv_{g}$ and
lapse-shift $(N,X)$ as a
solution to the Hamilton-Jacobi equations of the Hamiltonian (make $P=(K-kg)$)
${\mathcal{H}}=\int_{\Sigma}N(|P|^{2}-\frac{p^{2}}{2}-R)-2X.(\nabla.P)dv_{g},$ 
generated by the {\it energy and momentum constraints functions} 
${\mathcal{E}}=|K|^{2}-k^{2}-R$ and ${\mathcal{P}}=-2\nabla.(K-kg)$, which are
well know to be conserved along the flow
and identically zero (in vacuum).}. It is convenient to include the second fundamental form of the time foliation
as part of the definition of the Einstein flow, thus ${\bf g}_{\phi}=((g,K),(N,X))(t)$. The {\it Einstein flow equations}
in vacuum are
\beq\label{1.1}
R=|K|^{2}-k^{2},
\eeq
\beq\label{2.1}
\na K-dk=0,
\eeq
\beq\label{3.1}
\dot{g}=-2NK+{\mathcal{L}}_{X}g,
\eeq
\beq\label{4.1}
\dot{K}=-\nabla\nabla N+N(Ric +kK-2KK)+{\mathcal{L}}_{X}K.
\eeq

\n Equations (\ref{1.1}) and (\ref{2.1}) are the {\it constraint equations} and equations (\ref{3.1}) and (\ref{4.1}) the {\it Hamilton-Jacobi} 
equations of the flow. Contracting (\ref{4.1}) and using the constraints we get the {\it lapse equation}
\beq\label{le1t}
-\Delta N+|K|^{2}N=\dot{k}-X(k)=N\bn_{T} k.
\eeq

\n This equation can be seen too as the second variation of volume at the leaves of the time foliation.

It is well known that the constraint
equations comprise the necessary and sufficient
conditions on initial states $(g,K)$ to have solutions to the Cauchy
problem for the Einstein flow equations. Given a solution ${\bf
(M,g)}$ of the Einstein
equations and a diffeomorphism $\phi$, then the flow ${\bf g}_{\phi}$ is a solution of the Einstein flow equations, 
in particular two different diffeomorphisms $\phi_{1}$ and $\phi_{2}$ that
coincide and have same differential over an initial Cauchy hypersurface and therefore 
inducing the same initial states $((g_{0},K_{0}),(N_{0},X_{0}))$ give rise to two different solutions 
${\bf g}_{\phi_{1}}$ and ${\bf g}_{\phi_{2}}$ on
$\Sigma\times {\field R}$. Their associated space-time metrics from equation (\ref{metric}) are however isometric. 
This is the only freedom, i.e. module diffeomorphisms, solutions are unique. Note that we can always adjust the diffeomorphism over an initial Cauchy
hypersurface to make N and X have any prescribed values (with $N^{2}>|X|^{2}$).
We now state the well known theorem of well posedness of the 
$C^{\infty}$ Cauchy problem (see for instance \cite{HE},\cite{W} and references therein).

\vspace{0.2cm}
\begin{T} {\rm (Well posedness of the $(C^{\infty})$ Cauchy problem)} (Existence)
Given and initial state $(g_{0},K_{0})$ satisfying the energy and momentum
constraints and arbitrary initial lapse-shift
$(N_{0},X_{0})$, there is a solution to the Einstein flow equations
(\ref{1.1})-(\ref{4.1}) over a
short period of a parametric time.
(Uniqueness) Two globally hyperbolic solutions ${\bf g}_{1}$, ${\bf g}_{2}$ to the
Einstein flow equations with the same initial state $(g_{0},K_{0})$ and same
initial lapse and shift $(N_{0},X_{0})$ are isometric i.e.
there exists ${\bf (M,g)}$ and diffeomoprhisms $\phi_{1}$ and $\phi_{2}$ such
that ${\bf g}=\phi^{*}({\bf g_{1}})=\phi^{*}({\bf g_{2}})$.
\end{T}
\vspace{0.2cm}

\noindent The freedom in the choice of $\phi$, i.e. the invariance of solutions
under diffeomorphisms is called {\it gauge freedom}. A choice of $\phi$ from ${\bf g}$ in such a way that if
${\bf g}_{1}$ is isometric to ${\bf g}_{2}$ then $\phi({\bf g}_{1})=\phi({\bf g}_{2})$ is said {\it a choice of gauge}.

{\center \subsection{The CMC gauge.}\label{2.2}}

A state $(g,K)$ satisfying the energy and momentum
constraints with $k=tr_{g}K$ constant is called a CMC state. A cosmological
solution on $\Sigma\times {\field R}$ is in the CMC
gauge if the extrinsic mean curvature is constant restricted to the leaves $\Sigma\times \{t\}$ of
the time foliation.

In itself the CMC gauge is only temporal, i.e. it fixes only the time foliation but does not fix the freedom by
diffeomorphisms leaving the CMC foliation invariant. 

Assume that a vacuum cosmological solution $({\bf M,g})$ has a CMC Cauchy
hypersurface. Then, unless the solution is flat and of the form ${\bf g}=-dt^{2}+g_{F}$ where $g_{F}$ is a flat metric
over a three-manifold $\Sigma$, it must happen: a) there is a CMC foliation in at least a small
neighborhood around it and b) if two CMC Cauchy hypersurfaces have different
constant
mean curvatures then they are disjoint and there is a CMC foliation inside the
enclosed region on which $k$ varies monotonically (for a proof see for instance \cite{Ge}, \cite{Ba1}, for a discussion see \cite{Re}). 

Let us explain how property a) is proved as that is relevant for the rest of the article. 
The basic tool is the technique of barriers \cite{Ba1}, it guarantees 
the existence of a CMC slice of constant mean curvature $k$ between two slices with mean curvature 
functions $k_{1}$ and $k_{2}$ with $sup\ k_{1}<k<inf\ k_{2}$. This property settles the existence of a CMC foliation
around a CMC slice with $k\neq 0$. In fact starting at the given CMC slice $\Sigma$ consider the foliation
around it provided by the Gauss gauge with zero shift, i.e. with $N=1,X=0$. From equation (\ref{le1t}) 
one gets $\dot{k}=|K|^{2}>\frac{k^{2}}{3}$.
Thus $k$ is strictly monotonic and we can conclude the existence of two barriers having mean curvatures 
$k_{1}$ and $k_{2}$ with $sup\ k_{1}<k< inf\ k_{2}$. To show the existence of barriers around a state $\Sigma$ 
with mean curvature zero proceed as follows. If $\int_{\Sigma}|K|^{2}dv_{g}\neq 0$ at the initial slice then we can
solve the lapse equation $-\Delta N+|K|^{2}N=1$. In this case consider the normal field $\tilde{T}=NT$ and consider the flow of
geodesics starting at the slice and having velocities $\tilde{T}$ there. This provides a gauge at least on a small time
interval. The key fact is that at the initial slice we have $\dot{k}=1$ and therefore $k$ is strictly monotonic on a small
time interval around the initial slice and in that gauge. This provides suitable barriers and therefore a CMC foliation around the
initial CMC slice. If instead $K=0$ at the initial slice consider again the Gauss gauge starting from it. Let us restrict
to the future direction. Unless $K=0$ over a small time interval (and in the Gauss gauge) we have from the equation 
$\dot{k}=|K|^{2}$ 
that any other slice except the initial it is $K\neq 0$ and $k\geq 0$. We can solve therefore the lapse equation at a time after the 
initial time 
and proceeding as in the previous case get a slice with $inf\ k>0$. Doing the same 
in the opposite time direction we guarantee the existence of barriers and therefore the existence of a 
CMC foliation around the initial state. If it happens that $K=0$ on a small time interval (in the Gauss gauge) then
from equation (\ref{4.1}) we get that $Ric=0$ on a small time interval and therefore the solution is flat and of the form 
${\bf g}=-dt^{2}+g_{F}$ where $g_{F}$ is a flat metric over a three-manifold $\Sigma$.      

Fact a) above shows that
any CMC state (with $k\neq 0$) induces a (at least short time) solution
of the CMC flow, while b) shows that the CMC flow is unique and the maximal range of $k$ forms a connected
open interval over which $k$ varies monotonically. The range of $k$ depends on the topology of $\Sigma$ and more in particular on the Yamabe
invariant $Y(\Sigma)$ of $\Sigma$\footnote{The Yamabe invariant $Y(\Sigma)$ of a three-manifold $\Sigma$ is defined as the supremum of
the scalar curvatures of unit volume Yamabe metrics, where Yamabe metrics are metrics of constant scalar curvature minimizing the Yamabe functional. 
The Yamabe invariant is also called the {\it sigma constant} \cite{FM1},\cite{A5}.}. It is
conjectured \cite{Re} that if $Y(\Sigma)\leq 0$ then the range of $k$ is $(-\infty,0)$ while if $Y(\Sigma)>0$ then it
is $(-\infty,+\infty)$. The CMC foliation is conjectured to cover the maximal globally hyperbolic solution if 
$Y(\Sigma)>0$
while if $Y(\Sigma)\leq 0$ it is conjectured to cover the maximal globally hyperbolic solution towards the past but
not necessarily towards the future. In this last case the CMC foliation should avoid the singularities but nothing else than that 
(every point lying outside the CMC foliation should have all its future inextendible time-like geodesics extinct in a 
uniform interval of proper time). 

Over a CMC foliation the function $k$ is smooth except possibly at a maximal slice (i.e. a slice with $k=0$). 
The mean curvature $k$ is a smooth function at a maximal slice if $\int_{\Sigma}|K|^{2}dv_{g}\neq 0$ at the slice. 
If we take the mean curvature $k$ as time (when $k\neq 0$) we get the lapse equation
\beq\label{le2}
-\Delta N+|K|^{2}N=1.
\eeq

\n which is independent of the shift.

Finally a note about the set of CMC solutions. Not every cosmological solution $({\bf M,g})$ admits a 
Cauchy hypersuface of constant mean curvature \cite{CIP} although there are sufficient
conditions for that to be \cite{Ba2}. On
the other hand every three-manifold admits so called {\it Yamabe}
CMC initial states, i.e. states $(g_{Y}, K_{Y})$ with $R_{g_{Y}}=-6,\
K_{Y}=-g$. Therefore there are CMC solutions in any $\Sigma\times {\field R}$ topology.

A CMC solution over a manifold with non-positive Yamabe invariant and with the range of $k$ containing an interval of the form
$(a,0)$ is said a {\it long time CMC solution} (towards the future $k$ exhausts its potential range). 

\vspace{0.2cm}
{\center \subsection{The CMC flow Equations in a general material setting.}\label{2.3}}

When matter is present, the Einstein equations 
\ben
{\bf Ric}-\frac{1}{2}{\bf R g}=8\pi G {\bf T},
\een

\noindent in flow form are
\beq
R-|K|^{2}+k^{2}=16\pi G\rho,
\eeq
\beq
\nabla K=-8\pi G J,
\eeq
\beq\label{flowmatter}
\dot{g}=-2NK+{\mathcal{L}}_{X}g,
\eeq
\beq\label{flowmatter2}
\dot{K}=-\nabla\nabla N+N(Ric +kK-2KK)+{\mathcal{L}}_{X}K-N8\pi G({\bf T}-\frac{1}{2}tr_{g}{\bf T}{\bf g}),
\eeq

\noindent where ${\bf T}(T,T)=\rho$ and ${\bf T}(T,.)=J$ and $p$ is the average of the principal pressures (thus
$-\rho+3p=tr_{{\bf g}}{\bf T}$). The last term in the right hand side (RHS) of equation (\ref{flowmatter2}) (involving ${\bf T}$) 
should be restricted to $\Sigma$. The lapse equation for the CMC time $t=k$ is
\ben
-\Delta N+(4\pi G(\rho+3p)+|K|^{2})N=1.
\een
\vspace{0.2cm}
\begin{center}
{\bf\large Some useful terminology.}
\end{center}

\vspace{0.2cm}
We have been using informally the notion of ``control". It is convenient to use this terminology indeed, and we will appeal to 
it many times in the rest of the article. In an informal manner, ``A" ``controls" ``B" if B cannot degenerate when A is not degenerate. For instance,
$\nu$, ${\mathcal{V}}$ and $Q_{0}$ control the ``geometry" of CMC states. Quantitatively, (and being precise), a set of quantities $A_{1},\ldots A_{n}$ control
a quantity B if for every constant $M$ there is $N$ such that $|B|\leq N$ if $|A_{1}|\leq M,\ldots,|A_{n}|\leq M$. Another more familiar 
way to mean control is through the expression 
$|B|\leq C(A_{1},\ldots,A_{n})$. Still we will abuse slightly this definition an use it in a 
more flexible way. For instance we may say: ``$\nu$, ${\mathcal{V}}$, $Q_{0}$ control $r_{2}$", although properly speaking one must 
say: ``$1/\nu$, ${\mathcal{V}}$ and $Q_{0}$ control $r_{2}+1/r_{2}$. The reader should find evident the meaning of each expression. 

{\center \subsection{Manifolds and Sobolev spaces.}\label{2.4}}

Given a subset $\Omega$ of $\field{R}^{n}$ covered by a chart $\{x^{k}\}$, we denote by $H^{s}_{\{x\}}(\Omega)$ the $s$-Sobolev space 
of distributions with derivatives in $L^{2}$ until order $s$, and where the derivatives are the standard partial derivatives in the coordinate
system $\{x^{k}\}$.  
   
We say that the pair $(\Sigma,{\mathcal{A}})$ is a $H^{i+1}${\it-manifold} if and only if ${\mathcal{A}}$ is an atlas covering $\Sigma$
with transition functions in $H^{i+1}$ (more precisely if $\{x_{\alpha}\}$ is a chart from on a domain $\Omega$\footnote{$\Omega$ will always be an open set with smooth boundary.} 
and $x_{\beta}$ is
a chart from a domain $\Omega'$ then $x^{i}_{\alpha}(x^{j}_{\beta})$ are functions on $H^{i+1}_{\{x_{\beta}\}}(\Omega\cap \Omega')$). We say that
$(\Sigma,{\mathcal{A}},g)$ is a $H^{i+1}$-{\it Riemannian manifold} if the entrances $g_{ij}$ in any coordinate chart $\{x_{\alpha}\}$ 
are in $H^{i}_{\{x_{\alpha}\}}(\Omega)$ where $\Omega$ is the domain of the chart. 

Let $\{B(o_{\alpha},r),\alpha=1,\ldots,m\}$ be a covering of $\Sigma$ such that every one of the balls $B(o_{\alpha},r)$ is inside
one of the charts in ${\mathcal{A}}$. Say $\{\xi_{\alpha},\alpha=1,\ldots,m\}$ is a partition of unity subordinate to the covering 
$\{B(o_{\alpha},r)\}$ (i.e. $\xi_{\alpha}(p)=1$ if $p\in B(o_{\alpha},r/2)$, $\xi_{\alpha}(p)=0$ if $p\in B(o_{\alpha},r)^{c}$,
$\xi_{\alpha}\geq 0$ and $\sum_{\alpha}\xi_{\alpha}(p)=1$ for all $p\in \Sigma$). Recall that in $\field{R}^{n}$ the Sobolev space 
$H^{s}_{\field{R}^{n}}$ is defined as those distributions for which the
Fourier transform $\hat{u}(\varsigma)$ satisfies that $\int_{\field{R}^{n}}(1+|\varsigma|)^{2s}|\hat{u}(\varsigma)|^{2}d\varsigma$ is finite. 
Thinking that the charts $\{x_{\alpha}\}$ are coordinates in (a subset of) $\field{R}^{n}$, consider the tensors $U$ of rank $(l,l')$  and 
having entrances in $H^{s}$ for $s\leq i$ and $i\geq 2$. Then the Sobolev space $H^{s}_{{\mathcal{A}}}$ of tensors of rank $(l,l')$ is 
defined as the set of $U=\sum_{\alpha=1}^{\alpha=m}\xi_{\alpha}U_{\alpha}$ where $U_{\alpha}$ is a tensor as mentioned before. 
The $H^{s}_{{\mathcal{A}}}$ norm
is defined by $\|U\|_{H^{s}_{\mathcal{A}}}=\sum_{\alpha=1}^{\alpha=m}\|\xi_{\alpha}U_{\alpha}\|_{H^{s}_{\{x_{\alpha}\}}}$. As topological
spaces they are independent of the atlas ${\mathcal{A}}$ (as long as the atlas are compatible). 

The $H^{i}$-{\it harmonic radius} $r_{i}(o)$ at $o$ in a $H^{i+1}$-Riemannian three-manifold $(\Sigma,{\mathcal{A}},g)$, $i\geq 2$, 
is defined as the supremum of the radius $r$ for which there is a coordinate chart $\{x\}$ covering $B(o,r)$ 
and satisfying
\begin{equation}\label{cc}
\frac{3}{4}\delta_{jk}\leq g_{jk}\leq \frac{4}{3}\delta_{jk},
\end{equation}
\begin{equation}\label{ccc}
\sum_{\alpha=2}^{\alpha=i}r^{2\alpha-3}(\sum_{|I|=\alpha,j,k}\int_{B(o,r)}|\frac{\partial^{I}}{\partial x^{I}}g_{jk}|^{2}dv_{x})\leq 1,
\end{equation}

\noindent where in the sum above $I$ is the multindex $I=(\alpha_{1},\alpha_{2},\alpha_{3})$, and as usual 
$\partial^{I}/\partial x^{I}=(\partial_{x^{1}})^{\alpha_{1}}(\partial_{x^{2}})^{\alpha_{2}}(\partial_{x^{3}})^{\alpha_{3}}$. 
Both expressions above are invariant under the simultaneous scaling $\tilde{g}=\lambda^{2}g$, $\tilde{x}^{\mu}=\lambda x^{\mu}$ and 
$\tilde{r}=\lambda r$. Observe that if $j>i\geq 2$ then $r_{j}(o)\leq r_{i}(o)$. 
Define the $H^{i}$-{\it harmonic radius} $r_{i}$ to be the infimum of $r_{i}(o)$ for all $o$ in $\Sigma$. 
The harmonic radius marks the scale at which the metric is seen in balls of radius one controlled in 
$H^{i}_{\{x\}}$ by one. Atlas made of harmonic charts will be called {\it harmonic atlas}. Still one wants to have a more standardized notion
of harmonic atlas. For that we define a $H^{i}$-{\it canonic harmonic atlas} as one having the least number of 
harmonic charts for which: every chart $\{x_{\alpha}\}$ is defined on a ball $B(o,r_{i}/2)$ and can be extended to a chart on $B(o,r_{i})$
having the properties (\ref{cc})-(\ref{ccc}). Also every ball $B(o',(3/8)r_{i})$ (for arbitrary $o'$) 
must lie inside one of the charts. 

When we consider norms with respect to a canonic harmonic atlas $\atlas$, it is necessary, in order for them to be useful, that the 
partition function with respect to which the norms are defined (see above) is not unnecessarily degenerate. More precisely the 
$H^{i+1}_{\{x_{\alpha}\}}$-norm of a function $\xi_{\alpha}$ with support in a chart $\{x_{\alpha}\}$ in a $H^{i}$-canonic harmonic atlas 
$\atlas$ does not have to be big if it can be small. One way to implement the criteria is to consider functions $\xi_{\alpha}$ 
defined canonically as follows. Pick a $C^{\infty}$ non-negative function $\chi$ of one variable, being one inside the interval $[0,3/4]$ and 
zero inside the interval $[1,\infty]$. Given a chart $\{x_{\alpha}\}$ in a $H^{i}$-canonic harmonic atlas $\atlas$, define 
$\chi_{\alpha}(x)=\chi(\frac{r_{i}}{2}|x|)$, where $|x|$ is the radial coordinate from the center $o_{\alpha}$. Finally define 
\ben
\xi_{\alpha}(x)=\frac{\chi_{\alpha}(x)}{\sum_{\beta} \chi_{\beta}(x)}.
\een

Let $(\Sigma,g)$ be a $H^{i+3}$-Riemannian manifold, $i\geq 0$. As it is usual one can define the $H^{j}$-Sobolev norm of a tensor field $U$ intrinsically using 
the covariant derivative $\nabla$ associated to $g$ by
$\|U\|^{2}_{H^{j}_{g}}=\int_{\Sigma}\sum_{m=0}^{m=j}|\nabla^{m}U|^{2}dv_{g}$. The $H^{j}_{\atlas}$ and the $H^{j}_{g}$ Sobolev spaces are
equivalent for $i\geq 0$.\footnote{The case $i=0$ needs a bootstrapping argument to show that if $\nabla U=\partial U +\Gamma*U$
and $U$ are in $L^{2}$ then $\partial U$ is also in $L^{2}$.} 

All through the article we will consider pointwise products of tensors lying in different Sobolev spaces and whose product we would like to
think in a third Sobolev space. All the mathematics we will need in this respect is the following fact (see \cite{AM1} page 5 and references therein).
Let $U$ and $V$ be two tensors over a $H^{i+3}$-Riemannian three-manifold $(\Sigma,g)$. Assume $U$ is in $H^{s_{1}}_{\atlas}$ and $V$ is in $H^{s_{2}}_{\atlas}$, 
with $i+2\geq s_{1}\geq 0$, $i+2\geq s_{2}\geq 0$\footnote{The condition that $s_{k}\leq i+2$ for $k=0,1$ is to guarantee that the Sobolev 
norms $H^{s_{k}}_{\atlas}$ are well defined globally over $\Sigma$.} and one of the $s_{1}$ or $s_{2}$ is strictly greater than zero. Say 
$s\leq min(s_{1},s_{2},s_{1}+s_{2}-3/2)$. Then
\beq\label{SP}
\|U*V\|_{H^{s}_{\atlas}}\leq C(r_{i+2},Vol)\|U\|_{H^{s_{1}}_{\atlas}}\|V\|_{H^{s_{2}}_{\atlas}},
\eeq

\n where $\atlas$ is a $H^{i+2}$-canonic harmonic atlas. Thus for example if $s_{1}\geq 2$ we have independently of $s_{1}$, that the 
pointwise product $U*V$
is in $H^{s_{2}}_{\atlas}$ and its norm can be estimated from above by (\ref{SP}). On the other hand if $s_{1}=s_{2}=1$ the pointwise 
product of $U$ and $V$ lies in $H^{0}_{\atlas}$ and its norm can be estimated by (\ref{SP}). The reader should be aware these product 
properties are used intensively (specially inside the Section \ref{3.3.2}), and most of the time without explicit mention.
 
We will recur to elliptic estimates in a number of occasions. Except for standard elliptic estimates \cite{GT} and 
the elliptic estimates stated in Proposition \ref{EEII} later in the article, all the rest will fall under the following proposition. 

\vspace{0.2cm}
\begin{Prop}\label{Pe1} Say $r<r_{i+2}(o)$, $i\geq 0$, and let $\{x\}$ be a harmonic coordinate chart covering $B(o,r_{i+2}(o))$ and satisfying (\ref{cc}) and (\ref{ccc}). 
We will consider elliptic estimates for first and second order differential equations separately.

I. (Second order elliptic operators). Consider the differential equation
\beq\label{so}
g^{ij}\partial_{i}\partial_{j}U^{m}+A^{mi}_{n}\partial_{i}U^{n}+B^{m}_{n}U^{n}=f^{m}.
\eeq

\n with $A$ in $H^{i+1}_{\{x\}}(B(o,r_{i+2}(o)))$, $B$ in $H^{i}_{\{x\}}(B(o,r_{i+2}(o)))$ and $f$ in $H^{j}_{\{x\}}(B(o,r_{i+2}(o)))$ and 
with $j\leq i$.
Suppose in addition that $\|A\|_{H^{i+1}_{\{x\}}(B(o,r_{i+2}(0))}+\|B\|_{H^{i}_{\{x\}}B(o,r_{i+2}(o))}\leq \bar{C}(r)$. 
Under these hypothesis we have that if $U$ is a $H^{j+2}$-strong solution of (\ref{so}) then
\beq\label{operar}
\|U\|_{H^{j+2}_{\{x\}}(B(o,r/2))}\leq C(r,\bar{C}(r))(\|U\|_{H^{1}_{\{x\}}(B(o,r))}+\|f\|_{H^{j}_{\{x\}}(B(o,r))}).
\eeq

\n If $B=0$ we can get an estimate of one degree higher and (\ref{operar}) is valid for $j=i+1$ as well. When $i\geq 2$ the norm 
$\|U\|_{H^{1}}$
on the RHS of equation (\ref{operar}) can be replaced by the $L^{2}$-norm of $U$.

II. (First order elliptic operators). Consider the differential equation
\beq\label{fo}
G^{mi}_{n}\partial_{i}U^{n}+A^{m}_{n}U^{n}=f^{n}.
\eeq

\n Assume $G^{mi}_{n}\partial_{i}U^{n}$ is a first order elliptic operator with coefficients
$G^{mi}_{n}$ in 

\n $H^{i+2}_{\{x\}}(B(o,r_{i+2}(o)))$ and ellipticity constants controlled by $r$. Assume also $A$ is in $H^{i+1}_{\{x\}}(B(o,r_{i+1}(o)))$
and $f$ is in $H^{j}_{\{x\}}(B(o,r_{i+2}(o)))$ with $j\leq i+1$ and finally that $\|G\|_{H^{i+2}_{\{x\}}(B(o,r_{i+2}(o)))}+
\|A\|_{H^{i+1}_{\{x\}}(B(o,r_{i+2}(o)))}\leq \bar{C}(r)$. Under these conditions we have that if $U$ is a $H^{j+1}$-strong solution, then 
\beq\label{operarII}
\|U\|_{H^{j+1}_{\{x\}}(B(o,r/2))}\leq C(r,\bar{C}(r))(\|U\|_{H^{1}_{\{x\}}(B(o,r))}+\|f\|_{H^{j}_{\{x\}}(B(o,r))}).
\eeq

\n If $i\geq 1$ then the norm $\|U\|_{H^{1}}$ on the RHS of equation (\ref{operarII}) can be replaced with the $L^{2}$-norm of $U$.
\end{Prop}

\begin{Remark} {\rm The elliptic estimates above can be made global in a concrete way. Indeed let $(\Sigma,\atlas,g)$ be a 
$H^{i+3}$-Riemanian three-manifold, where $\atlas$ is a $H^{i+2}$-canonic harmonic atlas. Then, if we take Sobolev norms with respect
to the atlas $\atlas$ we have, for instance the elliptic estimates
\beq
\|U\|_{H^{j+2}_{\atlas}}\leq C(r_{i+2},\bar{C},Vol)(\|U\|_{H^{1}_{\atlas}}+\|f\|_{H^{j}_{\atlas}}).
\eeq

\n for second order elliptic equations of the form (\ref{so}). The constant $\bar{C}$ bounds the norms of the coefficients $A$ and $B$ in 
$H^{i+2}_{\atlas}$ and $H^{i+1}_{\atlas}$ respectively. Proving that the constant $C$ in the elliptic estimate above can be made dependent 
only on $\bar{C}$, $r_{i+2}$ and $Vol$ is not difficult. In fact $r_{i+2}$ and $Vol$ control the
number of charts in a canonical atlas and the $H^{i+3}_{\{x_{\alpha}\}}$-norms of $\xi_{\alpha}$. It follows from the definition of 
$H^{s}_{\atlas}$ given above
that the global elliptic estimates are a direct consequence of the local elliptic estimates inside each one of the harmonic charts.} 
\end{Remark}

\n {\bf Proof:}

I. Say $i\geq 0$. If we prove the case $j=0$ we can proceed by induction to prove the proposition for any $j\leq i$. Indeed 
assuming it is proved for $j=j_{0}\leq i-1$, differentiate (\ref{so}) with respect to $\partial_{k}$ for $k=1,2,3$, 
use the result for $j=j_{0}$ and with $\partial_{k}U$ as $U$. Similarly the estimate when $B=0$ can be obtained from the estimate for $j=i$ by
differentiating the equation (\ref{so}) with respect to $\partial_{k}$, $k=1,2,3$ and applying the result when $j=i$ 
to the function $\partial_{k}U$ as $U$.  

We proceed with the proof when $j=0$. Let us divide the proof according to the cases $i\geq 2$, $i=1$
and $i=0$. We will forget including the domains where norms are defined. When it corresponds, for example when using standard elliptic
estimates, one must restrict domains to proper subsets. We will also forget about the subindex $\{x\}$ in the norms. We write for instance
$L^{2}$ instead of $L^{2}_{\{x\}}$. Finally $C(r)$ will denote a generic function depending on $r$.

{\it Case 1}, $i\geq 2$: inspecting the coefficients $A$ and $B$ the result follows from the standard elliptic estimates (\cite{GT}, Theorem 9.11). 
In this case we can replace the $H^{1}$-norm of $U$ on the RHS of equation (\ref{so}) by its $L^{2}$-norm.

{\it Case 2}, $i=1$: note that $\|A*\partial U\|_{L^{2}}\leq C(r)\|A\|_{H^{2}}\|U\|_{H^{1}}$ and $\|B*U\|_{L^{2}}\leq C(r)\|B\|_{H^{1}}\|U\|_{H^{1}}$.
Now apply standard elliptic estimates to the equation 
\ben
g^{ij}\partial_{i}\partial_{j}U=-A^{mi}_{n}\partial_{i}U^{n}-B^{m}_{n}U^{n}+f,
\een

\n with the RHS as a non-homogeneous term and use the estimates before. This gives
\ben
\|U\|_{H^{2}}\leq C(r,\bar{C}(r))(\|U\|_{H^{1}}+\|f\|_{L^{2}}),
\een

\n as desired.

{\it Case 3}, $i=0$: this case follows by bootstrapping. First note the following three facts

{\it 1}. If $U$ is in $L^{\alpha}$ ($\alpha>2$) then H\"older inequalities give
\ben
\|B*U\|_{L^{\beta}}\leq\|B\|_{L^{2}}\|U\|_{L^{\alpha}},\ \beta=\frac{2\alpha}{(2+\alpha)}.
\een
\ben
\|A*\partial U\|_{L^{\beta}}\leq \|A\|_{L^{6}}\|\partial U\|_{L^{\alpha_{1}}},\ \beta=\frac{6\alpha_{1}}{6+\alpha_{1}}.
\een

\vspace{0.2cm}
{\it 2}. Standard $L^{p}$ elliptic estimates applied to the equation 
\ben
g^{ij}\partial_{i}\partial_{j}U^{m}=-A^{mi}_{n}\partial_{i}U^{n}-B^{m}_{n}U^{n}+f,
\een

\n with the RHS as a non-homogeneous term, give 
\ben
\|U\|_{H^{2,\beta}}\leq C(r)(\|B*U\|_{L^{\beta}}+\|A*\partial U\|_{L^{\beta}}+\|U\|_{L^{\beta}}+\|f\|_{L^{\beta}}).
\een

\vspace{0.2cm}
{\it 3}. Sobolev embeddings give \cite{GT}:

\vspace{0.2cm}
i) if $2\beta'<3$ then $\|U\|_{L^{\alpha}}\leq C(r)\|U\|
_{H^{2,\beta'}}$ with $\alpha=\frac{3\beta'}{3-2\beta'}$,

ii) if $3+\beta>2\beta>3$ then $\|U\|_{C^{0,2-\frac{3}{\beta}}}\leq C(r)\|U\|_{H^{2,\beta}}$,

iii) $\|\partial U\|_{L^{\alpha_{1}}}\leq C(r)\|U\|_{H^{2,\beta}}$ with $\alpha_{1}=\frac{3\beta}{(3-\beta)}$,

\vspace{0.2cm}
\n Start applying {\it 1} and {\it 2} with $\alpha=6$: from 1 we get $\beta=3/2$ and with that $\beta$ we get $\alpha_{1}=2$, 
using those values in {\it 2} we get 
\ben
\begin{split}
\|U\|_{H^{2,3/2}}&\leq C(r)(\|B\|_{L^{2}}\|U\|_{L^{6}}+\|A\|_{L^{6}}\|\partial U\|_{L^{2}}+\|U\|_{L^{3/2}}+\|f\|_{L^{3/2}})\\
&\leq C(r)((\|B\|_{L^{2}}+\|A\|_{L^{6}})\|U\|_{H^{1}}+\|U\|_{H^{1}}+\|f\|_{L^{2}}).
\end{split}
\een

\n Now apply {\it 3} i) with $\beta'=\beta-1/6=4/3$ to get $\alpha=12$ and {\it 3} iii) with $\beta=3/2$ to get 
$\alpha_{1}=3$. Returning to {\it 1} and plugging in those values we get $\beta=12/7$ for $\alpha=12$ and $\beta=2$
for $\alpha_{1}=3$. We use next {\it 2} with the minimum of those $\beta$, i.e. $\beta=12/7$ to get
\ben
\|U\|_{H^{2,12/7}}\leq C(r)(\|B\|_{L^{2}}\|U\|_{L^{12}}+\|A\|_{L^{6}}\|\partial U\|_{L^{12/5}}+\|U\|_{H^{1}}+\|f\|_{L^{2}}),
\een

\n and in turn
\ben
\|U\|_{H^{2,12/7}}\leq C(r)((\|B\|_{L^{2}}+\|A\|_{L^{6}})\|U\|_{H^{2,4/3}}+\|U\|_{H^{1}}+\|f\|_{L^{2}}),
\een

\n giving thus
\ben
\|U\|_{H^{2,12/7}}\leq C(r, \bar{C}(r))(\|U\|_{H^{1}}+\|f\|_{L^{2}}).
\een

\n From {\it 3} ii) we get $\|U\|_{C^{0,1/12}}\leq C(r)\|U\|_{H^{2,12/7}}$. With that we get $\|B*U\|_{L^{2}}\leq C(r)\|B\|_{L^{2}}\|U\|_{H^{2,12/7}}$. We use
this in {\it 2} with $\beta=2$ to get
\ben
\|U\|_{H^{2,2}}\leq C(r)(\|B\|_{L^{2}}\|U\|_{H^{2,12/7}}+\|A\|_{L^{6}}\|\partial U\|_{L^{3}}+\|U\|_{H^{1}}+\|f\|_{L^{2}},
\een

\n giving
\ben
\|U\|_{H^{2,2}}\leq C(r,\bar{C}(r))(\|U\|_{H^{1}}+\|f\|_{L^{2}}),
\een

\n as desired. The elliptic estimate on first order elliptic operators follows exactly along the same lines as we 
did for second order elliptic operators.\ep

To illustrate the use of the estimates in Proposition \ref{Pe1} let us prove a well known fact that will be needed later.

\begin{Prop}\label{RICC} At any point $o$ on a compact $H^{i+3}$-Riemanninan three-manifold $\Sigma$ ($i\geq 0$), $r_{i+2}(o)$ and $\|Ric\|_{H^{i+1}_{g}(B(o,r_{i+2}(o)))}$ 
control $r_{i+3}(o)$ from below. As a consequence, for any compact $H^{i+3}$-Riemannian three-manifold $\Sigma$, 
$\|Ric\|_{H^{i+1}_{g}(\Sigma)}$ and $r_{i+2}$ control $r_{i+3}$ from below.
\end{Prop}

\n {\bf Proof:} 

Pick a harmonic coordinate chart $\{x\}$ covering $B(o,r_{i+2}(o))$ and having the properties (\ref{cc})-(\ref{ccc}). 
In harmonic coordinates the Ricci tensor $Ric$ has the expression
\ben
Ric_{\alpha\beta}=-\frac{1}{2}g^{ij}\partial_{i}\partial_{j}g_{\alpha\beta}+\partial_{*}g_{*}\partial_{*}g_{*}g^{*}g^{*}.
\een

\n The Elliptic estimates in Proposition \ref{Pe1} (where $A=\partial_{*}g^{*}g^{*}g^{*}$ and $B=0$) show that 
$\|g_{\alpha\beta}\|_{H^{i+3}_{\{x\}}(B(0,r_{i+2}(o)/2)))}$ is controlled by $\|Ric\|_{H^{i+1}_{g}(B(o,r_{i+2}(o)))}$ and 
$r_{i+2}(o)$, therefore $r_{i+3}(o)$ is controlled from below by them too.\ep 

Finally let us state the fundamental theorem of convergence in Riemannian geometry. Recall that the {\it volume radius} \cite{A4} at
a point $o$ is defined as 
\ben
\nu(o)=sup\{r/vol(B(p,s))/s^{3}\geq \mu,\ {\rm for\ all}\ B(p,s)\subset B(o,r)\},
\een 

\n where $\mu$ is an arbitrary small numeric constant. The volume radius of $\Sigma$ is defined as the infimum of $\nu(o)$
for all $o$ in $\Sigma$.

\begin{T} \label{teo:cc1}{\rm (Fundamental theorem of convergence (see \cite{Pet} and reff. therein)).} Say $(\Sigma,{\mathcal{A}},g)$ is a compact 
$H^{i+3}$-riemannian three-manifold. Then the $H^{i+2}$-harmonic radius is controlled from below by 
$Vol_{g}(\Sigma)$, $\|Ric\|_{H^{i}_{g}}$ and $\nu_{g}$. Any sequence of metrics with $Vol_{g}(\Sigma)$ and 
$\|Ric\|_{H^{i}_{g}}$ 
uniformly bounded from above and $\nu_{g}$ uniformly 
bounded from below has a subsequence converging (up to diffeomorphisms) to a $H^{i}$-metric in the weak 
$H^{i}$-topology. 
\end{T}

\begin{Remark} {\rm The use of the volume radius is not strictly necessary and it may well be substituted with other notions of
``local flatness" (scaling as a distance) better adapted to use in General Relativity and more specifically in its dynamics.}
\end{Remark}

{\center \subsection{The Bel-Robinson energy and the space-time curvature.}\label{2.5}}

In this section we introduce Weyl fields and Bel-Robinson energies. We do so without further explanations or proofs. The reader
can rely in \cite{CK} for a detailed account. We will follow it in the presentation below.

A Weyl field is a traceless $(4,0)$ space-time 
tensor field having the symmetries of the curvature tensor ${\bf Rm}$. We will denote them by ${\bf W}_{abcd}$ or simply 
${\bf W}$. As an example, the Riemann tensor in a vacuum solution of the Einstein equations is a Weyl field that we will be denoting by 
${\bf Rm}={\bf W}_{0}$ (we will use indistinctly either ${\bf Rm}$ or $\W_{0}$). The covariant derivative of a Weyl field $\bn_{X} {\bf W}$ for an arbitrary vector field $X$
is also a Weyl field. In particular $\bn_{X}^{i}{\bf W}_{0}={\bf W}_{i}$ are Weyl fields. We will be using the Weyl fields ${\bf W}_{i}$ 
with $X=T$, where $T$ is the future pointing unit normal field to the CMC foliation. 

Given a Weyl tensor ${\bf W}$ define the left and
right duals by $^{*}{\bf W}_{abcd}=\frac{1}{2}\epsilon_{ablm}{\bf W}^{lm}_{\ \ cd}$ and ${\bf W}^{*}_{abcd}={\bf W}_{ab}^{\ \ lm}
\frac{1}{2}\epsilon_{lmcd}$ respectively. It is $^{*}{\bf W}={\bf W}^{*}$ and $^{*}(^{*}{\bf W})=-{\bf W}$. Define the current ${\bf J}$ 
and its dual ${\bf J}^{*}$ by
\ben
\bn^{a}{\bf W}_{abcd}={\bf J}_{abc},
\een
\ben
\bn^{a}{\bf W}^{*}_{\ abcd}={\bf J}^{*}_{abc}.
\een

\n When ${\bf W}$ is the Riemann tensor in a vacuum solution of the Einstein equations the currents ${\bf J}$ and ${\bf J}^{*}$ are zero 
due to the Bianchi identities. This fact will be of fundamental importance latter.

The $L^{2}$-norm with respect to the foliation will be defined through the Bel-Robinson tensor. Given a Weyl
field ${\bf W}$ define the Bel-Robinson tensor by
\ben
Q_{abcd}({\bf W})={\bf W}_{alcm}{\bf W}_{b\ d}^{\ l\ m}+{\bf W}_{alcm}^{*}{\bf W^{*}}_{b\ d}^{\ l\ m}.
\een
    
\noindent The Bel-Robinson tensor is symmetric and traceless in all pair of indices and for any pair of timelike vectors $T_{1}$ and $T_{2}$,
the quantity $Q(T_{1}T_{1}T_{2}T_{2})$ is positive (provided ${\bf W}\neq 0$). 

The electric and magnetic components of ${\bf W}$ are defined as
\beq\label{EB1}
E_{ab}={\bf W}_{acbd}T^{c}T^{d},
\eeq
\beq\label{EB2}
B_{ab}=^{*}{\bf W}_{acbd}T^{c}T^{d}.
\eeq

\noindent $E$ and $B$ are symmetric, traceless and null in the $T$ direction. It is also the case that $\W$ can be reconstructed from them 
(see \cite{CK}, page 143). If ${\bf W}$ is the Riemann tensor in a vacuum solution we have
\begin{equation}\label{de}
E_{ab}=Ric_{ab}+kK_{ab}-K_{a}^{\ c}K^{c}_{\ b},
\end{equation}
\begin{equation}\label{de2}
\epsilon_{ab}^{\ \ l}B_{lc}=\nabla_{a}K_{bc}-\nabla_{b}K_{ac}.
\end{equation}

\n The components of a Weyl field with respect to the CMC foliation are given by ($i,j,k,l$ are spatial indices) 
\beq\label{EB3}
{\bf W}_{ijkT}=-\epsilon_{ij}^{\ \ m}B_{mk},\ ^{*}{\bf W}_{ijkT}=\epsilon_{ij}^{\ \ m}E_{mk},
\eeq
\beq\label{EB4}
{\bf W}_{ijkl}=\epsilon_{ijm}\epsilon_{kln}E^{mn},\ ^{*}{\bf W}_{ijkl}=\epsilon_{ijm}\epsilon_{kln}B^{mn}.
\eeq

\noindent We also have 
\ben
Q(TTTT)=|E|^{2}+|B|^{2},
\een 
\ben
Q_{iTTT}=2(E\wedge B)_{i},
\een
\ben
Q_{ijTT}=-(E\times E)_{ij}-(B\times B)_{ij}+\frac{1}{3}(|E|^{2}+|B|^{2})g_{ij}.
\een

\n The operations $\times$ and $\wedge$ are provided explicitly later. The divergence of the Bel-Robinson tensor is
\ben
\begin{split}
\bn^{a}Q({\bf W})_{abcd}=&{\bf W}_{b\ d}^{\ m\ n}{\bf J}({\bf W})_{mcn}+{\bf W}_{b\ c}^{\ m\ n}{\bf J}({\bf W})_{mdn}\\
&+^{*}{\bf W}_{b\ d}^{m\ n}{\bf J}^{*}({\bf W})_{mcn}+^{*}{\bf W}_{b\ c}^{m\ n}{\bf J}^{*}(W)_{mcn}.
\end{split}
\een

\n We have therefore
\ben
\bn^{\alpha}Q({\bf W})_{\alpha TTT}=2E^{ij}({\bf W}){\bf J}({\bf W})_{iTj}+2B^{ij}{\bf J}^{*}({\bf W})_{iTj}.
\een

\noindent From that we get the {\it Gauss equation} which will be used several times during the article
\beq\label{Gausseq}
\begin{split}
\dot{Q}(\W)=&-\int_{\Sigma}2NE^{ij}({\bf W}){\bf J}({\bf W})_{iTj}+2NB^{ij}(\W){\bf J}^{*}({\bf W})_{iTj}\\
&+3NQ_{abTT}\dt^{ab}dv_{g}.
\end{split}
\eeq

\noindent $\dt_{ab}=\bn_{a}T_{b}$ is the {\it deformation tensor} and plays a fundamental role in the space-time 
tensor algebra. In components it is
\ben
\dt_{ij}=-K_{ij}, \ \ \ \dt_{iT}=0,
\een
\ben
\dt_{Ti}=\frac{\nabla_{i}N}{N}, \ \ \ \dt_{TT}=0.
\een

The next equations are essential when it comes to get elliptic estimates of Weyl fields.
\begin{equation}\label{eq1}
div E({\bf W})_{a}=(K\wedge B({\bf W}))_{a}+{\bf J}_{TaT}({\bf W}),
\end{equation}
\begin{equation}\label{eq2}
div B({\bf W})_{a}=-(K\wedge E({\bf W}))_{a}+{\bf J}^{*}_{TaT}({\bf W}),
\end{equation}
\begin{equation}\label{eq3}
curl B_{ab}(\W)=E(\bn_{T}{\bf W})_{ab}+\frac{3}{2}(E(\W)\times K)_{ab}-\frac{1}{2}kE_{ab}(\W)+{\bf J}_{aTb}(\W),
\end{equation}
\begin{equation}\label{eq4}
curl E_{ab}(\W)=B(\bn_{T}{\bf W})_{ab}+\frac{3}{2}(B(\W)\times K)_{ab}-\frac{1}{2}kB_{ab}(\W)+{\bf J}^{*}_{aTb}(\W).
\end{equation}

\noindent The operations $\wedge,\ \times$ and the operators $Div$ and $Curl$ are defined through
\ben
(A\times B)_{ab}=\epsilon_{a}^{\ cd}\epsilon_{b}^{\ ef}A_{ce}B_{df}+\frac{1}{3}(A\circ B)g_{ab}-\frac{1}{3}(tr A)(tr B)g_{ab},
\een
\ben
(A\wedge B)_{a}=\epsilon_{a}^{\ bc}A_{b}^{\ d}B_{dc},
\een
\ben
(div\ A)_{a}=\nabla_{b}A^{b}_{\ a},
\een
\ben
(curl\ A)_{ab}=\frac{1}{2}(\epsilon_{a}^{\ lm}\nabla_{l}A_{mb}+\epsilon_{b}^{\ lm}\nabla_{l}A_{ma}).
\een

\noindent Finally we mention a couple of formulas that will be used. If $V$ a symmetric traceless $(2,0)$ tensor 
with
\ben
(div\ V)_{a}=\nabla_{b}V^{b}_{\ a}=\rho,
\een
\ben
(curl\ V)_{ab}=\frac{1}{2}(\epsilon_{a}^{\ lm}\nabla_{l}V_{mb}+\epsilon_{b}^{\ lm}\nabla_{l}V_{ma})=\sigma,
\een

\noindent then 
\begin{equation}\label{Hodgeintegral}
\int_{\Sigma}|\nabla V|^{2}+3<Ric,V\circ V>-\frac{1}{2}R|V|^{2}=\int_{\Sigma} |\sigma|^{2}+\frac{1}{2}|\rho|^{2}.
\end{equation} 

\noindent We have also
\begin{equation}\label{Simons}
d^{\nabla *}d^{\nabla}(A)+2div^{*}div(A)=2\nabla^{*}\nabla A +{{\mathcal{R}}}(A).
\end{equation}

\n where ${\mathcal{R}}$ has the expression ${\mathcal{R}}(A)=Ric\circ A+A\circ Ric-2Rm\circ A$. The expression 
$Rm\circ A$ is defined as $(Rm\circ A)_{ab}=Rm_{acbd}A^{cd}$.

\vspace{0.2cm}
\begin{center}
{\bf \large The connection coefficients.}
\end{center}

Say ${\mathcal{F}}$ is the CMC foliation. Say $X$ and $Y$ are vector fields tangent to ${\mathcal{F}}$ and commuting with $NT$. 
In this setting we have
\ben
K(X,Y)=-<\bn_{X}T,Y>,
\een
\ben
\bn_{X}Y^{b}=\nabla_{X}Y^{b}-K(X,Y)T^{b},
\een
\ben
\bn_{X}T^{b}=-X^{a}K_{a}^{\ b},
\een
\ben
\bn_{T}T^{b}=-P^{ba}\frac{\bn_{a} N}{N},
\een

\n where $P_{ab}={\bf g}_{ab}+T_{a}T_{b}$ is the horizontal projector (it projects into ${\mathcal{F}}$),
\ben
\bn_{T}X^{b}=-X^{a}K_{a}^{\ b}+\frac{\nabla_{X} N}{N}T^{b},
\een

\n and
\ben
[X,T]=-\frac{\nabla_{X}N}{N}T.
\een

{\center \subsection{Scaling and Cosmological scaling.}\label{2.6}}

Given a solution ${\bf g}$ of the Einstein equations, we say that $\lambda^{2}{\bf g}$ is the
solution ${\bf g}$ at the scale of $\frac{1}{\lambda}$ and we call $\lambda$ the scale factor. We say that a CMC state 
$(g,K)$ is {\it cosmologically scaled or normalized} if $k=-3$. This condition has a cosmological interpretation. We will see later 
that the Hubble parameter ${\mathcal{H}}$ can be identified with $\frac{-k}{3}$ and in this sense a cosmological normalized state is one
for which its Hubble parameter is equal to one. In general, space-time tensors, scale as $\lambda^{s}U$ for some weight $s$. 
The table below shows the scaling rules for some common tensors.

\vspace{0.4cm}
\begin{center}
\begin{tabular}{|c|c|c|c|c|c|c|c|c|c|}
\hline
${\bf g}$&${\bf Ric}$&${\bf T}$&$g$&$K$&$k$&$N$&$\rho$&${\bf W}_{i}$&$Q_{i}$\\
\hline $\lambda^{2}{\bf g}$&${\bf Ric}$&${\bf T}$&$\lambda^{2}g$&$\lambda K$&$\frac{k}{\lambda}$&
$\lambda^{2}N$&$\lambda^{-2}\rho$&$\lambda^{-i+2}{\bf W}_{i}$&$\lambda^{-(2i+1)}Q_{i}$\\ \hline
\end{tabular}
\end{center}   

\vspace{0.2cm}
{\center \section{The CMC flow, the volume and the BR-energy.}\label{3}}

\n {\it Assumption:} from now and for the rest of the article we will assume CMC states $(g,K)$ have $k\neq 0$ and that CMC flows are defined 
on a range of $k$ of the form $(a,b)$ with $-\infty\leq a<b\leq 0$. Also the {\it future} will be the direction
in which the volume expands.

{\center \subsection{The reduced volume.}\label{3.1}}

In this section we assume that $Y(\Sigma)\leq 0$. The reduced volume that we will use was 
introduced and used systematically in \cite{FM1} as a reduced Hamiltonian\footnote{In \cite{FM1} the reduced volume is named the 
{\it reduced Hamiltonian}.} after passing to conformal variables\footnote{In the context of the long time a quantity similar to the 
reduced volume was also used in \cite{A1}. More precisely, the ``reduced volume" in \cite{A1} was defined dynamically: say $((g,K),(N,X))$
is a solution to the CMC flow and $t_{k}=dist(\Sigma_{k_{0}},\Sigma_{k})$ where $dist$ is the Lorentzian distance between the initial CMC
Cauchy hypersurface $\Sigma_{k_{0}}$ and $\Sigma_{k}$, then define ${\mathcal{V}}=\frac{Vol_{g}(\Sigma)}{t^{3}_{k}}$. It is proved 
\cite{A1} that it is a strictly monotonic quantity unless is constant in which case the solution is a flat cone. 
The reduced volume here defined has the advantage
that it is defined on the set of all CMC states whereas the one in [A] is defined along solutions.}. It is defined (up to a constant from \cite{FM1}) 
as
\ben
{\mathcal{V}}(g,K)=-(\frac{k}{3})^{3}Vol_{g}(\Sigma).
\een

\n Recall that in any flat cone the volume of the CMC slices grow as $1/(-k)^{3}$. With that in mind we may interpret the
reduced volume as a comparison between the volume of the particular solution and the volume of
a flat cone at a given $k$\footnote{The flat cones are the solutions with maximal rate of volume expansion.}. 
We enumerate below a list of properties of ${\mathcal{V}}$ \cite{FM1}.

\begin{enumerate}

\item The derivative of ${\mathcal{V}}$ with respect to $k$ is
\beq\label{monoton}
\frac{d{\mathcal{V}}}{d k}=-(\frac{k}{3})^{2}\int_{\Sigma}N|\hat{K}|^{2}dv_{g}=-(\frac{k}{3})^{2}\int_{\Sigma}(1-\frac{Nk^{2}}{3})dv_{g}.
\eeq

\noindent When ${\mathcal{V}}'=0$ then $N=\frac{3}{k^{2}}$, ${\mathcal{V}}''=0$ and 
${\mathcal{V}}'''=-k^{4}/9\int_{\sigma}N^{3}|\hat{Ric}|^{2}dv_{g}$.
 
\item From the property before we get that ${\mathcal{V}}$ is strictly monotonically decreasing along the future CMC flow unless is 
constant in which case the solution is a flat cone.

\item ${\mathcal{V}}$ is scale invariant. 

\item The infimum ${\mathcal{V}}_{inf}$ of ${\mathcal{V}}$ in the phase space of all CMC states is given by
\ben
{\mathcal{V}}_{inf}=inf\{{\mathcal{V}}(g,K)/(g,K)\ is \ CMC\}=(-\frac{1}{6}Y(\Sigma))^{\frac{3}{2}}.
\een

\noindent where $Y(\Sigma)$ is the Yamabe invariant of $\Sigma$.

\end{enumerate}

\n Under zero shift we have the remarkable fact that 
\beq\label{rf}
\begin{split}
\frac{1}{4}\int_{\Sigma}|\dot{g}|^{2}dv_{g}&\leq\int_{\Sigma}\frac{|\dot{g}|^{2}}{4\tilde{N}}dv_{g}=\int_{\Sigma}\tilde{N}|K|^{2}dv_{g}\\
&=\int_{\Sigma}\tilde{N}R+\tilde{N}k^{2}dv_{g}=\frac{k^{2}}{3}Vol_{g}=\frac{-9}{k}{\mathcal{V}}\\
&\leq 3 t {\mathcal{V}},
\end{split}
\eeq

\n where $\tilde{N}=\frac{Nk^{2}}{3}$ is the lapse associated to the {\it cosmological time} $t=-3/k$ and $\dot{g}=\partial g/\partial t$. 
To obtain the first inequality we have used the fact that $\tilde{N}\leq 1$ and to get the equality after the fourth term 
we have used the lapse equation (\ref{le2}) multiplied by $k^{2}/3$. As ${\mathcal{V}}$ is monotonic along the future direction in a CMC solution of
the Einstein vacuum equations, the RHS of
equation (\ref{rf}) is bounded above by $3t {\mathcal{V}}(t_{0})$ where $t_{0}$ is some initial time. As for the traceless part of the time
derivative of $g$ we have
\ben
\frac{1}{4}\int_{\Sigma}|\hat{\dot{g}}|^{2}dv_{g}\leq -3\frac{d{\mathcal{V}}}{dk},
\een

\n implying
\ben
\frac{1}{4}\int_{t_{0}}^{t_{1}}\frac{\|\hat{\dot{g}}\|^{2}_{L^{2}_{g}}}{t^{2}}dt\leq {\mathcal{V}}(t_{0})-{\mathcal{V}}(t_{1}).
\een

\subsection{The Friedman-Lema\^itre equations and a cosmological interpretation of the reduced volume 
monotonicity.}\label{3.2}

The Robertson-Walker cosmologies implement the {\it cosmological principle} in its perfect form. As such it characterizes the universe 
by a radius $a$, matter density $\rho$ and the pressure $p$. The space-time is modeled by a metric
\ben
{\bf g}=-dt^{2}+a^{2}(t)g_{{\mathcal{K}}},
\een

\noindent on a manifold $\Sigma\times \field{R}=\field{R}^{3}\times \field{R}$ and where $g_{\mathcal{K}}$ is
a metric of constant sectional curvature equal to ${\mathcal{K}}$. It follows that the stress 
energy-momentum tensor is of the form ${\bf T}_{ab}=(\rho+p)T_{a}T_{b}+p{\bf g}_{ab}$ with $\rho$ and $p$
depending only on time. The compact FLRW models are simply space-like compactification of the models above. 
In the
${\mathcal{K}}=0$ case compact Cauchy hypersurfaces obtained are flat while in the ${\mathcal{K}}=-1$ case are 
hyperbolic. The Einstein equations are equivalent to the Friedman-Lema\^itre equations that we present below in integral form
to make a closer connection with the Friedman-Lema\^itre equations for arbitrary solutions that we describe later 
\ben\label{1p}
{\mathcal{H}}^{2}= \frac{\int_{\Sigma}(16\pi G\rho)dv_{g}}{6V}+-{\mathcal{K}}(\frac{V_{0}}{V})^{\frac{2}{3}},
\een
\ben\label{2p}
\frac{a''}{a}=\frac{-\int_{\Sigma}(4\pi G(\rho+3p)dv_{g}}{3V}.
\een

\noindent In the formulas above $V_{0}$ is the volume of the manifold $\Sigma$ when it is given the unique hyperbolic metric
or some flat metric according to the type of the model. In a general cosmological solution (not perfectly 
homogeneous and isotropic) the cosmological 
parameters are defined in volume average \cite{Rei},\cite{B2}\footnote{Although the averages in \cite{Rei} and \cite{B2} 
both average in volume, they are not precisely the same.}. For instance the radius $a$ is defined as
\ben
a=(\frac{V}{V_{0}})^{\frac{1}{3}},
\een

\noindent and the proper time by
\ben
\frac{d\tau}{dk}=\frac{\int_{\Sigma}Ndv_{g}}{V}.
\een

\noindent The Hubble parameter ${\mathcal{H}}$ is computed as
\ben
{\mathcal{H}}=\frac{1}{V^{\frac{1}{3}}}\frac{dV^{\frac{1}{3}}}{d\tau}=\frac{1}{V^{\frac{1}{3}}}\frac{dV^{\frac{1}{3}}}{dk}\frac{dk}{d\tau}=
\frac{1}{V^{\frac{1}{3}}}\frac{1}{3}V^{-\frac{2}{3}}(\int_{\Sigma}-Nkdv_{g})\frac{V}{\int_{\Sigma}Ndv_{g}}=\frac{-k}{3}.
\een

\noindent This inspires to give the name of ``Hubble-gauge" to the CMC gauge, as observers at the same 
``instants of time"  would measure the same Hubble parameter ${\mathcal{H}}$ (i.e. it is constant along leaves of the CMC 
foliation). Define  
$\rho={\bf T}(T,T)$ and $-\rho+3p={\bf T}_{ab}{\bf g}^{ab}$ (note that $p$ is the average of the principal pressures). 
Then the Friedman-Lema\^itre equations in this general setting  are 
\begin{equation}\label{FL1}
{\mathcal{H}}^{2}=-\frac{\int_{\Sigma}Rdv_{g}}{6V}+\frac{\int_{\Sigma}(16\pi G\rho +|\hat{K}|^{2})dv_{g}}{6V},
\end{equation}
\begin{equation}\label{FL2}
\frac{a''}{a}=\frac{-\int_{\Sigma}{\mathcal{N}}(4\pi G(\rho+3p)+|\hat{K}|^{2})dv_{g}}{3V}.
\end{equation}

\noindent Where ${\mathcal{N}}=\frac{N}{\bar{N}}$ (bar denotes volume-average) and has average equal to one. 
All the derivatives above are with respect to the averaged proper time $\tau$. The first FL equation (equation (\ref{FL1})) 
is just the average of the energy constraint
\ben
16\pi \rho=R-|\hat{K}|^{2}+\frac{2}{3}k^{2}.
\een

\noindent To get the second FL equation (equation (\ref{FL2})) we observe that
\beq\label{hp1}
(\frac{a'}{a})'=\frac{a''}{a}-(\frac{a'}{a})^{2}=\frac{a''}{a}-{\mathcal{H}}^{2},
\eeq

\noindent and
\beq\label{hp2}
(\frac{a'}{a})'=\frac{d{\mathcal{H}}}{d\tau}=-\frac{1}{3}\frac{dk}{d\tau}=-\frac{V}{3\int_{\Sigma}Ndv_{g}}.
\eeq

\noindent On the other hand from the lapse equation we get after integration
\beq\label{hp3}
\int_{\Sigma}N(4\pi G(\rho+3p)+|\hat{K}|^{2})dv_{g}=V-3{\mathcal{H}}^{2}\int_{\Sigma}Ndv_{g}.
\eeq

\noindent Equations (\ref{hp1}), (\ref{hp2}) and (\ref{hp3}) together give the equation (\ref{FL2}). Up to a constant the reduced 
volume is ${\mathcal{V}}=(3aH)^{3}$ and we see that 
\ben
({\mathcal{V}}^{\frac{1}{3}})'=(3aH)'=3a''=-\frac{a\int_{\Sigma} {\mathcal{N}}(4\pi G(\rho+3p)+|\hat{K}|^{2})dv_{g}}{V},
\een

\noindent If there is no matter present, i.e. $\rho=p=0$, the previous equation is directly equivalent to the
monotonicity formula (\ref{monoton}) for the reduced volume. This implies that the monotonicity of the reduced volume and the
well known universe deceleration are equivalent properties. Note that the reduced volume is monotonic
in presence of matter at least if $\rho+3p\geq 0$. That is the case if the average of the principal pressures is
positive or if the mass density dominates over the (possible negative) pressure.

{\center \subsection{Controlling the states $(g,K)$ at a given time.}\label{3.3}}

This section is devoted to study the intrinsic quantities $\nu$, ${\mathcal{V}}$, 
${\mathcal{E}}_{\bar{I}}=\sum_{i=0}^{i=\bar{I}}Q(\W_{i})$ and $\rho$ (if matter satisfying the dominant energy condition 
is present) as variables controlling the states $(g,K)$ at a given time $t=k$. 

{\it Assumptions}. All the results presented in this Section (Section \ref{3.3}), 
except for some exceptions explicitly indicated, 
will be stated in the context of vacuum solutions of the Einstein equations. 
Also, all the analysis will be for general three-manifolds $\Sigma$, except when we indicate that a
result is restricted to manifolds with $Y(\Sigma)\leq 0$. We will assume
that all quantities (for instance $g$, $K$ or $\W_{i}$) come from a $C^{\infty}$ solution ${\bf g}$. In this way we avoid 
justifying certain operations on tensors that would require justification in the case of low regularity. Still the results
of this section extend to the natural regularity of each statement or proposition. For instance Proposition \ref{Q} is valid
for states $(g,K)$ with $H^{2}\times H^{1}$ regularity. The same holds in Proposition \ref{1/N}, where to exemplify, 
a footnote was added
on how to prove the result for states with $H^{2}\times H^{1}$ regularity.  All the time $C$ will represent a
generic constant (the same $C$ can represent different constants in different lines). If the constant $C$ is
numeric it will be stated explicitly.     
  
The main result in vacuum will be the following.

\vspace{0.2cm}
\begin{Lem} {\rm (Sobolev norms vs. Bel-Robinson norms)}\label{SvsBR} Say $\bar{I}\geq 0$ and say $\Sigma$ is a compact
three-manifold. Then the functional (defined over cosmological normalized states $(g,K)$)
\ben
\|(g,K)\|_{BR}=\frac{1}{\nu}+{\mathcal{V}}+{\mathcal{E}}_{\bar{I}},
\een

\n controls the $H^{\bar{I}+2}$-harmonic radius $r_{\bar{I}+2}$ and the 
$H^{\bar{I}+1}_{g}$-norm of $K$.
\end{Lem}

\vspace{0.2cm}
\noindent The proof of this lemma (at the end of this section) is conceptually divided into several propositions many of them having however independent value. 
As it turns out there are a number of intrinsic estimates, namely estimates involving only intrinsic norms
$H^{s}_{g}$, worth to be mentioned separately. Intrinsic estimates give more information than what elliptic estimates give. 
The elliptic estimates are stated in terms of the norms $H^{s}_{\atlas}$ where $\atlas$ are harmonic atlas. 
To highlight the difference between them, we break the section into two subsections.

\begin{center}
\subsubsection{Intrinsic estimates.}\label{3.3.1}
\end{center} 

We start studying the control of the first order Bel-Robinson energy $Q_{0}$ over states $(g,K)$.

\vspace{0.2cm}
\begin{Prop}\label{Q} Say $\Sigma$ is a compact three-manifold. Then $Q_{0}$ and $|k|^{2}\|\hat{K}\|^{2}_{L^{2}_{g}}$ control 
$\|\nabla\hat{K}\|_{L^{2}_{g}}^{2}$, $\|\hat{K}\|_{L^{4}_{g}}^{4}$ and $\|\hat{Ric}\|_{L^{2}}^{2}$. More in particular we have 
\begin{equation}\label{K-control2}
(\int_{M}2|\nabla\hat{K}|^{2}+|\hat{K}|^{4}dv_{g})^{\frac{1}{2}}\leq C(|k|\|\hat{K}\|_{L^{2}_{g}}+Q_{0}^{\frac{1}{2}}),
\end{equation}

\n where $C$ is a numeric constant.
\end{Prop}

\n Observe the absence of the volume in equation (\ref{K-control2}) and that all norms involved are intrinsic. 

\vspace{0.2cm}
\noindent {\bf Proof:} 

Substituting $Ric=E-kK+K\circ K$, $K=\hat{K}+\frac{k}{3}g$ and $V=\hat{K}$ in equation (\ref{Hodgeintegral}) we get
\ben
\int_{\Sigma}|\nabla\hat{K}|^{2}+\frac{5}{2}|\hat{K}|^{4}-k<\hat{K},\hat{K}\circ\hat{K}>-\frac{k^{2}}{3}|\hat{K}|^{2}+
3<E,\hat{K}\circ\hat{K}>dv_{g}=\int_{\Sigma}|B|^{2}dv_{g}.
\een

\n This equation gives the bound
\begin{equation}\label{K-control} 
\int_{\Sigma}|\nabla\hat{K}|^{2} +|\hat{K}|^{4}dv_{g}\leq C\int_{\Sigma}(|k|^{2}|\hat{K}|^{2}+|k||\hat{K}|^{3}
+|\hat{K}|^{2}|E|+|B|^{2})dv_{g},
\end{equation}

\noindent Observe now that the inequalities
\ben
\int_{\Sigma}|\hat{K}|^{2}(|E|^{2}+|B|^{2})^{\frac{1}{2}}dv_{g}\leq (\int_{\Sigma}|\hat{K}|^{4}dv_{g})^
{\frac{1}{2}}Q_{0}^{\frac{1}{2}},
\een
\ben
\int_{\Sigma}|\hat{K}|^{3}dv_{g}\leq (\int_{\Sigma}|\hat{K}|^{2}dv_{g})^{\frac{1}{2}}(\int_{\Sigma}
|\hat{K}|^{4}dv_{g})^{\frac{1}{2}},
\een

\noindent transform equation (\ref{K-control}) into
\ben
2\|\nabla\hat{K}\|^{2}_{L^{2}_{g}}+\|\hat{K}\|_{L^{4}_{g}}^{4}-C(|k|\|\hat{K}\|_{L^{2}_{g}}+Q_{0}^{\frac{1}{2}})\|\hat{K}\|_{L^{4}_{g}}^{2}- C(
|k|^{2}\|\hat{K}\|^{2}_{L^{2}_{g}}+Q_{0})\leq 0.
\een

\n Now make $x^{2}=2\int_{\Sigma}|\na \hat{K}|^{2}+|\hat{K}|^{4}dv_{g}$, $a=(|k|\|\hat{K}\|_{L^{2}_{g}}+Q_{0}^{\frac{1}{2}})$ in the 
last equation. We get
\ben
x^{2}-Cax-Ca^{2}\leq 0.
\een

\n Solving for $x$ in the inequality above we get equation (\ref{K-control2}) which finishes the proof.\ep

In presence of matter satisfying the dominant energy condition (see \cite{HE} for a discussion) 
we have the following modification of the last proposition.

\begin{Prop} Say $\Sigma$ is a compact three-manifold and say $(g,K)$ is a cosmological normalized state over a Cauchy-slice 
on a solution $({\bf M, g})$ of the
Einstein equations and with matter satisfying the dominant energy condition. Then $\|\hat{K}\|_{L^{2}_{g}}$, 
$Q_{0}$ and $\|G\rho\|_{L^{2}_{g}}$ control $\|\na \hat{K}\|_{L^{2}_{g}}$, 
$\|\hat{K}\|_{L^{4}_{g}}$ and $\|\hat{Ric}\|_{L^{2}_{g}}$. More precisely we have
\begin{equation}\label{first1}
(\int_{\Sigma}2|\nabla \hat{K}|^{2}+|\hat{K}|^{4}dv_{g})^{\frac{1}{2}}\leq C(\|\hat{K}\|_{L^{2}}+Q_{0}^{\frac{1}{2}}+\|G\rho\|_{L^{2}_{g}})
\end{equation}

\n where $C$ is a numeric constant.

\end{Prop}

\n {\bf Proof:} 

If the energy-momentum tensor of matter satisfies the dominant energy condition then ${\bf T}_{TT}\geq|{\bf T}_{ij}|$ for any $i,j=1,2,3$. 
Recall that the space-time curvature ${\bf Rm}$ is decomposed in terms of $\W_{0}$ and ${\bf Ric}$ (with an expression linear
on each) and that ${\bf Ric}=8\pi G({\bf T}-\frac{1}{2}tr_{\bf g}{\bf T}{\bf g})$. We have then 
that the $L^{2}_{g}$ norm of the components of ${\bf Rm}$ are controlled by $Q_{0}+\|G\rho\|^{2}_{L^{2}_{g}}$. Using these facts the proof goes exactly parallel
to the one of Proposition \ref{Q}, but this time using instead the identities
\ben
{\bf Rm}_{iTjT}=Ric_{ij}-K_{im}K^{m}_{\ j}-kK_{ij},
\een
\ben
{\bf Rm}_{mTij}=(d^{\nabla}K)_{mij}.
\een
\ben
R=16\pi G\rho +|\hat{K}|^{2}-\frac{2}{3}k^{2},
\een
\ben
\nabla K=-8\pi G J.
\een
\ep
 
The next proposition relates $\|\hat{K}\|_{L^{2}_{g}}$ with ${\mathcal{V}}-{\mathcal{V}}
_{inf}$ or ${\mathcal{V}}$ depending 
on the signature of the Yamabe invariant $Y(\Sigma)$.

\begin{Prop}\label{Lem33}. 
Say $\Sigma$ is a compact three-manifold. Then
\vspace{0.2cm}

$\begin{array}{ll}
{\rm i)}\ {\rm if}\ Y(\Sigma)>0,\ & |k|^{2}\int_{\Sigma}|\hat{K}|^{2}dv_{g}\leq C|k|^{\frac{1}{2}}
{\mathcal{V}}^{\frac{1}{2}}(\int_{\Sigma}|\hat{K}|^{4}dv_{g})^{\frac{1}{2}},\\
{\rm ii)}\ {\rm if\ Y(\Sigma)=0},\ & |k|^{2}\int_{\Sigma}|\hat{K}|^{2}dv_{g}\leq C
|k|^{\frac{1}{2}}({\mathcal{V}}-{\mathcal{V}}_{inf})^{\frac{1}{2}}(\int_{\Sigma}|\hat{K}|^{4}dv_{g})^{\frac{1}{2}},\\
{\rm iii)}\ {\rm if\ Y(\Sigma)<0},\ & |k|^{2}\int_{\Sigma}|\hat{K}|^{2}dv_{g}\leq 
C(|k|({\mathcal{V}}-{\mathcal{V}}_{inf})+\\
&\ \ \ \ \ \ \ \ \ \ \ \ \ \ \ \ \ \ \ \ \ \ \ \ +|k|^{\frac{1}{2}}({\mathcal{V}}-{\mathcal{V}}_{inf})^{\frac{1}{2}}(\int_{\Sigma}|\hat{K}|^{4}dv_{g})^{\frac{1}{2}})\label{fe3}.
\end{array}$

\vspace{0.2cm}
\n where $C$ is a numeric constant. 

\end{Prop}

\noindent{\bf Proof:} 

i) and ii) $(Y(\Sigma)>0\ {\rm or}\ Y(\Sigma)=0$). This is immediate from the formula 
\ben
|k|^{2}\int_{\Sigma}|\hat{K}|^{2}dv_{g}\leq |k|^{\frac{1}{2}}(|k|^{3}Vol(\Sigma))^{\frac{1}{2}}(\int_{\Sigma}
|\hat{K}|^{4}dv_{g})^{\frac{1}{2}}.
\een

 iii) $Y(\Sigma)<0$. Assume $k=-3$ and let $g_{Y}$ be the unique Yamabe metric of 
constant scalar curvature $R_{Y}=-6$ in the 
conformal class of $g$. If $g=\phi^{4}g_{Y}$ then $\phi$ is determined by
\begin{equation}\label{eq:y}
-\Delta_{g_{Y}}\phi+\frac{R_{Y}}{8}\phi -\frac{1}{8}\phi^{-3}|\hat{K}|_{Y}^{2}+\frac{1}{12}k^{2}\phi^{5}=0,
\end{equation}

\noindent where $\Delta=\nabla^{2}$. The maximum principle implies (putting the values $R_{Y}=-6$
and $k=-3$) that 
\ben
6(\phi_{min}^{5}-\phi_{min})\geq \phi_{min}^{-3}|\hat{K}|_{Y}^{2}\geq 0,
\een

\noindent which makes $\phi\geq 1$. Then observe that 
\ben
-Y(\Sigma)\leq -R_{Y}(\int_{\Sigma}1\ dv_{Y})^{\frac{2}{3}},
\een

\noindent where $dv_{Y}=dv_{g_{Y}}$. This gives
\ben
\begin{split}
0&\leq 6^{\frac{3}{2}}(\int_{\Sigma}\phi^{6}-1\ dv_{Y})\leq 6^{\frac{3}{2}}\int_{\Sigma}\phi^{6}dv_{Y}
-(-Y(\Sigma))^{\frac{3}{2}}\\ &=(\frac{2}{3}k^{2}Vol(\Sigma)^{\frac{2}{3}})^{\frac{3}{2}}
-(-Y(\Sigma))^{\frac{3}{2}}.
\end{split}
\een

\noindent Therefore
\ben
\int_{\Sigma}(\phi-1)^{k}dv_{Y}\leq C({\mathcal{V}}-{\mathcal{V}}_{inf}),
\een

\noindent for $k=1,\ldots,6$. Integrating equation (\ref{eq:y}), we get
\begin{equation}\label{fe1}
6\int_{\Sigma}(\phi^{5}-\phi) dv_{Y}=\int_{\Sigma}\phi^{-3}|\hat{K}|_{Y}^{2}\ dv_{Y}.
\end{equation}

\noindent Observe that
\begin{eqnarray}
&&\int_{\Sigma}\phi^{-2}|\hat{K}|_{Y}^{2}\ dv_{Y}=\int_{\Sigma}\phi\phi^{-3}|\hat{K}|_{Y}^{2}
\ dv_{Y}\nonumber\\ &=& \int_{\Sigma} \phi^{-3}|\hat{K}|_{Y}^{2}\ dv_{Y}+\int_{\Sigma}(\phi-1)\phi^{-3}|
\hat{K}|_{Y}^{2}\ dv_{Y}\nonumber,
\end{eqnarray}

\noindent and
\begin{eqnarray}\label{fe2}
&&\int_{\Sigma}(\phi-1)\phi^{-3}|\hat{K}|_{Y}^{2}\ dv_{Y}=\int_{\Sigma}(\phi-1)\phi^{2}\phi^{-5}
|\hat{K}|_{Y}^{2}\ dv_{Y}\\
&\leq&(\int_{\Sigma}(\phi-1)^{2}\phi^{4}\ dv_{Y})^{\frac{1}{2}}
(\int_{\Sigma}\phi^{-10}|\hat{K}|_{Y}^{4}\ dv_{Y})^{\frac{1}{2}}\nonumber.
\end{eqnarray}

\noindent On the other hand note that
\begin{equation}\label{eq:rh1}
|\int_{\Sigma}(\phi-1)^{2}\phi^{4}\ dv_{Y}|\leq \int_{\Sigma}|\phi^{6}-1|+2|\phi^{5}-1|+
|\phi^{4}-1|dv_{Y}\leq C({\mathcal{V}}-{\mathcal{V}}_{inf}).
\end{equation}

\noindent Putting together equations (\ref{fe1}),(\ref{fe2}) and (\ref{eq:rh1}) we get
\begin{equation}\label{eq:equarr}
\|\hat{K}\|^{2}_{L^{2}_{g}}\leq C(({\mathcal{V}}-{\mathcal{V}}_{inf})+({\mathcal{V}}-{\mathcal{V}}_{inf})^{\frac{1}{2}}\|\hat{K}\|^{2}_{L^{4}_{g}}),
\end{equation}

\noindent which after scaling finishes the proof.\ep

Combining Propositions \ref{Q} and \ref{Lem33} we get

\begin{Prop}\label{hatK4} Say $\Sigma$ is a compact three-manifold. Then if $Y(\Sigma)>0$ we have
\ben
\int_{\Sigma}2|\na \hat{K}|^{2}+|\hat{K}|^{4}dv_{g}\leq C(|k|{\mathcal{V}}+Q_{0}),
\een

\n while if $Y(\Sigma)\leq 0$ we have
\ben
\int_{\Sigma}2|\na \hat{K}|^{2}+|\hat{K}|^{4}dv_{g}\leq C(|k|({\mathcal{V}}-{\mathcal{V}}_{inf})+Q_{0}),
\een
 
\n where $C$ is a numeric constant.

\end{Prop}

We also get

\begin{Prop}\label{lem4} Say $\Sigma$ is a compact three-manifold. If $Y(\Sigma)>0$ we have
\ben
\int_{\Sigma}|k|^{2}|\hat{K}|^{2}dv_{g}\leq C(|k|{\mathcal{V}}+(|k|{\mathcal{V}}Q_{0})^{\frac{1}{2}}),
\een

\n while if $Y(\Sigma)\leq 0$ (same for $Y(\Sigma)=0$ than for $Y(\Sigma)<0$)
\ben
\int_{\Sigma}|k|^{2}|\hat{K}|^{2}dv_{g}\leq C(|k|({\mathcal{V}}-{\mathcal{V}}_{inf})+
(|k|({\mathcal{V}}-{\mathcal{V}}_{inf})Q_{0})^{\frac{1}{2}}).
\een

\n where $C$ is a numeric constant.
\end{Prop}

\noindent {\bf Proof:} 

Combine equations in Proposition \ref{Lem33} and equation (\ref{K-control2}). Making 
$x=|k|\|\hat{K}\|_{L^{2}}$ and $a=|k|^{\frac{1}{2}}({\mathcal{V}}-{\mathcal{V}}_{inf})^{\frac{1}{2}}$ if $Y(\Sigma)\leq 0$ 
or $a=|k|^{\frac{1}{2}}{\mathcal{V}}$ if $Y(\Sigma)>0$
we arrive at the inequality $x^{2}-Cax-Ca^{2}-CaQ_{0}^{\frac{1}{2}}\leq 0$. From it we get $x^{2}\leq C(a^{2}+aQ_{0}^{\frac{1}{2}})$.\ep

A direct consequence of the propositions above is

\begin{Prop}\label{Ric} Say $\Sigma$ is a compact three-manifold, then ${\mathcal{V}}$, $|k|$ and $Q_{0}$ control $\|\hat{Ric}\|_{L^{2}_{g}}$.
In particular we have
\ben
\|\hat{Ric}\|^{2}_{L^{2}_{g}}\leq C(|k|{\mathcal{V}}+Q_{0}),
\een

\n where $C$ is a numeric constant.

\end{Prop}

\n {\bf Proof:} 

Use $\hat{Ric}=E-\frac{k}{3}\hat{K}+\hat{K}\circ \hat{K}-\frac{1}{3}|\hat{K}|^{2}g$ together with the Propositions \ref{hatK4} and \ref{lem4}.\ep

Using the energy constraint $R=|\hat{K}|^{2}-\frac{2}{3}k^{2}$ and Proposition \ref{hatK4} we get

\begin{Prop}
Let $\Sigma$ be a compact three-manifold. Then, ${\mathcal{V}}$, $|k|$ and $Q_{0}$ control 
the scalar curvature in the following way
\ben
\int_{\Sigma}|\nabla R|^{\frac{4}{3}}+R^{2}dv_{g}\leq C(|k|{\mathcal{V}}+Q_{0}),
\een

\n where $C$ is a numeric constant. 
\end{Prop}

Note that $|\nabla R|^{\frac{4}{3}}$ and $R^{2}$ scale as a {\it distance}$^{-4}$. 

\n {\bf Proof:}

Squaring the energy constraint and integrating we obtain 
\ben
\int_{\sigma}R^{2}dv_{g}\leq \int_{\Sigma}|\hat{K}|^{4}+\frac{4}{9}k^{4}dv_{g}\leq C(|k|{\mathcal{V}}+Q_{0}).
\een

\n where in the last inequality we have used Proposition \ref{hatK4}. On the other hand, differentiating the energy constraint 
we have $|\nabla R|^{\frac{4}{3}}\leq C|\nabla \hat{K}|^{\frac{4}{3}}|\hat{K}|^{\frac{4}{3}}$. Integrating and applying the H\"older
inequality we obtain
\ben
\int_{\Sigma}|\nabla R|^{\frac{4}{3}}dv_{g}\leq C(\int_{\Sigma}|\nabla \hat{K}|^{2}dv_{g})^{\frac{2}{3}}(\int_{\Sigma}|\hat{K}|^{4}dv_{g})^{\frac{1}{3}},
\een

\n and if we apply Proposition \ref{hatK4} over each one of the factors on the RHS of the last equation we obtain
\ben
\int_{\Sigma}|\nabla R|^{\frac{4}{3}}dv_{g}\leq C(|k|{\mathcal{V}}+Q_{0}),
\een

\n as desired.\ep

Because of the monotonicity of the reduced volume in the future direction we have
 
\begin{Prop} Say $Y(\Sigma)\leq 0$. Then along the future CMC evolution $Q_{0}$ controls $\|Ric\|_{L^{2}_{g}}$, $\|\hat{K}\|_{H^{1}_{g}}$ and 
$\|\hat{K}\|_{L^{4}_{g}}$.
\end{Prop}

In the presence of matter satisfying the dominant energy condition we have
\begin{Prop}
Say $\Sigma$ is a compact three-manifold. Then (over cosmological normalized states) $Q_{0}$, $\|G\rho\|_{L^{2}_{g}}$ and 
${\mathcal{V}}-{\mathcal{V}}_{inf}$ control $\|\hat{K}\|_{H^{1}_{g}}$, $\|\hat{K}\|_{L^{4}_{g}}$ and 
$\|\hat{Ric}\|_{L^{2}_{g}}$. In particular if $Y(\Sigma)\leq 0$ we have the formula (same for $Y(\Sigma)=0$ than for $Y(\Sigma)<0$)
\ben
\int_{\Sigma}|\hat{K}|^{2}dv_{g}\leq C(({\mathcal{V}}-{\mathcal{V}}_{inf})
+({\mathcal{V}}-{\mathcal{V}}_{inf})^{\frac{1}{2}}(Q_{0}^{\frac{1}{2}}+\|G\rho\|_{L^{2}}))
\een

\n while if $Y(\Sigma)>0$ we have
\ben
\int_{\Sigma}|\hat{K}|^{2}dv_{g}\leq C({\mathcal{V}}+{\mathcal{V}}^{\frac{1}{2}}(Q_{0}^{\frac{1}{2}}+\|G\rho\|_{L^{2}_{g}})),
\een

\n where $C$ is a numeric constant.

\end{Prop}

\n {\bf Proof:}

$Y(\Sigma)\leq 0$: the proof proceeds in parallel to the proof of Proposition \ref{Lem33} but this time instead using the equation
\ben
-\Delta \phi +\frac{3}{4}(\phi^{5}-\phi)=\frac{1}{8}(|\hat{K}|^{2}+16\pi G\rho)\phi^{5},
\een

\n and then the equation (\ref{first1}).

$Y(\Sigma)>0$: use equation (\ref{first1}) in the formula
\ben
k|^{2}\int_{\Sigma}|\hat{K}|^{2}dv_{g}\leq C|k|^{\frac{1}{2}}{\mathcal{V}}^{\frac{1}{2}}(\int_{\Sigma}|\hat{K}|^{4}dv_{g})^{\frac{1}{2}},
\een

\n and after making $x=\|\hat{K}\|_{L^{2}_{g}}$ solve for $x$.\ep

The lapse has important natural properties that we describe in the proposition below.

\begin{Prop}\label{ln} Say $\Sigma$ is a compact three-manifold. Then, $\|N\|_{L^{\infty}}\leq 3/k^{2}$ and $\|N\|_{H^{1}_{g}}$ is 
controlled by $1/|k|$ and ${\mathcal{V}}$. In particular we have the bound
\beq\label{lb}
\int_{\Sigma}|\na N|^{2}+|\hK|^{2}N^{2}+\frac{k^{2}}{3}N^{2}dv_{g}=\int_{\Sigma} N\leq \frac{1}{9|k|^{5}}{\mathcal{V}}.
\eeq	

\end{Prop}

\n {\bf Proof:} 

From the maximum principle we have $N\leq 3/|k|^{2}$. To get equation (\ref{lb}) multiply the lapse equation by $N$, integrate, and use the 
estimate for $\|N\|_{L^{\infty}}$.\ep

\begin{center}
\subsubsection{Estimates of elliptic type.}\label{3.3.2}
\end{center}

{\it Assumptions in Section \ref{3.3.2}}. All through this section we will assume $(g,K)$ is a cosmologically normalized state.

We turn now our attention to the study of the influence of the higher order Bel-Robinson energies over states.

We need a notion of Sobolev norm for space-time tensors and with respect to the CMC foliation. We do that in the following way. Consider a 
space-time tensor ${\bf U}_{a_{1},\ldots,a_{l}}$ of rank $(l,0)$. For any subsequence $I=(i_{1},\ldots,i_{n})$ ($n\leq l$) of the sequence 
$(1,\ldots,l)$, and the obvious compliment subsequence $\bar{I}$, define 
\ben
T_{I}=T_{a_{i_{1}}}\ldots T_{a_{i_{n}}}
\een 

\n and
\ben 
P_{\bar{I}}=P_{a_{\bar{i}_{1}}}^{\ a_{\bar{i}_{1}}'}\ldots P_{a_{\bar{i}_{l-n}}}^{\ a_{\bar{i}_{l-n}}'},
\een

\n where $P_{a}^{\ a'}={\bf g}_{a}^{\ a'}+T_{a}T^{a'}$ is the horizontal projector. We can decompose ${\bf U}$ as
\ben
{\bf U}=\sum_{n=1,\ldots,l;\ |I|=n} (P_{\bar{I}}<{\bf U},T_{I}>)(T_{I}),
\een

\n where $<{\bf U},T_{I}>$ is the contraction of ${\bf U}$ and $T_{I}$. For each summand, the factors on the left are horizontal, while the 
factors on the
right are vertical. For instance, for ${\bf U}_{ab}$ we have
\ben
{\bf U}_{ab}=P_{a}^{\ a'}P_{b}^{\ b'}{\bf U}_{a'b'}-P_{a}^{\ a'}({\bf U}_{a'b'}T^{b'})T_{b}-P_{b}^{\ b'}({\bf U}_{a'b'}T^{a'})T_{a}
+({\bf U}_{a'b'}T^{a'}T^{b'})T_{a}T_{b}.
\een

\n The same decomposition holds for tensors of arbitrary rank $(l,l')$. Now the $H^{s}_{\star}$-norm of ${\bf U}$ on a slice $\Sigma$ of a CMC
foliation is defined as the sum of the $H^{s}_{\star}$-norms of the tensors $P_{\bar{I}}<{\bf U},T_{I}>$. We will be using this convention 
somehow implicitly all through and without further comments. 

From now
on ${\bf W}_{i}=\bn_{T}^{i}{\bf W}_{0}$ where $T$ is the future pointing unit normal to the CMC foliation. 

During the proof of the
propositions until the end of this section, we will use the notation $H^{i}_{r}$ instead of $H^{i}_{\{x\}}(B(o,r))$ which is the one used
inside the statements.

\vspace{0.2cm}
\begin{Prop}\label{NN2} Say $\Sigma$ is a compact three-manifold. Then, (the data) $\nu$, ${\mathcal{V}}$ and $Q_{0}$ control 
$\|N\|_{H^{2}_{\atlas}}$ where $\atlas$ is a $H^{2}$-canonic harmonic atlas.
\end{Prop}

\n {\bf Proof:}

By Proposition \ref{ln}, $\|N\|_{L^{1}_{g}}$ is controlled by ${\mathcal{V}}$ and $Q_{0}$. By Proposition \ref{hatK4}, $|\hat{K}|^{2}$ and 
therefore $|K|^{2}$ are controlled in $L^{2}_{g}$ by the data. The result then follows by Proposition \ref{Pe1} (I). \ep

\begin{Remark} One can get an estimate for the intrinsic norm $\|N\|_{H^{2}_{g}}$ in terms only of ${\mathcal{V}}$ and $Q_{0}$ (i.e. without involving the volume radius) 
if the Ricci curvature is bounded below. Indeed that follows from 
\ben
\int_{\Sigma}|\nabla\nabla N|^{2}+Ric(\nabla N,\nabla N)dv_{g}=\int_{\Sigma}(\Delta N)^{2}dv_{g},
\een

\n the use of the lapse equation in the RHS of the equation above and finally Propositions \ref{ln} and \ref{hatK4}. 
\end{Remark}

The proposition below shows that for cosmologically normalized states, $\nu$, ${\mathcal{V}}$ and $Q_{0}$ control the $H^{2}_{\atlas}$
norm of $1/N$, where ${\atlas}$ is a $H^{2}$-canonic harmonic atlas. This implies in particular that the infimum of the lapse is never zero even for 
states with low regularity\footnote{We will operate assuming a priori regularity of $1/N$. This is indeed guaranteed from the assumption 
made at the beginning of Section \ref{3.3}. To show that in this case the estimate also descend for states $(g,K)$ with $H^{2}\times H^{1}$
regularity proceed as follows. Smooth out the coefficient 
$|K|^{2}$ to get, from 
the maximum principle 
applied to the lapse equation, an upper bound on $1/N$. This gives the necessary regularly to operate. 
As the estimate on the $L^{\infty}$ norm
of $1/N$ that one obtains (having smoothed $|K|^{2}$) depends only on $\nu$, ${\mathcal{V}}$ and $Q_{0}$ it passes to the limit when the
smoothing is undone.}. As a corollary, we get that $\nu$, ${\mathcal{V}}$ and $Q_{0}$ control the $H^{1}_{\atlas}$ norm of the deformation
tensor $\dt$.  

\vspace{0.2cm}
\begin{Prop}\label{1/N} Let $\Sigma$ be a compact three-manifold. Then $\nu$, ${\mathcal{V}}$ and $Q_{0}$ control $\|1/N\|_{H^{2}_{\atlas}}$ 
where $\atlas$ is a $H^{2}$-canonic harmonic atlas. In particular they control $\|\dt\|_{H^{1}_{\atlas}}$.
\end{Prop}

\n {\bf Proof:} 

Multiplying the lapse equation by $1/N^{2}$ and integrating gives
\ben
\int_{\Sigma}2\frac{|\nabla N|^{2}}{N^{3}}+\frac{1}{N^{2}}dv_{g}=\int_{\Sigma}\frac{|K|^{2}}{N}dv_{g}
\leq(\int_{\Sigma}|K|^{4}dv_{g})^{\frac{1}{2}}(\int_{\Sigma}\frac{1}{N^{2}}dv_{g})^{\frac{1}{2}}.
\een
 
\n This shows in particular that $\|N\|_{H^{1}_{g}}$ is controlled by ${\mathcal{V}}$ and $Q_{0}$. We multiply now the lapse 
equation by $1/N^{3}$ and integrate, it gives
\ben
\int_{\Sigma}3\frac{|\nabla N|^{2}}{N^{4}}+\frac{1}{N^{3}}dv_{g}=\int_{\Sigma}\frac{|K|^{2}}{N^{2}}dv_{g}\leq (\int_{\Sigma}|K|^{4})^{\frac{1}{2}}(\int_{\Sigma}\frac{1}{N^{4}}dv_{g})^{\frac{1}{2}}.
\een

\n By the Sobolev embedding\footnote{It is important here that the norm of the embedding is controlled from above by $\nu$, 
${\mathcal{V}}$ and $Q_{0}$.} $H^{1}_{\atlas}\hookrightarrow L^{6}_{\atlas}$, the RHS 
is controlled by $\nu$, ${\mathcal{V}}$ and $Q_{0}$. We will use this estimate below. Consider the Laplacian of $1/N$. We compute
\ben
\Delta\frac{1}{N}=-\frac{\Delta N}{N}+\frac{|\nabla N|^{2}}{N^{2}}=-\frac{1}{N}-|K|^{2}+\frac{|\nabla N|^{2}}{N^{2}}.
\een

\n We have then the elliptic non-homogeneous equation for $1/N$
\ben
\Delta \frac{1}{N}+\frac{1}{N}+\frac{|\nabla N|^{2}}{N}\frac{1}{N}=-|K|^{2},
\een

\n where, to the effect of applying elliptic estimates, we are thinking $|\nabla N|^{2}/N$ as a factor in front of the variable $1/N$. 
We know that $|K|^{2}$ is controlled in 
$L^{2}_{\atlas}$. The result then follows by Proposition \ref{Pe1} (I), if we show that $|\nabla N|^{2}/N$ is controlled in $L^{2}_{\atlas}$. 
We compute  
\ben
\int_{\Sigma}(\frac{|\nabla N|^{2}}{N})^{2}dv_{g}=(\int_{\Sigma}\frac{|\nabla N|^{2}}{N^{4}}dv_{g})^{\frac{1}{2}}(\int_{\Sigma}|\nabla N|^{6}dv_{g})^{\frac{1}{2}}.
\een

\n As was shown above, the first factor in the RHS of the previous equation is controlled by $\nu$, ${\mathcal{V}}$ and $Q_{0}$.
The second factor is controlled by $\nu$, ${\mathcal{V}}$ and $Q_{0}$ by the embedding $H^{1}_{\atlas}\hookrightarrow L^{6}_{g}$ and 
Proposition \ref{NN2}.\ep 
\begin{Prop} \label{Lem3}Say $I\geq 0$. Let $\bar{r}<r<r_{I+2}(o)$, and say $\{x\}$ is a harmonic coordinate system covering 
$B(o,r_{I+2}(o))$ and satisfying (\ref{cc})-(\ref{ccc}). Then (the data) $\|{\bf W}_{0}\|_{H^{I}_{\{x\}}(B(o,r))}$, 
$\|{\bf W}_{1}\|_{H^{I}_{\{x\}}(B(o,r))}$, 
$\|\hat{K}\|_{L^{2}_{g}(B(o,r))}$, $\bar{r}$ and $r$ control $\|Ric\|_{H^{I+1}_{\{x\}}(B(o,\bar{r}))}$ 
and $\|\hat{K}\|_{H^{I+2}_{\{x\}}(B(o,\bar{r}))}$. In particular, they control $r_{I+3}(o)$ from below.
\end{Prop}

\n {\bf Proof:}

The proof proceeds studying the equation (\ref{Simons}) (making $\hat{K}=A$) to get the estimate on $\hat{K}$ and an appropriate elliptic
system ((\ref{eE1})-(\ref{eE2})) to get the estimate on $Ric$. From equation (\ref{de2}) we have
\ben
d^{\nabla} (\hat{K})_{ijk}=\nabla_{i}\hat{K}_{jm}-\nabla_{j}\hat{K}_{im}=\epsilon_{ij}^{\ \ \,l}B_{lm},
\een

\n and therefore
\begin{equation}\label{eq:54}
d^{\nabla *}d^{\nabla}\hat{K}=-\epsilon_{j}^{\ \,li}\nabla_{i}B_{lm}-
\epsilon_{m}^{\ \,li}\nabla_{i}B_{lj}=-2curl(B).
\end{equation}

\noindent From equation (\ref{eq3}) we have
\begin{equation}\label{eq:55}
curl(B)=E(\nabla_{T}W)+\frac{3}{2}(E\times K)-\frac{1}{2}kE.
\end{equation}

\noindent Equations (\ref{eq:54}), (\ref{eq:55}) and (\ref{Simons}) give the elliptic equation
\begin{equation}\label{K-hat-equation}
2\nabla^{*}\nabla\hat{K}=-{\mathcal{R}}(\hat{K})-2(E({\bf W}_{1})+\frac{3}{2}(E\times K)-\frac{1}{2}kE).
\end{equation}

\noindent Recall ${\mathcal{R}}(\hat{K})$ is a linear expression in $\hat{K}$ with coefficients involving only $Ric$. We consider
the case $I=0$ first. This case in turn is the only one that demands a special treatment. In order 
to apply Proposition \ref{Pe1} (I) to the elliptic equation (\ref{K-hat-equation}) we need first to
obtain control on the $H^{1}_{(\bar{r}+r)/2)}$-norm of $\hat{K}$. This estimate follows from standard elliptic estimates on
the elliptic system
\ben
d^{\nabla} (\hat{K})_{ijk}=\epsilon_{ij}^{\ \ \,l}B_{lm},
\een
\ben
\nabla^{j}\hat{K}_{ij}=0.
\een

\n We apply then Proposition \ref{Pe1} (I), on the equation (\ref{K-hat-equation}) to get the control on the 
$H^{2}_{\bar{r}}$-norm of $\hat{K}$. To get the estimate on the $H^{1}_{\bar{r}}$-norm of 
$Ric$ we get first an estimate for the $H^{1}_{(\bar{r}+r)/2}$-norm of $E$. From it and the equation 
$E=Ric+kK-K\circ K$ the estimate on $Ric$ follows. To get the estimate on $E$ apply standard\footnote{To apply {\it standard} elliptic
estimates we note that as was proved before the coefficients involving $K$ are controlled in $H^{2}_{\bar{r}+3(\bar{r}-r)/4}$.} 
elliptic estimates on the elliptic system
\beq\label{eE1}
curl E=-B({\bf W}_{1})-\frac{3}{2}(B\times K)+\frac{1}{2}kB,
\eeq
\beq\label{eE2}
div E=(K\wedge B).
\eeq

\n Now we treat the cases $I>0$. We note that as the harmonic chart $\{x\}$ satisfies (\ref{cc})-(\ref{ccc}) we have control on 
$\|Ric\|_{H^{I}_{r}}$
by the data, and as was mentioned before the coefficients of $\hat{K}$ in the expression ${\mathcal{R}}(\hat{K})$ involve (linearly) only 
$Ric$. Thus we can apply Proposition \ref{Pe1} (I) to the elliptic equation (\ref{K-hat-equation}) to get the desired control on 
$\|\hat{K}\|_{H^{I+2}_{\bar{r}}}$. 
Similarly applying Proposition \ref{Pe1} (II) to the elliptic system (\ref{eE1})-(\ref{eE2}) we get control on $\|E\|_{H^{I+1}_{\bar{r}}}$,
and therefore on $\|Ric\|_{H^{I+1}_{\bar{r}}}$ from the equation $E=Ric+kK-K\circ K$. \ep

\begin{Prop}\label{NEll} Say $I\geq 0$. Let $\bar{r}<r<r_{I+2}(o)$, and say $\{x\}$ is a harmonic coordinate system covering 
$B(o,r_{I+2}(o))$ and satisfying (\ref{cc})-(\ref{ccc}). Then (the data) $\|{\bf W}_{0}\|_{H^{I}_{\{x\}}(B(o,r))}$, 
$\|{\bf W}_{1}\|_{H^{I}_{\{x\}}(B(o,r))}$, 
$\|\hat{K}\|_{L^{2}_{g}(B(o,r))}$, $\|N\|_{L^{2}_{g}(B(o,r))}$, $\bar{r}$ and $r$ control $\|N\|_{H^{I+3}_{\{x\}}(B(o,\bar{r}))}$.
\end{Prop}

\n {\bf Proof:}

We consider the case $I=0$, the cases $I\geq 1$ easily follow by induction. By Proposition \ref{Lem3}, $\|Ric\|_{H^{1}_{\bar{r}+(r-\bar{r})/4}}$
is controlled by the data. Therefore $\|g\|_{H^{3}_{(\bar{r}+r)/2}}$ is controlled by the data. By Proposition \ref{Lem3} too, $|\hat{K}|^{2}$
is controlled in $H^{2}_{(\bar{r}+r)/2}$. We can apply then standard elliptic estimates on the lapse equation
\ben
-\Delta N+|K|^{2}N=1,
\een

\n to get that $\|N\|_{H^{3}_{\bar{r}}}$ is controlled by the data, as desired.\ep  

\begin{Prop}\label{Lem1} Say $\bar{r}<r\leq r_{I+1}(o)$ with $I\geq 1$ and say $\{x\}$ is a harmonic coordinate system satisfying 
(\ref{cc})-(\ref{ccc}) and covering
$B(o,r_{I+1}(o))$. Then for any $(i,j)$ satisfying $0\leq j\leq I$ and $1\leq i\leq I$, $\|{\bf W}_{j}\|_{H^{i}_{\{x\}}(B(o,\bar{r}))}$ is 
controlled by (the data) $\|{\bf W}_{j+1}\|_{H^{i-1}_{\{x\}}(B(o,r))}$, $\|{\bf W}_{j}\|_{H^{i-1}_{\{x\}}(B(o,r))}$, 
$\|{\bf W}_{0}\|_{H^{i-1}_{\{x\}}(B(o,r))}$, 
$\|{\bf W}_{1}\|_{H^{i-1}_{\{x\}}(B(o,r))}$, 
$\|{\bf J}({\bf W}_{j})\|_{H^{i-1}_{\{x\}}(B(o,r))}$, $\bar{r}$, $r$ and $\|\hat{K}\|_{L^{2}_{g}(B(o,r))}$. 
\end{Prop}

\n {\bf Proof:} 

Think the elliptic system (\ref{eq1})-(\ref{eq4}) as a first order elliptic system of the form (\ref{fo}) with 
$U=(E({\bf W}_{j}),B({\bf W}_{j}))$. By Proposition \ref{Lem3} the coefficients $A^{m}_{n}$ which involve only $\hat{K}$ are controlled
in $H^{i+1}_{\{x\}}(B(o,(\bar{r}+r)/2))$ by the data. The result then follows by applying Proposition \ref{Pe1} (II) to the elliptic system
(\ref{eq1})-(\ref{eq2}).\ep

\begin{Remark}\label{RR} {\rm Applying Proposition \ref{Lem1} when $j=0$ we deduce that $\|\W_{0}\|_{H^{i}_{\bar{r}}}$ is controlled by 
$\|\W_{0}\|_{H^{i-1}_{r}}$ and $\|\W_{1}\|_{H^{i-1}_{r}}$, $\bar{r}$, $r$ and $\|\hat{K}\|_{L^{2}_{r}}$. This tells essentially that 
one can replace the data $\|\W_{0}\|_{H^{i-1}_{r}}$, $\|\W_{1}\|_{H^{i-1}_{r}}$ by the data $\|\W_{0}\|_{H^{i}_{r}}$, $\|\W_{1}\|_{H^{i-1}_{r}}$
inside those statements whose hypothesis contain the data $r$ and $\|\hat{K}\|_{L^{2}_{g}}$. This Remark will be used later(sometimes implicitly).}
\end{Remark}

\n We prove next, in Proposition \ref{Current}, an inductive formula for 
the currents ${\bf J}({\bf W}_{i})$ and then in Proposition \ref{currentn2}, estimates on the time derivative of the 
deformation tensor $\dt$.  

\begin{Prop}\label{Current} Say $j\geq 1$. ${\bf J}({\bf W}_{j})$ has an expansion of the form
\begin{equation}\label{J1J}
{\bf J}({\bf W}_{j}) = \sum (\bn_{T}^{m_{1}} \dt)^{n_{1}}*\cdots*(\bn_{T}^{m_{s}} \dt)^{n_{s}}*\dt^{l}*
                 \bn {\bf W}_{k}
\end{equation}
\begin{equation}\label{J2}
\ \ \ \ \ \ \ \ \ \ \ +\sum (\bn_{T}^{\tilde{m}_{1}} \dt)^{\tilde{n}_{1}}*\cdots*(\bn_{T}^{\tilde{m}_{s}} 
                   \dt)^{\tilde{n}_{s}}*\dt^{\tilde{l}}*\bn^{q}_{T}(T*{\bf Rm}*{\bf W}_{\tilde{k}}),                
\end{equation}

\noindent where every asterisk $*$ is some tensor product and each expression of the form $(\bn_{T}^{m}\dt)^{n}$ 
is a ${*}$-product of $\bn_{T}\dt$ with itself n-times. The sum on the RHS of (\ref{J1J}) is among sequences $((m_{1},n_{1}),\ldots,(m_{s},n_{s}),l,k)$
with $k\leq j-1$, 
$m_{1}\geq1,\ldots, m_{s}\geq 1$ and $\sum_{j}n_{j}(1+m_{j})+l+k=j$, while
the sum (\ref{J2}) is among sequences  

\n $((\tilde{m}_{1},\tilde{n}_{1}),\ldots,(\tilde{m}_{s},\tilde{n}_{s}),\tilde{l},q,\tilde{k})$
with $\tilde{m}_{1}\geq1, \ldots, \tilde{m}_{s}\geq 1$ and 
$\sum_{j} \tilde{n}_{j}(1+\tilde{m}_{j})+\tilde{k}+\tilde{l}+q=j-1$.
\end{Prop} 

\n {\bf Proof:} 

First note that 
\beq\label{ind}
\begin{split}
{\bf J}({\bf W}_{j+1})_{bcd}&=\bn^{a}(\bn_{T}{\bf W}_{j,abcd})=(\bn^{a}T^{e})\bn_{e}{\bf W}_{j,abcd}+T^{e}\bn^{a}\bn_{e}{\bf W}_{j,abcd}\\
&=\dt*\bn {\bf W}_{j}+T*{\bf Rm}*{\bf W}_{j}+\bn_{T}{\bf J}({\bf W}_{j}).
\end{split}
\eeq

\n Now for $j=1$ we have ${\bf J}({\bf W}_{1})=\dt*\bn{\bf W}_{0}+T*{\bf Rm}*{\bf W}_{0}$ which agrees with the
form of the formula above ((\ref{J1J}),(\ref{J2})). For $j>1$ we proceed by induction. Assume the ${\bf J}({\bf W}_{j})$ 
has the desired expansion. The first
two terms in the RHS of equation (\ref{ind}) are of the desired form, so it remains to prove that $\bn_{T}{\bf J}({\bf W}_{j})$ 
is of the desired form. Terms in (\ref{J1J}) and (\ref{J2}) are characterized by sequences 
$((m_{1},n_{1}),\ldots,(m_{s},n_{s}),l,k)$ with
$\sum_{j}n_{j}(1+m_{j})+l+k=j$ and 
$((\tilde{m}_{1},\tilde{n}_{1}),\ldots,(\tilde{m}_{s},\tilde{n}_{s}),\tilde{l},q,\tilde{k})$ with 
$\sum_{j} \tilde{n}_{j}(1+\tilde{m}_{j})+\tilde{k}+\tilde{l}+q=j-1$. We show that the $\bn_{T}$ derivative of any term
of the form (\ref{J1J}) or (\ref{J2}) gives terms of the same form characterized by sequences adding $j+1$ for terms of the form
(\ref{J1J}) and $j$ for terms of the form (\ref{J2}). Let us consider derivatives of terms of the form (\ref{J1J}). 
The derivatives $\bn_{T}(\bn_{T}^{m}\dt)^{n}=n(\bn_{T}^{m+1}\dt)*(\bn_{T}^{m}\dt)^{n-1}$ transform the pairs $(m,n)$ (inside a sequence characterizing a term of the form \ref{J1J}) into pairs $(m+1,1),(m,n-1)$ 
but leaving the rest of the sequence unaltered. As $1(1+1+m)+(n-1)(1+m)=1+n(1+m)$ the new sequence adds $j+1$. Similarly 
the derivatives $\bn_{T} \dt^{l}=l\bn_{T}\dt*\dt^{l-1}$ transform the $l$ (inside a sequence) into the pair $(1,1)$ and the number $l-1$ 
but leaving the rest of the
sequence unaltered. Again in this case as $1(1+1)+l-1=l+1$, the new sequence adds $j+1$. Finally the derivatives $\bn_{T}\bn{\bf W}_{k}$ 
are 
\beq
\bn_{T}\bn {\bf W}_{k}=\bn {\bf W}_{k+1}+\dt*{\bf W}_{k}+T*{\bf Rm}*{\bf W}_{k},
\eeq

\noindent which transform the number $k$ (inside a sequence) into three new sequences. One with the new number $k+1$ and leaving the 
rest of the sequence unaltered, thus adding $j+1$.
A second with the numbers
$k$ and $l=1$ and leaving the rest unaltered, thus adding $j+1$. Finally a third of the kind (\ref{J2})
which has values $\tilde{l}=l$, $q=0$, $\tilde{k}=k$ and 
$((\tilde{m}_{1},\tilde{n}_{1}),\ldots,(\tilde{m}_{s},\tilde{n}_{s})=((m_{1},n_{1}),\ldots,(m_{s},n_{s}))$ adding $j$ as desired.
The analysis of the $\bn_{T}$ derivatives for terms of the form (\ref{J2}) proceeds exactly in the same fashion.\ep

\begin{Prop}\label{currentn2} Say $\bar{r}<r\leq r_{I+2}(o)$, $I\geq 0$. Let $\{x\}$ be a harmonic 
coordinate system satisfying the conditions (\ref{cc})-(\ref{ccc}) and covering $B(o,r_{I+2}(o))$. 
Then, for any $(m,i)$ satisfying, 
$m\geq 0$, $i\geq 0$ and $2\leq m+i\leq I+2$, $\|\bn_{T}^{m}\dt\|_{H^{i}_{\{x\}}(B(o,\bar{r}))}$ is controlled by (the data)
$\|\W_{k}\|_{H^{(m-k)+i-1}_{\{x\}}(B(o,r))}$, for $k=0,\ldots,m$, 

\n $\|\bn_{T}^{k}N\|_{H^{i+(m-k)+1}_{\{x\}}(B(o,r))}$, for $k=0,\ldots,m$, 
$\|\hK\|_{L^{2}_{g}(B(o,r))}$, $\bar{r}$ and $r$.
\end{Prop}

\n {\bf Proof:}

The proof proceeds by induction. We first observe that the cases comprising those $(m,i)$ such that 
$m=0$ and $2\leq i\leq I+2$ are proved by Propositions \ref{Lem3} and \ref{NEll}. 
The induction process will be as follows: {\it assume} the proposition is proved for all $(m,i)$ 
with $2\leq m+i\leq \bar{I}+2$ and $m\leq \bar{m}$. This assumption will be referred in what follows as the 
{\it inductive hypothesis} (IH). 
Under the inductive hypothesis we will prove that the proposition is valid for $m=\bar{m}+1$ and all $i$ with $2\leq i+\bar{m}+1\leq I+2$. 
In this way we cover all $(m,i)$ with $2\leq m+i\leq I+2$. From now on we assume $(m,i)=(\bar{m}+1, \bar{i})$ with $2\leq \bar{i}+\bar{m}+1\leq I+2$
and the data for $(m,i)=(\bar{m}+1,\bar{i})$. 

{\it Observation 1:} A crucial observation which is easily checked is the following: the hypothesis of the Proposition for $m=\bar{m}+1$ and $i=\bar{i}$, contain the hypothesis of the
Proposition for all $(m,i)$ with $m\leq \bar{m}$ and $2\leq i+m\leq \bar{m}+\bar{i}+1$. In this way we have that 
$\|\bn_{T}^{\delta}\dt\|_{H^{\bar{i}+(\bar{m}-\delta)+1}_{\bar{r}}}$, with $0\leq \delta\leq \bar{m}$, 
is controlled by the data of the Proposition for $(m,i)=(\bar{m}+1,\bar{i})$. 

The following commutation relation will be used recursively.
\beq\label{crel}
\bn_{T}(\bn_{a}U_{b})=\bn_{a}(\bn_{T}U_{b})+T^{c}U^{d}{\bf Rm}_{cabd}-\dt_{a}^{\ c}\bn_{c}U_{b}.
\eeq

\n From it, we get
\beq\label{number}
\bn_{T}^{\bar{m}+1}\dt_{ab}=\bn_{T}^{\bar{m}}(\bn_{a}(\bn_{T}T_{b})+T^{c}T^{d}{\bf Rm}_{cabd}-\dt_{a}^{\ c}\dt_{cb}).
\eeq

\n We treat each one of the three terms that appear on the RHS of the last equation separately. We treat first the last term. 
We make the expansion
\ben
\bn_{T}^{\bar{m}}(\dt_{a}^{\ c}\dt_{cb})=\sum_{\alpha+\beta=\bar{m}} \bn_{T}^{\alpha}\dt_{a}^{\ c} \bn_{T}^{\beta}\dt_{cb}.
\een

\n From {\it Observation 1}, we get that at each summand, $\bn_{T}^{\alpha}\dt$ and $\bn_{T}^{\beta}\dt$, are controlled in 
$H^{\bar{i}+(\bar{m}-\alpha)+1}_{\bar{r}}$ and $H^{\bar{i}+(\bar{m}-\beta)+1}_{\bar{r}}$ respectively. We get therefore that the full
expression is controlled in $H^{\bar{i}}_{\bar{r}}$. We treat next the second kind of terms in equation (\ref{number}). We make the 
expansion
\ben
\bn_{T}^{\bar{m}}(T^{c}T^{d}{\bf Rm}_{cabd})=\sum_{\alpha+\beta+\gamma=\bar{m}}(\bn_{T}^{\alpha}T^{c})(\bn_{T}^{\beta}T^{d})\W_{\gamma,cabd}.
\een

\n We will treat this term using the following {\it Fact}, which, as the other {\it Facts} to be stated later, are going to be proved after the main
argument is finished.

\begin{Fact}\label{Fact1}: terms of the form $\bn_{T}^{\delta}T$ for $1\leq \delta\leq \bar{m}+1$ 
are controlled in $H^{\bar{i}+(\bar{m}-\delta)+2}_{\bar{r}}$ by the data.
\end{Fact}

\n We know that $\|\W_{\gamma}\|_{H^{\bar{m}+\bar{i}-\gamma}_{\bar{r}}}$ is controlled by the data. It follows from this and {\it Fact} \ref{Fact1} 
that the second kind of terms in equation (\ref{number}) are also controlled in $H^{\bar{i}}_{\bar{r}}$. We discuss now the first kind of 
terms in equation (\ref{number}), namely the terms of the form
\ben
\bn_{T}^{\bar{m}}(\bn_{a}(\bn_{T}T_{b})).
\een

\n We would like to pass the $\bn_{T}$'s on the left of this expression to the right of $\bn_{a}$. We will show that every time a $\bn_{T}$ is moved
past of $\bn_{a}$ we generate a pair of terms that are seen to be controlled in $H^{\bar{i}}_{\bar{r}}$. We write
\beq\label{noo2}
\begin{split}
\bn_{T}^{\bar{m}}(\bn_{a}(\bn_{T}T_{b}))=&\bn_{T}^{\bar{m}-1}(\bn_{a}(\bn_{T}(\bn_{T}T_{b}))+T^{c}(\bn_{T}T^{d}){\bf Rm}_{cabd}\\
&-\dt_{a}^{\ c}(\bn_{c}(\bn_{T}T_{b}))).
\end{split}
\eeq

\n We state now {\it Fact \ref{Fact2}} and {\it Fact \ref{Fact3}} that treat the second and third kind of terms appearing in the right hand 
side of the previous equation.  

\begin{Fact}\label{Fact2}: terms of the form 
\ben
\bn_{T}^{\bar{m}-j}(T^{c}(\bn_{T}^{j}T^{d}){\bf Rm}_{cabd}),
\een

\n with $1\leq j\leq \bar{m}$ are controlled in $H^{\bar{i}}_{\bar{r}}$ by the data.
\end{Fact}

\begin{Fact}\label{Fact3}: terms of the form
\ben
\bn_{T}^{\Gamma}(\dt_{a}^{\ c}\bn_{c}(\bn_{T}^{j}T_{b})),
\een

\n with $1\leq j\leq \bar{m}$ and $0\leq\Gamma\leq \bar{m}-j$, are controlled in $H^{\bar{i}}_{\bar{r}}$ by the data. 
\end{Fact}

\n The first kind of terms in equation (\ref{noo2}) is reduced again moving $\bn_{T}$ past of $\bn_{a}$. We compute
\ben
\bn_{T}^{\bar{m}-1}(\bn_{a}(\bn_{T}^{2}T_{b}))=\bn_{T}^{\bar{m}-2}(\bn_{a}(\bn_{T}^{3}T_{b})+T^{c}\bn_{T}^{2}T^{d}{\bf Rm}_{cabd}-\dt_{a}^{\ c}\bn_{c}(\bn_{T}^{2}T_{b})).
\een

\n Again the second and third kind of terms in the previous equation are treated using {\it Fact 2} and {\it Fact 3}. We keep going like this
until we get a last term to be treated. This term has the form
\beq\label{fund}
\bn_{a}(\bn_{T}^{\bar{m}+1}T_{b}).
\eeq

\n We must treat this term following another route. We write
\ben
\bn_{a}(\bn_{T}^{\bar{m}+1}T_{b})=\bn_{a}(\bn_{T}^{\bar{m}}(-\frac{\nabla_{b}N}{N}))=-\bn_{a}(\bn_{T}^{\bar{m}}(P_{b}^{\ b'}\frac{\bn_{b'}N}{N})),
\een

\n where $P_{b}^{\ b'}$ is the horizontal projection. We make the expansion
\ben
\bn_{a}(\bn_{T}^{\bar{m}}(P_{b}^{\ b'}\frac{\bn_{b'}N}{N}))=\sum_{\alpha+\beta=\bar{m}}\bn_{a}((\bn_{T}^{\alpha}P_{b}^{\ b'})(\bn_{T}^{\beta}\frac{\bn_{b'}N}{N})).
\een

\n Let us consider the expression $\bn_{T}^{\alpha}P_{b}^{\ b'}$. We compute
\ben
\bn_{T}^{\alpha}P_{b}^{\ b'}=-\bn_{T}^{\alpha}(T_{b}T^{b'})=-\sum_{\gamma+\delta=\alpha}(\bn_{T}^{\gamma}T_{b})(\bn_{T}^{\delta}T^{b'}).
\een

\n By {\it Fact \ref{Fact1}} we know each summand is controlled in $H^{\bar{i}+2}_{\bar{r}}$ and therefore the full expression is. 
Note by this, that if $\bn_{a}$ is applied to them, 
we get by writing $\bn_{a}=P_{a}^{\ a'}\bn_{a'}-T_{a}\bn_{T}$, that the outcome is controlled in $H^{\bar{i}+1}_{\bar{r}}$. Let us consider now
the expression $\bn_{T}^{\beta}\frac{\bn_{b'}N}{N}$. We compute
\beq\label{fN1}
\bn_{T}^{\beta}\frac{\bn_{b'}N}{N}=\sum_{\gamma+\delta=\beta}(\bn_{T}^{\gamma}\frac{1}{N})(\bn_{T}^{\delta}\bn_{b'}N).
\eeq

\n Note that if we expand $\bn_{T}^{\gamma}1/N$ (using the quotient rule) we get using the data that this term is controlled at least in 
$H^{\bar{i}+2}_{\bar{r}}$. We consider next the expression $\bn_{T}^{\delta}\bn_{b'}N$. From the identity
\ben
\bn_{T}\bn_{b'}f=\bn_{b'}\bn_{T}f-\dt_{b'}^{\ c}\bn_{c}f,
\een
 
\n we compute
\beq\label{fN}
\bn_{T}^{\delta}\bn_{b'}N=\bn_{T}^{\delta-1}\bn_{b'}\bn_{T}N-\bn_{T}^{\delta-1}(\dt_{b'}^{\ c}\bn_{c}N).
\eeq

\n We use the next {\it Fact} ({\it Fact \ref{Fact4}}) to treat the second term in the RHS of the previous equation.

\begin{Fact}\label{Fact4}: Say $0\leq \delta-\bar{\delta}\leq \bar{m}-1$ and $0\leq \delta\leq \bar{m}-1$. Then, terms of the form 
\ben
\bn_{T}^{\delta-\bar{\delta}}(\dt_{b'}^{\ c}\bn_{c}(\bn_{T}^{\bar{\delta}}N)),
\een

\n are controlled in $H^{\bar{i}+2}_{\bar{r}}$ by the data.
\end{Fact}

\n If we keep moving $\bn_{T}$ past of $\bn_{b'}$ in the RHS of equation (\ref{fN}) and applying at each time the 
{\it Fact \ref{Fact4}}, we get that, except for the term 
\ben
\bn_{b'}\bn_{T}^{\bar{m}}N,
\een

\n occurring when $\beta=\bar{m}$, all the rest are controlled in $H^{\bar{i}+2}_{\bar{r}}$. Similarly if we apply $\bn_{a}$ over
the expression (\ref{fN1}) we get, after writing $\bn_{a}=P_{a}^{\ a'}\bn_{a'}-T_{a}\bn_{T}$, that when $\beta<\bar{m}$, the outcome 
is controlled in $H^{\bar{i}+1}_{\bar{r}}$ and when $\beta=\bar{m}$ it can be written as the term 
\ben\label{ant}
\bn_{a}\bn_{b'}\bn_{T}^{\bar{m}}N,
\een

\n plus a term controlled in $H^{\bar{i}+1}_{\bar{r}}$. Putting all together we conclude that the expression (\ref{fund}) is controlled in 
$H^{\bar{i}}_{\bar{r}}$ if we can prove that the expression
\ben
P_{b}^{\ b'}\bn_{a}\bn_{b'}\bn_{T}^{\bar{m}}N,
\een

\n is controlled in $H^{\bar{i}}_{\bar{r}}$. To see that write it in the form
\ben
\begin{split}
P_{b}^{\ b'}\bn_{b'}(P_{a}^{\ a'}\bn_{a'}\bn_{T}^{\bar{m}}N-T_{a}\bn_{T}^{\bar{m}+1}N)=&\nabla_{b}\nabla_{a}\bn_{T}^{\bar{m}}N-P_{b}^{\ b'}\dt_{b'a}\bn_{T}^{\bar{m}+1}N\\
&-T_{a}\nabla_{b}\bn_{T}^{\bar{m}+1}N,
\end{split}
\een

\n and use the data. To finish the proposition it remains to prove {\it Facts \ref{Fact1}-\ref{Fact4}} that we do next.

\n {\it Proof of Fact \ref{Fact1}:}

We prove it by induction. First note that $\bn_{T}T_{a}=T^{c}\dt_{ca}$ is controlled in $H^{\bar{i}+\bar{m}+1}_{\bar{r}}$ and thus the 
{\it Fact \ref{Fact1}} 
holds when $\delta=1$. {\it Assume} we have shown the {\it Fact \ref{Fact1}} is valid until $\delta=\delta_{0}$, we will show it is also valid 
when $\delta=\delta_{0}+1$. We compute
\ben
\bn_{T}^{\delta_{0}+1}T_{a}=\bn_{T}^{\delta_{0}}(T^{c}\dt_{ca})=\sum_{\alpha+\beta=\delta_{0}}(\bn_{T}^{\alpha}T^{c})(\bn_{T}^{\beta}\dt_{ca}).
\een

\n From the IH we know $\bn_{T}^{\beta}\dt$ is controlled in $H^{\bar{i}+(\bar{m}-\beta)+1}_{\bar{r}}$ and by the assumption $\bn_{T}^{\alpha}T^{c}$
is controlled in $H^{\bar{i}+(\bar{m}-\alpha)+2}_{\bar{r}}$. The two estimates imply the {\it Fact \ref{Fact1}}.\ep

\n {\it Proof of Fact \ref{Fact2}:}

We compute
\ben
\bn_{T}^{\bar{m}-j}(T^{c}(\bn_{T}^{j}T^{d}){\bf Rm}_{cabd})=\sum_{\alpha+\beta+\gamma=\bar{m}-j}(\bn_{T}^{\alpha}T^{c})(\bn_{T}^{\beta+j}T^{d})\W_{\gamma,cabd}.
\een

\n By hypothesis $\W_{\gamma}$ is controlled in $H^{\bar{m}+\bar{i}-\gamma}_{\bar{r}}$ and by {\it Fact \ref{Fact1}} any one of the other
two factors in the previous equation is controlled at least in $H^{\bar{i}+2}_{\bar{r}}$. The {\it Fact \ref{Fact2}} gets proved from both 
estimates.\ep

\n {\it Proof of Fact \ref{Fact3}:}

The proof of this fact follows by induction. First note that the {\it Fact \ref{Fact3}} is valid when $\Gamma=0$. This follows from {\it Fact \ref{Fact1}}
and writing $\bn_{c}=P_{c}^{\ c'}\bn_{c'}-T_{c}\bn_{T}$. {\it Assume} the {\it Fact \ref{Fact3}} is valid for all $(\Gamma,j)$ satisfying $\Gamma\leq \Gamma_{0}$, 
$\Gamma_{0}\leq \bar{m}-2$, $0\leq \Gamma\leq \bar{m}-j$ and $1\leq j\leq \bar{m}$. Then we show the {\it Fact} is valid when 
$\Gamma=\Gamma_{0}+1$ as well. We compute
\ben\label{f3}
\bn_{T}^{\Gamma}(\dt_{a}^{\ c}(\bn_{c}\bn_{T}^{j}T_{b}))=\sum_{\alpha+\beta=\Gamma}(\bn_{T}^{\alpha}\dt_{a}^{\ c})
(\bn_{T}^{\beta}\bn_{c}(\bn_{T}^{j}T_{b})).
\een

\n The factors $\bn_{T}^{\alpha}\dt$ are controlled in $H^{\bar{i}+(\bar{m}-\alpha)+1}_{\bar{r}}$ and therefore controlled in 
$H^{\bar{i}+2}_{\bar{r}}$. It is enough to prove then that the factors $\bn_{T}^{\beta}\bn_{c}(\bn_{T}^{j}T_{b})$ are controlled in 
$H^{\bar{i}}_{\bar{r}}$ by the data. We compute them in the form
\ben
\bn_{T}^{\beta}(\bn_{c}(\bn_{T}^{j}T_{b}))=\bn_{T}^{\beta-1}(\bn_{c}(\bn_{T}^{j+1}T_{b})+
T^{m}(\bn_{T}^{j}T^{n}){\bf Rm}_{mcbn}-\dt_{c}^{\ m}\bn_{m}(\bn_{T}^{j}T_{b})).
\een

\n The third kind of term in the RHS of the previous equation is controlled in $H^{\bar{i}}_{\bar{r}}$ by the data. 
The second kind is controlled in $H^{\bar{i}}_{\bar{r}}$ by the same argument that {\it Fact \ref{Fact2}} was proved. For 
the first kind of term, we move $\bn_{T}$ past of $\bn_{c}$ again. This generates two new terms which as we have shown for 
the two last terms in the RHS of the previous equation, are controlled in $H^{\bar{i}}_{\bar{r}}$ by the assumption and 
{\it Fact 2}. We keep operating like this until we get a last term $\bn_{c}\bn_{T}^{\beta+j}T_{b}$. Writing $\bn_{c}=P_{c}^{\ c'}\bn_{c'}-T_{c}\bn_{T}$. 
and using {\it Fact \ref{Fact1}} we get that this last term is also controlled in $H^{\bar{i}}_{\bar{r}}$, thus finishing the proof.\ep 
 
\n {\it Proof of Fact \ref{Fact4}:}

We prove this {\it Fact \ref{Fact4}} by induction. First note that the {\it Fact \ref{Fact4}} is valid when $\delta=\bar{\delta}$ for $\delta=0,\ldots,\bar{m}-1$.
This follows directly by writing $\bn_{c}=P_{c}^{\ c'}\bn_{c'}-T_{c}\bn_{T}$ and using the data. {\it Assume} now we have shown the {\it Fact \ref{Fact4}} is valid
for all $(\delta,\bar{\delta})$ with $0\leq \bar{\delta}\leq \delta\leq \bar{m}-1$ and $\delta-\bar{\delta}=L$, where $0\leq L\leq \bar{m}-2$.
We will show the {\it Fact \ref{Fact4}} is valid also when $\delta-\bar{\delta}=L+1$. We write
\beq\label{f4}
\bn_{T}^{L+1}(\dt_{b'}^{\ c}\bn_{c}(\bn_{T}^{\bar{\delta}}N))=\sum_{\alpha+\beta=L+1}(\bn_{T}^{\alpha}\dt_{b'}^{\ c})(\bn_{T}^{\beta}\bn_{c}(\bn_{T}^{\bar{\delta}}N)),
\eeq

\n The factors $\bn_{T}^{\alpha}\dt$ are controlled in $H^{\bar{i}+(\bar{m}-\alpha)+1}_{\bar{r}}$. We need to show the factors
$\bn_{T}^{\beta}(\bn_{c}(\bn_{T}^{\bar{\delta}}N))$ are controlled in $H^{\bar{i}+2}_{\bar{r}}$. We write
\ben
\bn_{T}^{\beta}(\bn_{c}(\bn_{T}^{\bar{\delta}}N))=\bn_{T}^{\beta-1}(\bn_{c}\bn_{T}^{\bar{\delta}+1}N-\dt_{c}^{\ d}\bn_{d}\bn_{T}^
{\bar{\delta}}N).
\een

\n The second kind of term in the RHS of the previous equation is controlled in $H^{\bar{i}+1}_{\bar{r}}$ by the assumption. 
For the first term we move again a $\bn_{T}$ past of $\bn_{c}$ and use the assumption. We keep moving $\bn_{T}'s$ past of $\bn_{c}$ until 
getting the last term
$\bn_{c}\bn_{T}^{\beta+\bar{\delta}} N$. It follows from the data that this term is controlled in $H^{\bar{i}+2}_{\bar{r}}$ 
(note that $\beta+\bar{\delta}\leq \bar{m}-1$). Putting
all together we get that the expression (\ref{f4}) is controlled in $H^{\bar{i}+2}_{\bar{r}}$ by the data.\ep

\begin{Prop}\label{Lem2} Say $\bar{r}<r\leq r_{I+2}$. For $(i,j)$ satisfying $0\leq i\leq I$ and $1\leq j\leq I$ and $i+j\leq I+1$, 
$\|{\bf J}({\bf W}_{j})\|_{H^{i}_{\{x\}}(B(o,\bar{r}))}$ is controlled by (the data) $\|{\bf W}_{j-k}\|_{H^{i+k}_{\{x\}}(B(o,r))}$, for 
$k=0,\dots,j$, 
$\|\bn_{T}^{j-k}N\|_{H^{i+2+k}_{x}(B(o,r))}$, for $k=0,\ldots,j$, $\bar{r}$, $r$ and $\|\hat{K}\|_{L^{2}_{g}(B(o,r))}$.
\end{Prop}

\n {\bf Proof:}

The proof is based analyzing the inductive formula for the current
\begin{equation}\label{J1J2}
{\bf J}({\bf W}_{j}) = \sum (\bn_{T}^{m_{1}} \dt)^{n_{1}}*\cdots*(\bn_{T}^{m_{s}} \dt)^{n_{s}}*\dt^{l}*
                 \bn {\bf W}_{k}
\end{equation}
\begin{equation}\label{J22}
\ \ \ \ \ \ \ \ \ \ \ +\sum (\bn_{T}^{\tilde{m}_{1}} \dt)^{\tilde{n}_{1}}*\cdots*(\bn_{T}^{\tilde{m}_{s}} 
                   \dt)^{\tilde{n}_{s}}*\dt^{\tilde{l}}*\bn^{q}_{T}(T*{\bf Rm}*{\bf W}_{\tilde{k}}),                
\end{equation}

\n where the indices satisfy $\sum_{j}n_{j}(1+m_{j})+l+k=j$, with $k\leq j-1$ for summands appearing on the RHS of (\ref{J1J2}) and 
$\sum_{j} \tilde{n}_{j}(1+\tilde{m}_{j})+\tilde{k}+\tilde{l}+q=j-1$
for summands of the form (\ref{J22}). We treat first the summands appearing on the RHS of (\ref{J1J2}). We will show that 
the terms $\bn_{T}^{m}\dt$ and $\dt$
are controlled in $H^{i+2}_{\bar{r}}$ and that the term $\bn{\bf W}_{k}$ is controlled in $H^{i}_{\bar{r}}$. The product that each 
summand of the form (\ref{J1J2}) represents is then controlled in $H^{i}_{\bar{r}}$. By Proposition \ref{currentn2}, $\bn_{T}^{m}\dt$ is 
controlled in $H^{i+2}_{\bar{r}}$ 
from $\|\W_{k}\|_{H^{m+i+1-k}_{r}}$, $k=0,\ldots,m$ and $\|\bn_{T}^{m-k}N\|_{H^{i+3+k}_{r}}$ $k=0,\ldots, m$ as long as $m+i\leq I$ (and 
$m+i+2\geq 2$ which is satisfied trivially). We get this condition from the hypothesis: indeed we have $1+m\leq j$, 
$i+j\leq I+1$ so we have $i+m\leq I$. On the other hand we have a priori control on $\|\W_{k}\|_{H^{i+j-k}_{r}}$ for $k=0,\ldots,j$ and 
$\|\bn_{T}^{j-k}N\|_{H^{i+2+k}_{r}}$ $k=0,\ldots,j$ from the data, which covers the condition on $\W_{k}$ and $N$ required 
before. Similarly we have control on $\|\dt\|_{H^{i+j+1}_{r}}$ and we know $i+j+1\geq i+2$. Let us consider now the factors $\bn \W_{k}$ where $k\leq j-1$. 
Writing $\bn_{a} \W_{k}=P_{a}^{\ a'}\bn_{a'} \W_{k}+T_{a}\W_{k+1}$ we get that the expression 
is controlled in $H^{i}_{\bar{r}}$ by the data. The proof that summands of the form (\ref{J22}) are also controlled in $H^{i}_{\bar{r}}$ is direct from
what we have shown and the data, after expanding $\na_{T}^{q}(T*{\bf Rm}*\W_{\tilde{k}})$ using
the product rule.\ep

We state below the global versions of Propositions \ref{NEll}, \ref{Lem3}, \ref{Lem1}, \ref{currentn2} and \ref{Lem2} and whose proof is straightforward. 
That will be useful in the proof of Lemma \ref{SvsBR}.

\begin{Prop}\label{AB}
Say $\Sigma$ is a compact $H^{I+3}$-Riemannian three-manifold, where $I\geq 0$. Then, $\|Ric\|_{H^{I+1}_{\atlas}}$, $\|\hat{K}\|_{H^{I
+2}_{\atlas}}$, and $r_{I+3}$ 
are controlled by (the data) $\|\W_{0}\|_{H^{I}_{\atlas}}$, $\|\W_{1}\|_{H^{I}_{\atlas}}$, $\|\hat{K}\|_{L^{2}_{g}}$, ${\mathcal{V}}$ and 
$r_{I+2}$, where $\atlas$ is a $H^{I+2}$-canonic harmonic atlas.
\end{Prop}

\begin{Prop}\label{ABC} Say $\Sigma$ is a compact $H^{I+3}$-manifold, where $I\geq 0$. Then, (the data) $\|{\bf W}_{0}\|_{H^{I}_{\atlas}}$, 
$\|{\bf W}_{1}\|_{H^{I}_{\atlas}}$, 
$\|\hat{K}\|_{L^{2}_{g}}$, $\|N\|_{L^{2}_{g}}$ and $r_{I+2}$ control $\|N\|_{H^{I+3}_{\atlas}}$, where $\atlas$ is a $H^{I+2}$-canonic 
harmonic atlas.
\end{Prop}

\begin{Prop}\label{BB}
Say $\Sigma$ is a compact $H^{I+2}$-Riemannian three-manifold, where $I\geq 1$. Then, for any $(i,j)$ satisfying $0\leq j\leq I$ and
$1\leq i\leq I$, $\|\W_{j}\|_{H^{i}_{\atlas}}$ is controlled by the data $\|\W_{j+1}\|_{H^{i-1}_{\atlas}}$, $\|\W_{j}\|_{H^{i-1}_{\atlas}}$, $\|\W_{0}\|_{H^{i-1}_{\atlas}}$,
$\|\W_{1}\|_{H^{i-1}_{\atlas}}$, $\|{\bf J}(\W_{j})\|_{H^{i-1}_{\atlas}}$, $\|\hat{K}\|_{L^{2}_{g}}$, ${\mathcal{V}}$ and $r_{I+1}$, where 
$\atlas$ is a $H^{I+1}$-cononic harmonic atlas.
\end{Prop}

\begin{Prop}\label{CC}
Say $\Sigma$ is a compact $H^{I+3}$-Riemannian three-manifold, $I\geq 0$. Then, for any $(m,i)$ satisfying $m\geq0$, $i\geq 0$ and $2\leq i+m\leq I+2$,
$\|\bn_{T}^{m}\dt\|_{H^{i}_{\atlas}}$ is controlled by $\|\W_{k}\|_{H^{i+(m-k)-1}_{\atlas}}$, for $k=0,\ldots,m$, $\|\bn_{T}^{k}N\|_{H^{(m-k)+i+1}_{\atlas}}$, 
for $k=0,\ldots,m$, $\|\hat{K}\|_{L^{2}_{g}}$, ${\mathcal{V}}$ and $r_{I+2}$, where $\atlas$ is a $H^{I+2}$-canonic harmonic atlas.
\end{Prop}

\begin{Prop}\label{DD}
Say $\Sigma$ is a compact $H^{I+3}$-Riemannian three-manifold, where $I\geq 0$. Then, for any $(i,j)$ satisfying $0\leq i\leq I$, 
$1\leq j\leq I$ and $i+j\leq I+1$, $\|{\bf J}(\W_{j})\|_{H^{i}_{\atlas}}$, is controlled by (the data) 
$\|\W_{j-k}\|_{H^{i+k}_{\atlas}}$, for $k=0,\ldots,j$, $\|\bn_{T}^{j-k}N\|_{H^{i+k+2}_{\atlas}}$, for $k=0,\ldots,j$, $\|\hat{K}\|_{L^{2}_{g}}$, 
${\mathcal{V}}$ and $r_{I+2}$, where $\atlas$ is a $H^{I+2}$-canonic harmonic atlas.
\end{Prop} 

In the next proposition we make a last step before proving Lemma \ref{SvsBR}. We will denote with an upper-index $(k)$ 
the $k$-th Lie derivatives in the time direction $\partial_{t}=NT$.

\begin{Prop}\label{globlap}
Say $\Sigma$ is a compact $H^{I+3}$-Riemannian three-manifold, $I\geq 0$. Then, for any $(m,i)$ satisfying $2\leq i+m\leq I+2$, $0\leq i$, 
$0\leq m$ (the data) 
${\mathcal{V}}$, $r_{I+2}$, $\|\W_{k}\|_{H^{i-1+(m-k)}_{\atlas}}$, for $k=0,\ldots,m$, control 
$\|\bn_{T}^{k}N\|_{H^{i+1+(m-k)}_{\atlas}}$,  $\|g^{(k)}\|_{H^{i+1+(m-k)}_{\atlas}}$ for $k=0,\ldots,m$, and 
$\|K^{(k)}\|_{H^{i+(m-k)}_{\atlas}}$ for $k=0,\ldots,m-1$ (if $m\neq 0$), where $\atlas$ is a $H^{I+2}$-canonic harmonic atlas.
\end{Prop}

\begin{Remark}\label{gcp}{\rm Note that the hypothesis made on the Weyl fields $\W_{k}$ in Proposition \ref{globlap} are the same as
the hypothesis made for the Weyl fields $\W_{k}$ inside the Proposition \ref{CC}. Also note that we may have used the norms 
$\|\bn_{T}^{k}N\|_{H^{i+1}_{\atlas}}$ instead of the norms $\|N^{(k)}\|_{H^{i+1}_{\atlas}}$ as we can see from the identity 
$N\dot{}=N\bn_{T}N$ that one set of norms control the other. Finally note that the conclusions extracted on the lapse in 
Proposition \ref{globlap} are exactly the hypothesis on the lapse inside Proposition \ref{CC}.}
\end{Remark}

\n {\bf Proof:}

The proof proceeds by induction. We treat first the case when $m=0$ and $2\leq i\leq I+2$. Note that by Proposition \ref{lem4}
$\|\hat{K}\|_{L^{2}_{g}}$ is controlled by $\|\W_{0}\|_{H^{i-1}_{\atlas}}$, and ${\mathcal{V}}$. Note also that standard 
elliptic estimates on the elliptic system
\ben
d^{\nabla} (\hat{K})_{ijk}=\epsilon_{ij}^{\ \ \,l}B_{lm},
\een
\ben
\nabla^{j}\hat{K}_{ij}=0.
\een

\n show that $\|\hat{K}\|_{H^{i}_{\atlas}}$ is controlled by $\|\W_{0}\|_{H^{i-1}_{\atlas}}$, $r_{I+2}$ and ${\mathcal{V}}$. This proves
the required control on $K$. Now, use this estimate for $K$ on equations (\ref{eq3}) and (\ref{eq4}) (that define $E(\W_{1})$ and $B(\W_{1})$)
to show that $\|\W_{1}\|_{H^{i-2}_{\atlas}}$ is controlled by $\|\W_{0}\|_{H^{i-1}_{\atlas}}$, $r_{I+2}$ and ${\mathcal{V}}$. Now use
Proposition \ref{ABC} to show that $\|N\|_{H^{i+1}_{\atlas}}$ is controlled by $\|\W_{0}\|_{H^{i-1}_{\atlas}}$, $r_{I+2}$ and ${\mathcal{V}}$.
Finally from Proposition \ref{AB} we get that $\|Ric\|_{H^{i-1}_{\atlas}}$ is controlled and therefore by Theorem 
\ref{teo:cc1} $\|g\|_{H^{i+1}_{\atlas}}$ is controlled too.

We treat next the case when $m=1$ and $1\leq i\leq I+1$. The fact that $\hat{K}$ is controlled in $H^{i+1}_{\atlas}$ follows from Proposition 
\ref{AB}. This in turn implies the estimate on $g\dod=-2NK$. It remains to show that $N$ is controlled in $H^{i+2}_{\atlas}$ and 
$N\dod$ in $H^{i+1}_{\atlas}$ by the data. The first estimate follows from Proposition \ref{ABC}. For the second estimate we need to 
differentiate the lapse equation. For convenience we write it in the form
\ben
-\Delta N+(R_{g}+k^{2})N=1.\
\een

\n Note from this form of the lapse equation that the $N$ is a function of the metric $g$ only. 
Differentiating with respect to time $(t=k)$ we get
\beq\label{DDD}
-\Delta N\dod+|K|^{2}N\dod=\Delta\dod N+(R_{g}\dod+2k)N.
\eeq

\n For the derivative of the Laplacian we have \cite{Bess}
\ben
\Delta\dod f=g\dod^{ab}\nabla_{a}\nabla_{b}f-g^{ab}(\nabla_{a}f)(\nabla^{c}g\dod_{cb}+\frac{1}{2}\nabla_{b}(g^{de}g\dod_{de})).
\een

\n and for the derivative of the scalar curvature we have \cite{Bess}
\ben
R_{g}\dod=-\Delta tr_{g}g\dod+\delta_{g}\delta_{g}g\dod-<Ric,g\dod>.
\een

\n We can write then 
\beq\label{Ninn1}
\Delta\dod N=-2NK^{ab}\nabla_{a}\nabla_{b}N-g^{ab}(\nabla_{a}N)(-2(\nabla^{c}N)K_{cb}-k\nabla_{b}N),
\eeq

\n and
\beq\label{Ninn2}
R_{g}\dod=2k(-1+N|K|^{2})-2k(\nabla_{a}\nabla_{b}N)K^{ab}+2NkRic_{ab}K^{ab}.
\eeq

\n A direct inspection of equations (\ref{Ninn1}) and (\ref{Ninn2}) shows that the RHS of equation (\ref{DDD}) is controlled
in $H^{i}_{\atlas}$ by the data. The elliptic estimates of Proposition \ref{Pe1} (I) applied to the equation (\ref{DDD}) would show that
$N\dod$ is controlled in $H^{i+2}_{\atlas}$ if we can show that the $H^{1}_{\atlas}$-norm of $N\dod$ is controlled by the data. To get that
estimate multiply equation (\ref{DDD}) by $N\dod$ and integrate (denote the right hand of equation (\ref{DDD}) as $F$). We get
\ben
\int_{\Sigma}|\nabla N\dod|^{2}+|K|^{2}(N\dod)^{2}dv_{g}=\int_{\Sigma}FN\dod dv_{g}\leq (\int_{\Sigma}F^{2}dv_{g})^{\frac{1}{2}}(\int_{\Sigma}(N\dod) ^{2}dv_{g})^{\frac{1}{2}}. 
\een

\n It is apparent from this that $\|N\dod\|_{H^{1}_{g}}$ is controlled by the data. 

{\it Assume} now that the proposition has been proved for all $(m,i)$ with $m\leq \bar{m}$ and $2\leq i+m\leq I+2$. We will show it is valid  
when $(m,i)=(\bar{m}+1,\bar{i})$ with $2\leq \bar{i}+\bar{m}+1\leq I+2$ as well. Observe that the estimates we want to prove for the metric 
$g$ follows from those we want to prove for the lapse $N$ and the second fundamental form $K$ using $g\dot{}=-2NK$. Observe too, that the 
hypothesis when $(m,i)=(\bar{m}+1,\bar{i})$ contains the hypothesis when $(m,i)=(\bar{m},\bar{i}+1)$. As a consequence of this, 
we have automatic control over $\|\bn_{T}^{k}N\|_{H^{\bar{i}+\bar{m}-k+2}_{\atlas}}$, for $k=0,\ldots,\bar{m}$ and 
$\|K^{(k)}\|_{H^{\bar{i}+\bar{m}+1-k}_{\atlas}}$, 
for $k=0,\ldots,\bar{m}-1$ from the data (for $(m,i)=(\bar{m}+1,\bar{i})$). It remains to prove therefore that the data (for $(m,i)=(\bar{m}+1,\bar{i})$) 
controls also 
$\|K^{(\bar{m})}\|_{H^{\bar{i}+1}_{\atlas}}$ and 
$\|\bn_{T}^{\bar{m}+1}N\|_{H^{\bar{i}+1}_{\atlas}}$. We show first the control over $K$ and then we show it over $N$. 
It follows from the Remark \ref{gcp} and Proposition \ref{CC} that $\|\bn_{T}^{k}\dt\|_{H^{\bar{m}+\bar{i}+1-k}_{\atlas}}$, for
$k=0,\ldots, \bar{m}$ are controlled by the data. If $U_{ab}$ is a $T$-null and symmetric, space-time tensor, the time derivative of $U$ 
has the expression
\beq\label{dttd}
U\dod_{ab}=N\bn_{T}U_{ab}+N(U_{ac}\dt_{b}^{\ c}+U_{bc}\dt_{a}^{\ c}),
\eeq

\n It is easy to see that $U\dot{}$ is $T$-null and symmetric. Choose $U_{ab}=K_{ab}=P_{a}^{\ a'}P_{b}^{\ b'}\dt_{a'b'}$. Using recursively equation (\ref{dttd})
together with the fact that $\|\bn_{T}^{k}\dt\|_{H^{\bar{m}+\bar{i}+1-k}_{\atlas}}$, for
$k=0,\ldots, \bar{m}$ is controlled by the data, it is direct to see that $\|K^{(\bar{m})}\|_{H^{\bar{i}+1}_{\atlas}}$ is controlled by the 
data too.

We prove now that $\|\bn_{T}^{\bar{m}+1}N\|_{H^{\bar{i}+1}_{\atlas}}$ is controlled by the data too. This will come from differentiating the
lapse equation $\bar{m}+1$-times. We will rely on the following fact whose proof will not be included and is straightforward
\begin{Fact}:\label{Fact5} the $\bar{m}+1$-th time derivative of the lapse equation has an expression of the form
\ben
-\Delta N^{(\bar{m}+1)}+|(R_{g}+k^{2})N^{(\bar{m}+1)}=F,
\een

\n where $F$ is an expression controlled in $H^{i}_{\atlas}$ by the data (for $(m,i)=(\bar{m}+1,\bar{i})$).
\end{Fact}

\n With respect to the proof of this fact, we only mention that it can be proved differentiating equation (\ref{DDD}) $\bar{m}$-times, 
using the expressions (\ref{Ninn1}) and (\ref{Ninn2}) for the
RHS of (\ref{DDD}). The derivative of the Ricci tensor is given by \cite{Bess}
\ben
Ric\dod=\frac{1}{2}\Delta_{L}g\dod-\delta^{*}_{g}(\delta_{g}g\dod)-\frac{1}{2}D_{g}d(tr_{g}g\dod),
\een

\n where $\Delta_{L}$ is the Lichnerowicz Laplacian (see \cite{Bess}). The time derivative of the connection $\na$ is calculated by the 
following formula. 
Denote $\na^{t}$ the covariant derivative of $g(t)$ and
pick an arbitrary time independent vector field $U^{a}$. Then
\beq\label{dert}
\na^{t}U^{b}=\na^{t_{*}}_{a}U^{b}+\tilde{\Gamma}^{b}_{ac}U^{c},
\eeq	

\n where $\tilde{\Gamma}$ is
\beq\label{cont}
\tilde{\Gamma}^{b}_{ac}=\frac{1}{2}(\na_{a}^{t_{*}}g(t)_{bd}+\na_{b}^{t_{*}}g(t)_{ad}-\na^{t_{*}}_{d}g(t)_{ab})g(t)^{dc}.
\eeq

\n Now $\na_{a}\dod$ at $t=t_{*}$ is calculated by differentiating (\ref{cont}) with respect to time and evaluating at 
$t=t_{*}$ (note that the time derivative of the first term in (\ref{dert}) vanishes). 

From {\it Fact \ref{Fact5}} and Proposition \ref{Pe1} we get therefore that $\|\bn_{T}^{\bar{m}+1}N\|_{H^{\bar{i}+1}_{\atlas}}$ is controlled by the data.\ep

We are ready to prove Lemma \ref{SvsBR}. 

\vspace{0.2cm}
\n {\bf Proof (of Lemma \ref{SvsBR}):} 

We prove first that the $\bar{I}+2$ harmonic radius is controlled from below by the BR-functional $\|(g,K)\|_{BR}$. We consider the case
$\bar{I}=0$ first and then the cases $\bar{I}>0$. 

{\it Case} $\bar{I}=0$. This case follows by Proposition \ref{Ric} and Theorem \ref{teo:cc1}.

{\it Case} $\bar{I}>0$. The proof in this case proceeds as follows. We prove first that $r_{\bar{I}+2}$ is controlled by 
$r_{\bar{I}+1}$ and $\|(g,K)\|_{BR}$.
Indeed, the proof of that shows, more generally, that for any $1\leq J\leq \bar{I}$, $r_{J+2}$ is controlled by 
$r_{J+1}$ and $\|(g,K)\|_{BR}$. As a consequence, 
we deduce that $r_{\bar{I}+2}$ is controlled by $r_{2}$ and $\|(g,K)\|_{BR}$. By {\it case} $\bar{I}=0$, $r_{2}$ is controlled by 
$\|(g,K)\|_{BR}$. The 
{\it Case} $\bar{I}>0$ would then follow.
We prove now that $r_{\bar{I}+2}$ is controlled by $r_{\bar{I}+1}$ and $\|(g,K)\|_{BR}$. Note that by Proposition \ref{hatK4}, $\|\hat{K}\|_{L^{2}_{g}}$ 
is controlled by the
BR-functional. We can then replace $\|\hat{K}\|_{L^{2}_{g}}$ by $\|(g,K)\|_{BR}$ in the hypothesis of Propositions \ref{AB}, \ref{BB} and   
\ref{DD} that we are going to use in what follows. By Proposition \ref{AB} we know that $r_{\bar{I}+2}$ is controlled by 
$\|\W_{0}\|_{H^{\bar{I}-1}_{\atlas}}$, $\|\W_{1}\|_{H^{\bar{I}-1}_{\atlas}}$, $r_{\bar{I}+1}$ and $\|(g,K)\|_{BR}$, where $\atlas$ is a 
$H^{\bar{I}+1}$-canonic harmonic atlas. If
$\bar{I}-1=0$ we are done, as we have that $r_{3}$ is controlled by $r_{2}$ and $\|(g,K)\|_{BR}$. If instead $\bar{I}>1$ we apply consecutively
Propositions \ref{BB} and \ref{DD}. From Proposition \ref{BB} applied with $I=\bar{I}$, $i=\bar{I}-1$ and $j=0,1$ we deduce that 
$\|\W_{0}\|_{H^{\bar{I}-1}_{\atlas}}$ and $\|\W_{1}\|_{H^{\bar{I}-1}_{\atlas}}$ are controlled by $\|\W_{0}\|_{H^{\bar{I}-2}_{\atlas}}$, 
$\|\W_{1}\|_{H^{\bar{I}-2}_{\atlas}}$, $\|\W_{2}\|_{H^{\bar{I}-2}_{\atlas}}$, $\|{\bf J}(\W_{1})\|_{H^{\bar{I}-2}_{\atlas}}$, $r_{\bar{I}+1}$ and
$\|(g,K)\|_{BR}$. From Proposition \ref{DD} applied with $I=\bar{I}-1$, $j=1$ and $i=\bar{I}-2$ we deduce that $\|{\bf J}(\W_{1})\|_{H^{\bar{I}-2}_{\atlas}}$
is controlled by $\|\W_{0}\|_{H^{\bar{I}-1}_{\atlas}}$, $\|\W_{1}\|_{H^{\bar{I}-2}_{\atlas}}$ and $\|(g,K)\|_{BR}$. To organize visually 
the iteration, we have included the Figure \ref{fig1}. The norms included at a given column control the norms at the previous column (there 
may be more information at a given column than what is actually needed to control the norms on a previous column). 
The step we have just done represents the control of the second column over the first.
Now, applying consecutively Propositions \ref{BB} and \ref{DD} over each new column we get at each step, we reach a last column 
with the norms $\|\W_{0}\|_{H^{0}_{\atlas}},\ldots,\|\W_{\bar{I}}\|_{H^{0}_{\atlas}}$\footnote{Note that $r_{\bar{I}+1}$, ${\mathcal{V}}$ and ${\mathcal{E}}_{\bar{I}}$ control
$\|\W_{j}\|_{H^{0}_{\atlas}}$ for $j=0,\ldots,\bar{I}$ and where $\atlas$ is a $H^{\bar{I}+1}$-canonic harmonic atlas}. 
As a result of this we get that $r_{\bar{I}+2}$ 
is controlled by $r_{\bar{I}+1}$ and $\|(g,K)\|_{BR}$ as desired. 

To show control over $\|\hat{K}\|_{H^{\bar{I}+1}_{g}}$ observe that as we have shown above 
$\|\W_{0}\|_{H^{\bar{I}-1}_{\atlas}}$ and $\|\W_{1}\|_{H^{\bar{I}-1}_{\atlas}}$
are controlled by $\|(g,K)\|_{BR}$ where $\atlas$ is a $H^{\bar{I}-1}$-canonic harmonic atlas. By Proposition \ref{AB} $\|\hK\|_{H^{\bar{I}+1}_{\atlas}}$ is
controlled by $\|(g,K)\|_{BR}$. It follows that $\|\hat{K}\|_{H^{\bar{I}+1}_{g}}$ is controlled by $\|(g,K)\|_{BR}$ too.\ep 

\begin{figure}
\begin{center}
\begin{tabular}{|c|c|c|c|c|c|c|c|c|c|c|}
\hline
&1&2&3&4&5&$\ldots$&$I+1$&\ldots&\ldots\\
\hline
${\bf W}_{0}$&$H^{I+1}$&$H^{I}$&$H^{I}$&$H^{I}$&$H^{I}$&\ldots&$H^{I}$&\ldots&$H^{0}$\\
\hline
${\bf W}_{1}$&$H^{I}$&$H^{I-1}$&$H^{I-1}$&$H^{I-1}$&$H^{I-1}$&\ldots&$H^{I-1}$&\ldots&$H^{0}$\\
\hline
${\bf W}_{2}$&&$H^{I-1}$&$H^{I-2}$&$H^{I-2}$&$H^{I-2}$&\ldots&$H^{I-2}$&\ldots&$H^{0}$\\
\hline
${\bf W}_{3}$&&&$H^{I-2}$&$H^{I-3}$&$H^{I-3}$&\ldots&\ldots&\ldots&$H^{0}$\\
\hline
${\bf W}_{4}$&&&&$H^{I-3}$&$H^{I-4}$&\ldots&\ldots&\ldots&$H^{0}$\\
\hline
$\vdots$&&&&&$H^{I-4}$&\ldots&$H^{1}$&\ldots&$H^{0}$\\
\hline
$\vdots$&&&&&&\ldots&$H^{0}$&\ldots&$H^{0}$\\
\hline
${\bf W}_{I+1}$&&&&&&&$H^{0}$&\dots&$H^{0}$\\
\hline
\end{tabular}
\end{center}
\caption{Iteration of control of the Sobolev norms of ${\bf W}_{0}$ and ${\bf W}_{1}$. Each column contains the kind of norms 
that control the norms on the previous column. In the table $I=\bar{I}-1$.}
\label{fig1}
\end{figure}

To finish this section let us prove a proposition on the structure of the $L^{2}_{g}$-norm of the current ${\bf J}(\W_{i})$ that will be of
use in the initial value formulation.

\begin{Prop}\label{inductionE}
Let $\Sigma$ be a compact three-manifold, $(g,K)$ a cosmological normalized state and $j\geq 2$. Then the $L^{2}_{g}$-norm of the current 
${\bf J}(\W_{j})$ can be bounded as
\beq\label{BIVF}
\|{\bf J}(\W_{j})\|_{L^{2}_{g}}\leq C({\mathcal{E}}_{j-1},\nu,{\mathcal{V}})Q_{j}^{\frac{1}{2}}+C({\mathcal{E}}_{j-1},\nu,{\mathcal{V}}).
\eeq
\end{Prop}

\n {\bf Proof:}

The proof is better divided in two cases, when $j=2$ and when $j\geq 3$. Before going into the analysis of these cases let us 
make the following observation.

{\it Observation 1}. As an outcome of the proof of Lemma \ref{SvsBR} it can be seen that 
$\|\bn_{T}^{m}\dt\|_{H^{2}_{g}}$ for $m\leq j-2$, $\|\bn_{T}^{j-1}\dt\|_{H^{1}_{g}}$, 
$\|\bn \W_{0}\|_{H^{1}_{g}}$, $\|\bn \W_{k}\|_{H^{0}_{g}}$ for $k=0,\ldots,j-2$ 
and $\|{\bf J}(\W_{j-1})\|_{H^{0}_{g}}$ are controlled by ${\mathcal{E}}_{j-1}$, 
$\nu$ and ${\mathcal{V}}$. 

\n {\it Case} $j=2$. When $j=2$ we can write schematically (see for instance Proposition \ref{Current})
\beq\label{JW2}
\begin{split}
{\bf J}(\W_{2})=&\dt*\dt*\bn\W_{0}+\bn_{T}*\bn\W_{0}+\dt*\bn\W_{1}+T+{\bf Rm}*\W_{1}\\
&+\dt*T*{\bf Rm}*\W_{0}.
\end{split}
\eeq

\n We need to prove that the $L^{2}_{g}$-norm of any one of the terms on the RHS of the last equation can be
bounded by an expression of the form (\ref{BIVF}). Let us treat them case by case. By {\it Observation 1} 
in the first term in the RHS of equation (\ref{JW2}) the coefficients $\dt$ of $\bn \W_{0}$ are controlled in $H^{2}_{g}$ and 
the term $\bn \W_{0}$ itself is controlled in $H^{0}_{g}$ by ${\mathcal{E}}_{1},\nu$, and ${\mathcal{V}}$. Altogether the
$L^{2}_{g}$-norm can therefore be bounded by $C({\mathcal{E}}_{1},\nu,{\mathcal{V}})$. Let us treat next the term 
$\bn_{T}\dt*\bn \W_{0}$. We can write\footnote{Note that for any $U$ an $V$ we have $\|U*V\|_{L^{2}_{g}}\leq \|U\|_{L^{4}_{g}}\|V\|_{L^{4}_{g}}$,
and still by Sobolev embeddings less or equal than $C({\mathcal{E}}_{1},\nu,{\mathcal{V}})\|V\|_{H^{1}_{g}}\|U\|_{H^{1}_{g}}$.}
\ben
\|\bn_{T}\dt*\bn\W_{0}\|_{L^{2}_{g}}\leq C({\mathcal{E}}_{1},\nu,{\mathcal{V}})
\|\bn_{T}\dt\|_{H^{1}_{g}}\|\bn\W_{0}\|_{H^{1}_{g}}.
\een

\n By the {\it Observation 1} the norm $\|\bn_{T}\dt\|_{H^{1}_{g}}$ is controlled by ${\mathcal{E}}_{1},\ \nu$ and ${\mathcal{V}}$. We need
to estimate the norm $\|\bn\W_{0}\|_{H^{1}_{g}}$. Write
\ben
\bn_{a}\W_{0}=P_{a}^{\ a'}\bn_{a'}\W_{0}-T_{a}\W_{1}.
\een

\n Using formulas (\ref{EB1})-(\ref{EB4}) we can write schematically 
\ben
\W_{0}=\epsilon*\epsilon*E+\epsilon*B*T+E*T*T.
\een

\n Using this and the connection formulas we compute (schematically)
\ben
\begin{split}
P_{a}^{\ a'}\bn_{a'}\W_{0}=&K*\epsilon*\epsilon*E+\epsilon*\epsilon*\nabla E+K*\epsilon*B*T+\epsilon*\nabla B*T+\epsilon*B*K\\
&+\nabla E*T*T+E*T*K.
\end{split}
\een

\n From this we can write
\beq
\|P_{a}^{\ a'}\bn_{a'}\W_{0}\|_{L^{2}_{g}}\leq C({\mathcal{E}}_{1},\nu,{\mathcal{V}})(\|E_{0}\|_{H^{1}_{g}}+\|B_{0}\|_{H^{1}_{g}}).
\eeq

\n In the same way one can prove
\beq\label{QQ1}
\|\bn \W_{0}\|_{H^{1}_{g}}\leq C({\mathcal{E}}_{1},\nu,{\mathcal{V}})(\|E_{0}\|_{H^{2}_{g}}+\|B\|_{H^{2}_{g}}).
\eeq

\n From the elliptic regularity of Proposition \ref{Pe1} applied to the elliptic system (\ref{eq1})-(\ref{eq4}) we have
\beq\label{QQ2}
\|(E_{1},B_{1})\|_{H^{1}_{g}}\leq C({\mathcal{E}}_{1},\nu,{\mathcal{V}})(\|(E_{2},B_{2})\|_{H^{0}_{g}}+\|(E_{1},B_{1})\|_{H^{0}_{g}}+
\|{\bf J}(\W_{1})\|_{H^{0}_{g}}),
\eeq

\n and
\beq\label{QQ3}
\|(E_{0},B_{0})\|_{H^{2}_{g}}\leq C({\mathcal{E}}_{1},\nu,{\mathcal{V}})(\|(E_{1},B_{1})\|_{H^{1}_{g}}+\|(E_{0},B_{0})\|_{H^{0}_{g}}).
\eeq

\n It follows from equations (\ref{QQ1}), (\ref{QQ2}) and (\ref{QQ3}) that 
\ben
\|\bn_{T}\dt*\bn \W_{0}\|_{L^{2}_{g}}\leq C({\mathcal{E}}_{1},\nu,{\mathcal{V}})Q_{2}^{\frac{1}{2}}+C({\mathcal{E}}_{1},\nu,{\mathcal{V}}).
\een

\n Let us treat consider now the term $\dt*\bn \W_{1}$ in the RHS of equation (\ref{JW2}). The coefficient $\dt$ is controlled in $H^{2}_{g}$
by ${\mathcal{E}}_{1},\ \nu$ and ${\mathcal{V}}$. Now use (\ref{QQ2}) to get a bound on $\|\dt*\bn \W_{1}\|_{L^{2}_{g}}$ of the desired form. 
The $L^{2}_{g}$-norm of the last two terms $T*{\bf Rm}*\W_{1}$ and $\dt*T*{\bf Rm}*\W_{0}$ are treated similarly.  

{\it Case} $j\geq 3$. This time we appeal to the full structure of the current ${\bf J}(\W_{j})$ provided by Proposition \ref{Current}. Recall
\begin{equation}\label{J1J2}
{\bf J}({\bf W}_{j}) = \sum (\bn_{T}^{m_{1}} \dt)^{n_{1}}*\cdots*(\bn_{T}^{m_{s}} \dt)^{n_{s}}*\dt^{l}*
                 \bn {\bf W}_{k}
\end{equation}
\begin{equation}\label{J22}
\ \ \ \ \ \ \ \ \ \ \ +\sum (\bn_{T}^{\tilde{m}_{1}} \dt)^{\tilde{n}_{1}}*\cdots*(\bn_{T}^{\tilde{m}_{s}} 
                   \dt)^{\tilde{n}_{s}}*\dt^{\tilde{l}}*\bn^{q}_{T}(T*{\bf Rm}*{\bf W}_{\tilde{k}}),                
\end{equation}

\noindent The sum on the RHS of (\ref{J1J2}) is among sequences $((m_{1},n_{1}),\ldots,(m_{s},n_{s}),l,k)$
with $k\leq j-1$, 
$m_{1}\geq1,\ldots, m_{s}\geq 1$ and $\sum_{j}n_{j}(1+m_{j})+l+k=j$, while
the sum (\ref{J22}) is among sequences $((\tilde{m}_{1},\tilde{n}_{1}),\ldots,(\tilde{m}_{s},\tilde{n}_{s}),\tilde{l},q,\tilde{k})$
with $\tilde{m}_{1}\geq1, \ldots, \tilde{m}_{s}\geq 1$ and 
$\sum_{j} \tilde{n}_{j}(1+\tilde{m}_{j})+\tilde{k}+\tilde{l}+q=j-1$.

Let us treat the terms on the RHS of equation (\ref{J1J2}). Say $k=0$. In this case the coefficients in front of $\bn \W_{0}$
are controlled in $H^{2}_{g}$ except when there is only one coefficient of the form $\bn_{T}^{j-1}\dt$. Now by {\it Observation 1}
either $\bn \W_{0}$ or $\bn_{T}^{j-1}\dt$ are controlled in $H^{1}_{g}$ by ${\mathcal{E}}_{j-1},\nu,{\mathcal{V}}$. It follows that the
norm $\|\bn_{T}^{j-1}\dt*\bn \W_{0}\|_{L^{2}_{g}}$ is controlled by ${\mathcal{E}}_{1}$, $\nu$ and ${\mathcal{V}}$. Say now that 
$k=1,\ldots,j-2$. In these cases the coefficients in front of $\bn W_{k}$ are controlled in $H^{2}_{g}$ by ${\mathcal{E}}_{1}$, $\nu$ and
${\mathcal{V}}$. It follows from this and {\it Observation 1} that when $k=0,\ldots, j-2$ the $L^{2}_{g}$-norm of the expression in the 
RHS of equation (\ref{J1J2}) is controlled by ${\mathcal{E}}_{j-1}$, $\nu$ and ${\mathcal{V}}$. Say finally that $k=j-1$.
In this case there is only one possibility, namely the expression on the RHS of equation (\ref{J1J2}) is of the form
$\dt*\bn \W_{j-1}$. From elliptic regularity applied to the elliptic system (\ref{eq1})-(\ref{eq4}) we know
\ben
\|(E,_{j-1},B_{j-1})\|_{H^{1}_{g}}\leq C({\mathcal{E}}_{1},\nu,{\mathcal{V}})(\|(E_{j},B_{j})\|_{H^{0}_{g}}+\|(E_{j-1},B_{j-1})\|_{H^{0}_{g}}+
\|{\bf J}(\W_{j-1})\|_{H^{0}_{g}}).
\een

\n It follows that
\ben
\|\bn \W_{j-1}\|_{L^{2}_{g}}\leq C({\mathcal{E}}_{1},\nu,{\mathcal{V}})(Q_{j}^{\frac{1}{2}}+C({\mathcal{E}}_{j-1},\nu,{\mathcal{V}})),
\een

\n as desired. This finishes the treatment of the terms of the form (\ref{J1J2}). The terms of the form (\ref{J22}) are treated similarly.\ep

{\center \subsection{Controlling the flow $(g,K)$ along evolution.}\label{3.4}}

In this section we introduce the functional space in which solutions to the Einstein flow equations will lie. In the
same vein as Section \ref{3.3}, we introduce a BR-functional (see later)
\ben
\|\phi_{*}{\bf g}\|_{BR}=\|g\|_{C^{0}(I,2)(H_{{\mathcal{A}}})}+\|K\|_{C^{0}(I,1)(H_{{\mathcal{A}}})}+\sum_{k=0}^{k=i-2}\|(E_{k},B_{k})\|_{C^{0}(I,0)},
\een

\n and discuss how it controls: i. the set of flow solutions, ii. a dynamical smooth structure on the space-time manifold, 
and iii. the space-time metric as a metric over the space-time manifold with the dynamical
smooth structure. The results will be used in the next section when we discuss the initial value formulation for the CMC gauge.
  
We start giving the definition of {\it admissible} gauge. We then give the definitions of {\it space-time solutions} and
{\it flow solutions}. After that we introduce the BR-norm and present the main results of the section.

\vspace{0.2cm}
\begin{Def} Say $I$ is an interval (for the time coordinate $t=k$). Define 

\n $C(I,\alpha,\beta)(H_{\star})=\cap_{j=0}^{j=\alpha}C^{j}(I,H^{\beta-j}_{\star})$ where the 
subindex $\star$ indicates the structure with respect to which the Sobolev space is defined. Each space
$C^{j}(I,H_{\star}^{\beta-j})$ is provided with its usual $sup$ norm. We may allow $\beta$ to be less than $\alpha$.
\end{Def}

\begin{Def}\label{Admissible}
Say $(\Sigma,{\mathcal{A}}_{\infty})$ is a $C^{\infty}$-manifold. A differentiable function 

\n $X:C^{\infty}({\mathcal{S}}(\Sigma))\times C^{\infty}({\mathcal{TS}}(\Sigma))\rightarrow 
C^{\infty}({\mathcal{T}}(\Sigma))$ which is diffeomorphism invariant i.e. $X(\phi^{*}g,\phi^{*}K)=\phi^{*}(X(g,K))$ for any diffeomorphism 
$\phi:\Sigma\rightarrow \Sigma$ (the classes $C^{\infty}(\star)$ are defined with respect to ${\mathcal{A}}_{\infty}$) is an admissible 
shift iff 

\begin{enumerate}

\item for any $H^{j+1}$ ($j\geq 2$) atlas ${\mathcal{A}}$ which is compatible with ${\mathcal{A}}_{\infty}$ 
at the $j+1$ level of regularity, $X$ can be 
extended uniquely to a differentiable function (also
denoted by $X$) $X:H^{j}_{{\mathcal{A}}}\times H^{j-1}_{{\mathcal{A}}}\rightarrow H^{j}_{{\mathcal{A}}}$ 
with the property that:

\item $\|X\|_{H^{j}_{\atlas}}$ is controlled by $\|g\|_{H^{j}_{\atlas}}$ and $\|K\|_{H^{j-1}_{\atlas}}$ and

\item for every path $(g,K)(t)$, such that $(g',K')$ is in 
$C(I,j-k,j)(H_{{\mathcal{A}}})\times C(I,j-k,j-1)(H_{{\mathcal{A}}})$ ($0\leq k\leq j$) and with the norms
$\|(g,K)\|_{C^{0}(I,j)\times C^{0}(I,j)}$ and 

\n $\|(g',K')\|_{C(I,j-k,j)\times C(I,j-k,j-1)}$ 
bounded by $\Lambda$, $X'$ is in $C(I,j-k,j)(H_{{\mathcal{A}}})$ with norm controlled by 
$\Lambda$.

\end{enumerate}
\end{Def}

\n {\bf Examples.} 1. The zero shift is an example of an admissible gauge. 

2. Definition \ref{Admissible} of admissible gauges share many of its properties with the Andersson-Moncrief gauge \cite{AM1} on 
hyperbolic manifolds. Although we do not claim that the Andersson-Moncrief  gauge is admissible we would like to explain how some of 
the characteristics of admissible gauges are actually present in the it. Let $g_{H}$ be a hyperbolic metric on $\Sigma$. For every $g$ 
in $\Sigma$ perform a diffeomorphism $\phi$ in $\Sigma$ in such a way that the identity is a harmonic map between $\phi^{*}(g)$ and 
$g_{H}$. The identity is harmonic iff the vector field $V^{k}=g^{ij}e^{k}(\na_{i} e_{i}-\na^{H}_{i} e_{j})$ is zero. The CMCSH (Constant 
mean curvature - spatially harmonic) fixes the shift $X$ in such a way that at every time the identity map $id:(\Sigma,g(t))\rightarrow 
(\Sigma,g_{H})$ is harmonic. With this condition the equation for the lapse and shift are the following \cite{AM1}
\beq\label{am1}
-\Delta N+|K|^{2}N=1,
\eeq
\beq\label{am2}
\begin{split}
\Delta X^{i}+Ric^{i}_{\ f}X^{f}-{\mathcal{L}}_{X}V^{i}=&(-2NK^{mn}+ 2\na^{m}X^{n})e^{i}(\na_{m} e_{n} -\na^{H}_{m}e_{n})\\
&+2\na^{m}N K^{i}_{\ m}-\na^{i}N k.
\end{split}
\eeq

\n Thus, when the identity is a harmonic map, $X$ is defined through (\ref{am1}) and (\ref{am2}) and otherwise it is defined by making 
it diffemorphism invariant. Consider the $C^{\infty}$ atlas 
${\mathcal{A}}_{\infty}$ for which $g_{H}$ is $C^{\infty}$. As the differential 
operators (\ref{am1}) and (\ref{am2}) defining $(N,X)$ are elliptic
its is clear by elliptic regularity that if $(g,K)$ are $C^{\infty}$ then $(N,X)$ are too. Therefore 
$X:C^{\infty}({\mathcal{S}}(\Sigma))\times C^{\infty}({\mathcal{T}}{\mathcal{S}}(\Sigma))\rightarrow 
C^{\infty}({\mathcal{T}}{\mathcal{S}}(\Sigma))$. In the same way if ${\mathcal{A}}$ is a $H^{j+1}$ atlas
which is compatible with ${\mathcal{A}}_{\infty}$ at the $j$-level of regularity, by elliptic regularity 
$X$ extends to differentiable functions on 
$H^{j}_{\mathcal{A}}\times H^{j-1}_{\mathcal{A}}\rightarrow H^{j}_{\mathcal{A}}$.  On the other hand
Lemma 3.2 in \cite{AM1} shows that $\|X\|_{H^{j}_{\atlas}}$ is controlled by $\|g\|_{H^{j-1}_{\atlas}}$ and $\|K\|_{H^{j-2}_{\atlas}}$. 
This says that $X$ satisfies {\it item 2} in Definition \ref{Admissible} (in fact it represents an improvement). Also from Lemma 3.2 in 
\cite{AM1} we have that, for a path $(g,K)(\lambda)$, the norm $\|X'\|_{H^{j}_{\atlas}}$ is controlled by $\|g\|_{H^{j-1}_{\atlas}}$, 
$\|K\|_{H^{j-2}_{\atlas}}$, $\|g'\|_{H^{j-1}_{\atlas}}$, $\|K'\|_{H^{j-2}_{\atlas}}$. The shift $X$ satisfies therefore {\it item 3} 
in Definition \ref{Admissible} with $k=j$ (in fact it represents and improvement).   

\begin{Remark} {\rm Observe the following property of admissible gauges. Say $(\Sigma,\atlas)$ is a $C^{\infty}$-three-manifold and 
suppose $X$ is admissible. Suppose $\atlas_{1}$ is another $C^{\infty}$ atlas in $\Sigma$ compatible with $\atlas$ at least at the third 
level of regularity.  It is known there is a $C^{\infty}$-diffeomorphism $\phi:(\Sigma,\atlas)\rightarrow (\Sigma,\atlas_{1})$ and therefore 
the properties of $X$ over $\atlas$ pull back over $\atlas_{1}$.}
\end{Remark}

\begin{Def}\label{STsol} A $H^{i}$-space-time solution $({\bf M},{\bf g})$ of the Einstein equations (in vacuum) is a 
$H^{i+1}$-Lorentzian four-manifold $({\bf M},{\bf g})$ satisfying ${\bf Ric}=0$.
\end{Def}

\begin{Def}\label{g}
A $H^{i}$-flow solution of the Einstein CMC flow equations in the admissible spatial gauge $X(g,K)$, 
is a space-time solution 
$({\bf M}, {\bf g})$ (in vacuum) for which there is a $H^{i}$-diffeomorphism $\phi:{\bf M}\rightarrow \Sigma\times I$ 
where $(\Sigma,{\mathcal{A}})$ is a $H^{i+1}$-three-manifold and $\Sigma\times I$ is supplied with the product structure, 
such that 

\begin{enumerate}

\item $\phi_{*}({\bf g})$ is (therefore) a $H^{i-1}$-space-time solution of the Einstein equations in vacuum.

\item The time foliation is CMC.

\item The components $g,N,X$ of the $3+1$ splitting of $\phi_{*}{\bf g}$ in $\Sigma\times I$ satisfy: $g$ is 
in $C(I,H^{i-1}_{{\mathcal{A}}})$ and $g'$ is in $C(I,i-1,i-1)(H_{{\mathcal{A}}})$. $N,X$ are in 
$C(I,i-1,i-1)(H_{{\mathcal{A}}})$, and $K$ is in $C(I,i-1,i-1)(H_{{\mathcal{A}}})$. These fields
moreover satisfy the CMC flow equations (\ref{1.1})-(\ref{4.1}).

\item For every $k=0,\ldots i-2$, $\phi_{*}({\bf W}_{k})$ has electric-magnetic decomposition 
$(E_{k},B_{k})$ in $C(I,i-2-k,i-2-k)(H_{{\mathcal{A}}})$, satisfying the equations (\ref{eq1})-(\ref{eq4}).

\item There is a dynamical $H^{i+1}$-atlas ${\mathcal{A}}(t)$ with dynamical charts $\{x_{\alpha}^{k}(t)\}$ 
for which the set of space-time charts $\{{\bf x}_{\alpha}=(x_{\alpha}^{k}(t),t)\}$ form a $H^{i+1}$-atlas of 
$\Sigma\times I$ making 
$\phi$ a $H^{i+1}$-diffeomorphism. Also the transition functions $x_{\alpha}(x_{\beta})$ are in 
$C(I,i+1,i+1)(H_{x_{\beta}})$
and $(g,N,X_{{\bf x}_{\beta}})$ are in $C(I,i,i)(H_{x_{\beta}})$, where $X_{{\bf x}_{\beta}}$ is the shift
vector of the coordinate system ${\bf x}_{\beta}$.

\end{enumerate}
\end{Def}

A comment is in order. In {\it item 5} of Definition \ref{g}, the time derivatives required to define the spaces 
$C(I,\star,\star)(H_{x_{\beta}})$ are with respect to the time coordinate of ${\bf x}_{\beta}$. 

In practice (for $i=3$) the dynamical atlas ${\mathcal{A}}(t)$ will be given out of the following construction of harmonic
coordinates. Let $\{\tilde{x}^{k}\}$ be a coordinate chart in a $H^{3}$-Riemannian manifold $(\Sigma,\tilde{g})$. 
Say $B(o,\alpha_{1})\subset B(o,\alpha_{2})$ are two balls inside the chart. Pick a smooth non-negative function $\xi$
being one in $B(o,\alpha_{1})$ and zero in $B(o,\alpha_{2})$ and define for any given metric $g$ in $\Sigma$, 
the metric $\bar{g}_{ij}=\xi g_{ij}+(1-\xi)\delta_{ij}$. Think the coordinates $\{\tilde{x}^{k}\}$ as coordinates 
on a three-torus $T^{3}$ with metric $\bar{g}$ over the chart and extended to be flat on the rest. Extend the coordinate 
$\tilde{x}^{k}$ smoothly to the rest of the torus in such a way that $\int_{T^{3}}\tilde{x}^{k}dv_{\bar{\tilde{g}}}$ where
$\bar{\tilde{g}}$ is $\bar{\tilde{g}}_{ij}=\xi \tilde{g}_{ij}+(1-\xi)\delta_{ij}$.
Define the function $h^{k}$ to be zero on the chart and equal to $\Delta_{\bar{g}} \tilde{x}^{k}$
on the rest of the torus. Then we can solve uniquely for $x^{k}$ in $\Delta_{\bar{g}}x^{k}=h^{k}$ if we impose the 
condition that $\int_{T^{3}}x^{k}dv_{\bar{g}}=0$. Now suppose $\tilde{\atlas}=\{\{\tilde{x}_{\alpha}\},\alpha=1,\ldots,n\}$ is a 
$H^{2}$-canonic harmonic atlas for the manifold $(\Sigma,\tilde{g})$ 
where each $\{\tilde{x}_{\alpha}\}$ is defined over a ball $B(o_{\alpha},r_{2}/2)$ ($r_{2}(\tilde{g})$). Pick $\delta<1$ but close to one. Choose
$\alpha_{1}=\delta r_{2}$ and $\alpha_{2}=r_{2}$. Then for $\epsilon$ sufficiently small, if 
$\|g-\tilde{g}\|_{H^{2}_{\tilde{\atlas}}}\leq \epsilon$ the functions $\{x^{k}_{\alpha},k=1,2,3\}$ defined over 
$B(o_{\alpha},r_{2}/2)$ are harmonic coordinates for $g$. Also if $\epsilon$ is sufficiently small, the charts $\{x_{\alpha}\}$
extend to $B(o_{\alpha},\delta r_{2})$ and satisfying
\beq\label{delhrad1}
\frac{\delta 3}{4}\delta_{jk}\leq g_{jk}\leq \frac{4}{\delta 3}\delta_{jk},
\eeq
\beq\label{delhrad2}
r_{2}(\sum_{|I|=2,j,k}\int_{B(o_{\alpha},\delta r_{2})}|\frac{\partial^{I}}{\partial x^{I}}g_{jk}|^{2}dv_{x})\leq \frac{1}{\delta}.
\eeq

\n Thus we have constructed a new harmonic atlas for the metric $g$ that we will call $\delta$-{\it canonic}. It is clear that all the
results we have discussed so far in Section \ref{3.3} hold if instead of using a canonic atlas we use $\delta$-canonic atlas for a fixed 
$\delta$\footnote{We have said what a $H^{2}$-$\delta$-canonic is. It should be clear how to define $H^{i}$-$\delta$-canonic.}.
Now suppose $g(t)$ is a path of metrics in $C^{0}(I,2)(H_{\tilde{\atlas}})$ and suppose $\|g(0)-\tilde{g}\|_{H^{2}_{\tilde{\atlas}}}\leq 
\epsilon/2$ (with $\epsilon$ as before), then for any $t$ in a subinterval $I'\subset I$ the harmonic atlas 
$\{x_{\alpha}(t)\}=\atlas(t)$ is well defined. We will call $\atlas(t)$ the 
{\it dynamical atlas} constructed out of $(\tilde{g},\tilde{\atlas},g(t))$ and denote the
dynamical charts by $\{x_{\alpha}(t)\}$.    

\begin{Remark} {\rm As constructed the atlas $\atlas(t)$ is only an $H^{3}$-harmonic dynamical atlas and not, as is required, a 
$H^{4}$-harmonic dynamical atlas. We will see later (Proposition \ref{lem3}) that in fact this construction gives us the desired 
$H^{4}$-harmonic atlas.}
\end{Remark}

We describe now a useful $BR$-norm (Bel-Robinson norm) on the space of metrics on $\Sigma\times I$ with the properties 
in Definition \ref{g} above and then move to explain how it controls the space of flow solutions.

\begin{Def}\label{def2}
Let $\phi_{*}{\bf g}$ be a flow metric on $\Sigma\times I$ as explained in Definition \ref{g}. Define the BR-norm of ${\bf g}$ 
out of $\atlas$ by
\ben
\|\phi_{*}{\bf g}\|_{BR}=\|g\|_{C^{0}(I,2)(H_{{\mathcal{A}}})}+\|K\|_{C^{0}(I,1)(H_{{\mathcal{A}}})}+
\sum_{j=0}^{j=i-2}\|(E_{j},B_{j})\|_{C^{0}(I,0)(H_{{\mathcal{A}}})}
\een
\end{Def}

\noindent Note that the definition of BR-norm does not need the condition in {\it item 4} of Definition \ref{g} to be imposed on 
$\phi_{*}{\bf g}$. The importance of the $BR$-norm is that it is easy to handle in the 
Einstein equations and is intended
to measure (some) ``$H^{i}_{{\mathcal{A}}\times I}$-norm" of $\phi_{*}({\bf g})$ without being in $H^{i}_{{\mathcal{A}}\times I}$. 
At the same time as explained later it does controls (some) $H^{i}_{{\bf x}_{\beta}}$-norms with respect to the coordinates 
${\bf x}_{\alpha}$. 

From now on we fix $i=3$. We note however that the treatment we make may well be systematized to any regularity $i\geq 4$. 

We will need elliptic estimates for scalar equations of the form
\ben
\Delta \phi=h,
\een

\n where $h$ has an expression of the form
\ben
h=T_{0}\circ V_{0}+T_{1}\circ \na V_{1}+T_{2}\circ \na^{2} V_{2},
\een

\n for tensors $T_{m}$ and $V_{m}$, $m=0,1,2$ of arbitrary rank and where the operation $\circ$
is any full contraction (like $T_{2}^{abc}\nabla_{a}\nabla_{c}V_{2,b}$). In particular 
we would like to obtain elliptic estimates in terms of the $L^{2}_{\atlas}$-norm of $V_{i}$, $i=1,2,3.$ (see the hypothesis inside the Proposition). Thus the non-homogeneous term $h$ will lie, in general, in a negative Sobolev space. In this context, the elliptic estimates that we will obtain (except {\it item 1} in the proposition below) do not follow from standard elliptic estimates or Proposition \ref{Pe1} and deserve a special treatment\footnote{We haven't found these
estimates in the standard references on PDE.} due to the low regularity of the metric $g$ which is assumed in $H^{3}$.
    
\begin{Prop}\label{EEII} 
Let $(\Sigma,\atlas,g)$ be a $C^{\infty}$-Riemannian manifold and say $\atlas$ is a $H^{3}$-canonic harmonic atlas. Let $T_{m}$, $V_{m}$, $m=0,1,2$ be
$C^{\infty}$-tensors of arbitrary rank. Say $\phi$ is a 
solution of
\beq\label{lapest}
\Delta\phi=h,
\eeq

\n with $\int_{\Sigma}\phi dv_{g}=0$. Then, we have the elliptic estimates

\begin{enumerate}

\item For $j=0,1,2$ we have 
\ben
\|\phi\|_{H^{j+2}_{\atlas}}\leq C(r_{3},Vol)\|h\|_{H^{j}_{\atlas}},
\een

\item If $h=T_{1}\circ \na V_{1}+T_{0}\circ V_{0}$, we have  
\ben
\|\phi\|_{H^{1}_{\atlas}}\leq C(r_{3},Vol,\|T_{1}\|_{H^{2}_{\atlas}},\|T_{0}\|_{H^{1}_{\atlas}},\|V_{1}\|_{H^{0}_{\atlas}},\|V_{0}\|_{H^{0}_{\atlas}}),
\een

\item If $h=T_{0}\circ V_{0}+T_{1}\circ \na V_{1}+T_{2}\circ \na^{2} V_{2}$ we have
\ben
\|\phi\|_{H^{0}_{\atlas}}\leq C(r_{3},Vol,\|T_{1}\|_{H^{1}_{\atlas}},\|T_{2}\|_{H^{2}_{\atlas}},\|T_{0}\|_{H^{0}_{\atlas}},\|V_{1}\|_{H^{0}_{\atlas}},
\|V_{2}\|_{H^{0}_{\atlas}},\|V_{0}\|_{H^{0}_{\atlas}}).
\een

\end{enumerate}

\n With low regularity the estimates extend in the following sense.  Say 
$(\Sigma,{\atlas_{i}},g_{i})$ is a sequence converging in $H^{4}$ (for $\atlas_{i}$) and $H^{3}$ (for $g_{i}$) 
to $(\Sigma,\atlas_{\infty},g_{\infty})$ where $(\Sigma,\atlas_{\infty},g_{\infty})$ is 
$H^{4}$-Riemannian manifold, with $\atlas_{\infty}$ a $H^{3}$-canonic harmonic atlas. Say too
$h_{i}$, $T_{m,i}$ and $V_{m,i}$ are sequences with $h_{i}\rightarrow h_{\infty}$ in $H^{j}$ for $j$ either $0,\ 1$ or $2$,
$T_{m,i}\rightarrow T_{m,\infty}$ in $H^{m+1}$ (for {\it item 2}) and $H^{m}$ (for {\it item 3}) and $V_{m,i}\rightarrow V_{m,\infty}$ in 
$H^{0}$. Finally say $\phi_{i}$ is the
sequence of solutions to (\ref{lapest}). Then,  (for {\it item 1}) $\phi_{i}\rightarrow \phi_{\infty}$ in $H^{j}$ for $j$ either $0,\ 1$ or 
$2$ (for {\it item 2}) $\phi_{i}\rightarrow \phi_{\infty}$ in $H^{1}$ and (for {\it item 3}) $\phi_{i}\rightarrow \phi_{\infty}$ in $H^{0}$.
\end{Prop}

\n {\bf Proof:}

Recall the fact that $\Delta:H^{2}\rightarrow H^{0}$
is Fredholm of kernel the constants and index zero. Moreover if we consider the operators 
$\Delta:H^{2}_{\atlas}\rightarrow H^{0}_{\atlas}$ and $\Delta^{-1}:H^{0}_{\atlas}\rightarrow H^{2}_{\atlas}$ 
as operators from the orthogonal complement of the constants
into the orthogonal complement of the constants, their norms are controlled by $r_{3}$ and 
$Vol$\footnote{The boundedness of $\Delta$ is direct form elliptic estimates. The boundedness of $\Delta^{-1}$ follows 
by first proving the image of $\Delta$ is the orthogonal complement of the constants and then using the open mapping theorem. 
The fact that the image of $\Delta$ is the orthogonal complement of the constants follows by expanding $H^{0}$ in terms of 
eigenfunctions of $\Delta$.}.   

{\it Item 1}. This case follows from standard elliptic estimates (or Proposition \ref{Pe1}) and the observation above.  

{\it Item 3}. Pick $\xi$ with $\int_{\Sigma}\xi dv_{g}=0$. Observe that in any contraction of the form $T\circ \na V$ or $T\circ \na^{2} V$ 
we can always integrate by parts once or twice respectively. Multiply equation (\ref{lapest}) by $\xi$ and integrate. We get 
\ben
\int_{\Sigma}\xi\Delta \phi dv_{g}=\int_{\Sigma} (\na (\xi T_{1}))\circ V_{1}+(\na^{2}(\xi T_{2}))\circ V_{2}+(\xi T_{0})\circ V_{0} dv_{g}.
\een

\n Expanding the covariant derivatives on the RHS of the last equation and using Sobolev embeddings and H\"older inequalities 
we get
\beq\label{it3}
\begin{split}
\int_{\Sigma} (\Delta \xi) \phi dv_{g}\leq&C(r_{3},Vol)(\|\xi\|_{H^{2}_{\atlas}}(\|V_{0}\|_{H^{0}_{\atlas}}\|T_{0}\|_{H^{0}_{\atlas}}+
\|T_{2}\|_{H^{2}_{\atlas}}\|V_{2}\|_{H^{0}_{\atlas}})\\
&+\|\xi\|_{H^{1}_{\atlas}}\|T_{2}\|_{H^{2}_{\atlas}}\|V_{2}\|_{H^{0}_{\atlas}}+ 
\|\xi\|_{H^{2}_{\atlas}}\|T_{1}\|_{H^{1}_{\atlas}}\|V_{1}\|_{H^{0}_{\atlas}}).
\end{split}
\eeq

\n Make $\bar{\xi}=\Delta \xi$. Then from $\|\xi\|_{H^{2}_{\atlas}}\leq C(r_{3},Vol)\|\bar{\xi}\|_{H^{0}_{\atlas}}$ we get that
$\bar{\xi}\rightarrow <\bar{\xi},\phi>$ is a bounded linear map from $L^{2}_{g}$ into $\field{R}$. In particular $\|\phi\|_{L^{2}_{g}}$
and so $\|\phi\|_{H^{0}_{\atlas}}$ are controlled by $r_{3}$, $Vol$ and the respective norms of $T_{m}$ and $V_{m}$ for $m=0,1,2$.

{\it Item 2}. Multiply equation (\ref{lapest}) by $\phi$ and integrate by parts. We get 
\ben
\int_{\Sigma}|\na \phi|^{2}dv_{g}=\int_{\Sigma}\na \phi\circ T_{1}\circ V_{1} +\phi \na T_{1}\circ V_{1}+\phi T_{0}\circ V_{0}.
\een

\n Using again Sobolev embeddings and Holder inequalities we have 
\beq\label{it2}
\int_{\Sigma}|\na \phi|^{2}dv_{g}\leq C(r_{3},Vol)(\|\phi\|_{H^{1}_{\atlas}}\|T_{1}\|_{H^{2}_{\atlas}}\|V_{1}\|_{H^{0}_{\atlas}}+
\|\phi\|_{H^{1}_{\atlas}}\|T_{0}\|_{H^{1}_{\atlas}}\|V_{0}\|_{H^{0}_{\atlas}}).
\eeq

\n From {\it item 3} we have control on $\|\phi\|_{H^{0}_{\atlas}}$. Using it and equation (\ref{it2}) we get control on $\|\phi\|_
{H^{1}_{\atlas}}$ from $r_{3},Vol$ and the respective norms of $T_{m}$ and $V_{m}$ for $m=0,1,2$ as desired.
 
We prove now (sketchily) the last part of the proposition. Assume then we have a sequence $g_{i},\phi_{i},h_{i},V_{m,i},T_{m,i}$ as 
indicated in the statement of the proposition.

{\it Item 1} Although the proof of this case is trivial, we will follow one that can be applied in the proofs of {\it items 2} and 
{\it 3} too. For a metric $g$ and a field $\phi$ write
\ben
\Delta_{g}\phi=\Delta_{g_{\infty}}\phi+g_{\infty}^{ab}\Gamma^{c}_{ab}\nabla_{c}\phi+(g^{ab}-g_{\infty}^{ab})\nabla_{a}\nabla_{b}\phi.
\een

\n where $\nabla-\nabla^{\infty}=\Gamma$. Note that $\|\Gamma_{i}\|_{H^{2}_{\atlas_{\infty}}}\rightarrow 0$ and $\|g^{ab}_{i}-g_{\infty}^{ab}\|_{H^{3}_{\atlas_{\infty}}}\rightarrow 0$. 
Using this formula rewrite the subtraction of equation (\ref{lapest}) for the fields $\phi_{l},g_{l},h_{i}$ to the equation 
(\ref{lapest}) for the fields $\phi_{k},g_{k},h_{k}$, as
\ben
\Delta_{g_{\infty}}(\phi_{l}-\phi_{k})=h_{lk},
\een

\n with $\|h_{lk}\|_{H^{j}_{\atlas_{\infty}}}\rightarrow 0$ and $\int_{\Sigma}\phi_{l}-\phi_{k}dv_{g_{\infty}}\rightarrow 0$. It follows 
that $\|\phi_{l}-\phi_{k}\|_{H^{j+2}_{\atlas_{\infty}}}\rightarrow 0$, thus the sequence $\{\phi_{i}\}$ is Cauchy in 
$H^{j+2}_{\atlas_{\infty}}$ and therefore convergent. 

{\it Items 2 and 3}. This item follows in the same way as {\it item 1} was proved. Observe that the RHS of equations 
(\ref{it3}) and (\ref{it2}) tends to zero (thus making the norms $\|\phi\|_{H^{0}_{\atlas}}$ and
$\|\phi\|_{H^{1}_{\atlas}}$ tend to zero too) if, simultaneously, the (respective) norm of one of the fields in each one of the pairs 
$(T_{0},V_{0})$, $(T_{1},V_{1})$ and $(T_{2},V_{2})$ tends to zero. In the same way as in {\it item 1} subtract equation 
(\ref{lapest})
for the fields $\phi_{l}, g_{l}, T_{m,l}, V_{m,l}$ for $m=0,1,2$ to the equation (\ref{lapest}) for the fields $\phi_{k}, g_{k}, T_{m,k}, 
V_{m,k}$ for
$m=0,1,2$. Rearranging the result conveniently and using the observation above we get that $\phi_{l}-\phi_{k}$ is a Cauchy sequence in 
$H^{1}_{\atlas_{\infty}}$ (for {\it item 2}) and in $H^{0}_{\atlas_{\infty}}$ (for {\it item 3}). The claim then follows.\ep

\begin{Remark}\label{RRA} {\rm One can obtain exactly the same kind of estimates as in {\it items 1} and {\it 2} in 
Proposition \ref{EEII} for equations of the form
\ben
-\Delta \phi + f\phi=h,
\een

\n where $f$ is in $H^{2}$ and $f>f_{inf}>0$. This time the estimates depend also on the $H^{2}_{\atlas}$-norm of $f$ and $f_{inf}$. 
The estimates, in particular, will be required to estimate the $H^{0}_{\atlas}$-norm of $N\dooo$ (the third time derivative of the lapse).}
\end{Remark}

\begin{Prop}\label{lem1} Let $\phi_{*}{\bf g}$ be a flow solution in $\Sigma\times I$ with BR-norm out of a $H^{4}$-atlas $\atlas$ 
bounded by $\Lambda$. Then, the norms of
$g,g'$ and $K$ in $C^{0}(I',H_{\atlas}^{2})$, $C(I',2,2)(H_{\atlas})$ and $C(I',2,2)(H_{\atlas})$ respectively and 
the norms of $N$, $X$ and $X\dod$ in $C(I',3,3)(H_{\atlas})$, $C^{0}(I',2)(H_{\atlas})$ and $C(I',2,2)(H_{\atlas})$ 
respectively, are
controlled by $\Lambda$, where $I'$ is a subinterval of $I$ with size controlled from below by $\Lambda$.
\end{Prop}  

\noindent {\bf Proof:} 

The idea for the proof is to look at the equations
\begin{equation}\label{1}
g\dod =-2NK+{\mathcal{L}}_{X}g,
\end{equation}
\begin{equation}\label{2}
K\dod=-\nabla^{2}N+N(E - K\circ K)+{\mathcal{L}}_{X}K,
\end{equation}
\begin{equation}
-\Delta N +|K|^{2}N=1,
\end{equation}
\begin{equation}\label{4}
E\dod=N Curl B-(\nabla N\wedge B)-\frac{5}{2}N(E\times K)-\frac{2}{3}N(E.K)g-\frac{1}{2}kNE,
+{\mathcal{L}}_{X}E,
\end{equation}
\begin{equation}\label{5}
B\dod=-N Curl E-(\nabla N\wedge E)-\frac{5}{2}N(B\times K)-\frac{2}{3}N(B.K)g-\frac{1}{2}kNB
+{\mathcal{L}}_{X}B,
\end{equation}

\noindent and observe that the time derivatives of $g$ and $K$ can be calculated using recursively those equations 
and its time derivatives. There are delicate points however that have to de addressed with care. 

Without loss of generality we will prove the proposition 
with $\atlas$ a canonical
atlas for the initial metric $g(0)$\footnote{The initial metric $g(0)$ is always assumed to be in $H^{3}$}. We will make 
that assumption on $\atlas$ from now on.  

{\it Observation 1.} By the definition of the $BR$-norm, $g$ is in $C^{0}(I,2)(H_{\atlas})$ with norm 
controlled by $\Lambda$. Therefore, given $\epsilon>0$, there is an interval $I'\subset I$ with a 
size controlled from below by $\epsilon$ and $\Lambda$ such that the atlas 
${\mathcal{A}}(t)$ is well defined for all $t\in I'$ and every chart satisfies 
$\|x^{k}_{\alpha}(t)-x^{k}_{\alpha}(0)\|_{C(I',3)(H_{\mathcal{A}})}\leq \epsilon$. 
As a result, if $\epsilon$ is chosen sufficiently small, then for any $t\in I'$ the identity 
$id:(\Sigma,{\mathcal{A}}(0))\rightarrow (\Sigma,{\mathcal{A}}(t))$
is a $H^{3}$-diffeomorphism\footnote{The fact that $id$ is a diffeomorphism is somehow standard and 
can be elaborated from the following fact. Say 
$\phi:\field{R}^{3}\rightarrow \field{R}^{3}$
is a $H^{3}_{\field{R}^{3}}$ map being the identity on $B(0,1)^{c}$ and with $\|\phi-id\|_{H^{3}_{\field{R}^{3}}}\leq \epsilon$. 
Then if $\epsilon$ is sufficiently small $\phi^{-1}$ exists, is in $H^{3}_{\field{R}^{3}}$ and has 
$H^{3}_{\field{R}^{3}}$ norm controlled by $\epsilon$}. Moreover, for any tensor field $U$, the norm $\|U\|_{H^{s}_{\atlas(t)}}$ controls
the norm $\|U\|_{H^{s}_{\atlas}}$ and vice versa, for any $t\in I'$. 
 
{\it Observation 2.} As an application of the {\it Observation 1}, note that the norms $\|g\|_{C^{0}(I,2)(H_{{\mathcal{A}(t)}})}$, 
$\|K\|_{C^{0}(I,1)(H_{\atlas(t)})}$, $\|(E_{0},B_{0})\|_{C^{0}(I,0)(H_{\atlas(t)})}$,
$\|(E_{1},B_{1})\|_{C^{0}(I,0)(H_{\atlas(t)})}$ are controlled by $\La$. Therefore, by the results in Section \ref{3.3.2}, for any 
$t\in I'$ the norms $\|g\|_{H^{3}_{\atlas(t)}}$, $\|K\|_{H^{2}_{\atlas(t)}}$, $\|N\|_{H^{3}_{\atlas(t)}}$, $\|(E_{0},B_{0})\|_{H^{1}_{\atlas(t)}}$, 
$\|(E_{1},B_{1})\|_{H^{0}_{\atlas(t)}}$ are controlled by $\La$. From the {\it Observation 1} above so are controlled the same norms
but with respect to the atlas $\atlas$. 

The proof goes in a ladder-like scheme. We follow the order  
\ben
g,K,N,X\Rightarrow g\dod,K\dod,N\dod,X\dod\Rightarrow g\doo,K\doo,N\doo,X\doo\Rightarrow g\dooo,N\dooo,X\dooo.
\een 

\n Namely, proceeding from left to right, once we prove that a field is controlled (in its respective norm) we use that control to prove
control on the next field (in its respective norm). 

\n 1. $\|g\|_{C^{0}(I',2)(H_{\atlas})}$ {\it controlled by $\La$}. Direct from the definition of BR-norm.

\n 2. $\|K\|_{C^{0}(I',2)(H_{\atlas})}$ {\it controlled by $\La$}. Direct from {\it Observation 2}. 

\n 3. $\|N\|_{C^{0}(I',3)(H_{\atlas})}$ {\it controlled by $\La$}. Direct from {\it Observation 2}.

\n 4. $\|X\|_{C^{0}(I',2)(H_{\atlas})}$ {\it controlled by $\La$}. As $X$ is admissible we get, applying {\it item 1} in 
the Definition  \ref{Admissible} with $j=3$ and with atlas $\atlas(t)$, that $\|X\|_{H^{3}_{\atlas(t)}}$
is controlled by $\Lambda$. This {\it item (4)} then follows by {\it Observation 1}. 

\vspace{0.2cm}
\n 5. $\|g\dod\|_{C^{0}(I',2)(H_{\atlas})}$ {\it controlled by $\La$}. From equation (\ref{1}) and {\it Observation 2} 
(or {\it items 2} and {\it 3})
we see that we need to control the $H^{2}_{\atlas}$-norm of ${\mathcal{L}}_{X}g$. This follows from the formula 
\ben
({\mathcal{L}}_{X}g)_{ab}=\na_{a}X_{b}+\na_{b}X_{a},
\een

\n and {\it Observation 1}. 

\n 6. $\|K\dod \|_{C^{0}(I',1)(H_{\atlas})}$ {\it controlled by $\La$}. From equation (\ref{2}) and {\it Observation 2} we see that
we need to control the $H^{1}_{\atlas}$-norm of ${\mathcal{L}}_{X}K$. This follows from the formula
\beq\label{Lieder}
{\mathcal{L}}_{X}U_{ab}=\na_{X}U_{ab}+U_{ac}\na_{b}X^{c}+U_{bc}\na_{a} X^{c}.
\eeq 

\n with $U=K$, {\it Observation 1} and {\it item 4}.

\n 7. $\|N\dod\|_{C^{0}(I',2)(H_{\atlas})}$ {\it controlled by $\La$}. Differentiating the lapse equation with respect to time 
we get 
\begin{equation}\label{Nder}
-\Delta N\dod+|K|^{2}N\dod=-(|K|^{2})\dod N+\Delta\dod N.
\end{equation}

\n Recall that the derivative of the Laplacian has the expression
\begin{equation}\label{lapder}
\Delta\dod f=g\dod^{ab}\nabla_{a}\nabla_{b}f-g^{ab}(\nabla_{a}f)(\nabla^{c}g\dod_{cb}+\frac{1}{2}\nabla_{b}(g^{de}g\dod_{de})).
\end{equation}

\n Using (\ref{lapder}) (with $f=N$) in (\ref{Nder}) and the previous {\it items}, we see that the RHS of (\ref{Nder}) is 
controlled in $H^{1}_{\atlas(t)}$. The elliptic estimates of Proposition \ref{EEII} (see the Remark \ref{RRA}) show that $N\dod$ is 
controlled in $H^{2}_{\atlas(t)}$
(in fact in $H^{3}_{\atlas(t)}$) by $\La$ and therefore by {\it Observation 1} so in $H^{2}_{\atlas}$.

\n 8. $\|X\dod\|_{C^{0}(I',2)(H{\atlas})}$ {\it controlled by $\La$}. As $X$ is 
admissible we can apply {\it item 2} in Definition \ref{Admissible} with $j=2$, $k=2$ and atlas $\atlas$ to get that $X\dod$ 
is controlled in $H^{2}_{\atlas}$ by $\La$.

\vspace{0.2cm}
\n 9. $\|g\doo\|_{C^{0}(I',1)(H_{\atlas})}$ {\it controlled by $\La$}. Differentiating equation (\ref{1}) in time we get
\begin{equation}\label{g2d}
g\doo=-2N\dod K-2N(K)\dod +({\mathcal{L}}_{X}g)\dod.
\end{equation}

\n The first two terms in the previous equation are controlled in $H^{1}_{\atlas}$ by the {\it items 6} and {\it 7}. 
Let us consider 
the derivative in time of the Lie derivative of $g$ with respect to $X$. Differentiating 
${\mathcal{L}}_{X}g_{ab}=\nabla_{a}X_{b}+\nabla_{b}X_{a}$ with respect to time we get
\begin{equation}\label{bored}
({\mathcal{L}}_{X}g)_{ab}\dod=\nabla_{a}\dod X_{b}+\nabla_{b}\dod X_{a}+\nabla_{a}X\dod_{b}+\nabla_{b}X\dod_{a}.
\end{equation}

\n Recalling how the time derivative of $\nabla$ was calculated in Proposition \ref{globlap}, a 
direct inspection of all the terms of $\na\dod_{a}X_{b}$ show that $\na_{a}\dod X_{b}$
is controlled in $H^{1}_{\atlas}$. The control in $H^{1}_{\atlas}$ of the last two terms in equation (\ref{bored}) follows 
from {\it item 8}.  

\n 10. $\|K\doo\|_{C^{0}(I',0)(H_{\atlas})}$ {\it controlled by $\La$}. Differentiate equation (\ref{2}) with respect to time. We get
\begin{equation}
K\doo=-(\nabla^{2})\dod N-\nabla^{2}N\dod+N\dod(E-K\circ K)+N(E\dod-(K\circ K)\dod)+({\mathcal{L}}_{X}K)\dod.
\end{equation}

\n A direct inspection of this equation and use of the previous {\it items} shows that all the terms in it, except perhaps $NE\dod$ and $({\mathcal{L}}_{X}K)\dod$,
are controlled in $H^{0}_{\atlas}$ by $\Lambda$. To check that the term $NE\dod$ is controlled in $H^{0}_{\atlas}$ examine equation (\ref{4}) 
and use the previous {\it items}. Recalling that ${\mathcal{L}}_{X}K=\na_{X}K_{ab}+K_{ac}\na_{b}X^{c}+K_{bc}\na_{a}X^{c}$ 
and using the previous {\it items}, we see that the term $({\mathcal{L}}_{X} K)\dod$ is controlled in $H^{0}_{\atlas}$ by $\La$.

\n 11. $\|N\doo\|_{C^{0}(I',1)(H_{\atlas})}$ {\it controlled by $\La$}. Differentiate (\ref{Nder}) once more with respect to time to get
\beq\label{Nsecder}
-\Delta N\doo +|K|^{2}N\doo=2\Delta\dod N\dod-2(|K|^{2})\dod N\dod+\Delta\doo N+(|K|^{2})\doo N.
\eeq

\n We want to apply Proposition \ref{EEII} (see Remark \ref{RRA}). We check that the RHS
of equation (\ref{Nsecder}) is controlled in $H^{0}_{\atlas}$ by $\La$. The first term $\Delta\dod N\dod$ 
in the RHS of equation (\ref{Nsecder}) is calculated using
equation (\ref{lapder}) with $f=N\dod$. A direct inspection of its terms using {\it items 5} and {\it 7} show that they are 
controlled in $H^{0}_{\atlas}$ by $\La$. Now let us consider the time derivative of $|K|^{2}$. Writing 
$|K|^{2}=K_{ab}K_{cd}g^{ac}g^{bc}$, the time derivative is 
\beq\label{Kder1}  
|K|^{2}\dod=2K_{ab}K_{cd}g\dod^{ac}g^{bd}+2K\dod_{ab}K_{cd}g^{ac}g^{bd}.
\eeq

\n Differentiating once more we get
\beq\label{Kder2}
\begin{split}
|K|^{2}\doo=&2K\doo_{ab}K_{cd}g^{ac}g^{bd}+2K\dod_{ab}K\dod_{cd}g\dod^{ac}g^{db}+4K\dod_{ab}K_{cd}g\dod^{ac}g^{bd}\\
&4K\dod_{ab}K_{cd}g^{ac}g^{bd}+2K_{ab}K_{cd}g\doo^{ac}g^{bd}+2K_{ab}K_{cd}g\dod^{ac}g\dod^{bd}.
\end{split}
\eeq  
 
\n Inspecting equations (\ref{Kder1}) and (\ref{Kder2}) and using {\it items 5,6,9} and {\it 10} one gets 
that $|K|^{2}\dod$ and $|K|^{2}\doo$ are
controlled in $H^{1}_{\atlas}$ and $H^{0}_{\atlas}$ respectively by $\La$. This shows that the second and fourth terms in equation
(\ref{Nsecder}) are controlled in $H^{0}_{\atlas}$. Finally, let us consider the term $\Delta\doo N$. For this we differentiate 
$\Delta\dod$ in equation (\ref{lapder}) once more (leaving $f=N$ invariant). We get
\ben
\begin{split}
\Delta\doo N=&g\doo^{ab}\na_{a}\na_{b}N+g\dod^{ab}\na\dod_{a}\na_{b}N-g\dod^{ab}\na_{a}N(\na^{c}g\dod_{ab}+\frac{1}{2}\na_{a}(g\dod^{de}g\dod_{de})\\
&+g^{ab}\na_{a}N(\na\dod^{c}g\dod_{cb}+\na^{c}g\doo_{cb}+\frac{1}{2}\na_{b}(g\dod^{dc}g\dod_{de}+
g^{de}g\doo_{de})).
\end{split}
\een

\n A direct inspection of this equation using {\it items 1, 3, 5} and {\it 9} and the expression of $\na\dod$ found before 
shows that $\Delta\doo N$ 
is also controlled in $H^{0}_{\atlas}$ by $\La$.   

\n 12. $\|X\doo\|_{C^{0}(I',1)(H_{\atlas})}$ {\it controlled by $\La$}. As $X$ is admissible this {\it item} follows applying item two in 
Definition \ref{Admissible} with $j=2$, $k=1$ and atlas $\atlas$.

\vspace{0.2cm}
\n 13. To control $\|g\dooo\|_{C^{0}(I',0)(H_{\atlas})}$, $\|N\dooo\|_{C^{0}(I',0)(H_{\atlas})}$ and 
$\|X\dooo\|_{H^{0}_{\atlas}}$ proceed in the same fashion as in the previous {\it items}.\ep.

Let $\{x_{\alpha}\}$ be the $\delta-$canonic dynamical as was defined after Definition \ref{g}. Note as it was defined, it is only a 
$H^{3}$-atlas and not a $H^{4}$-atlas as required in Definition \ref{g}. In the proposition below we discuss this and further properties
of $\atlas(t)$ in relation with Definition \ref{g}. Recall the coordinates $x_{\alpha}$ are defined over $B(o_{\alpha},r_{2}/2)$ where
$r_{2}$ is the $H^{2}$-harmonic radius of the metric $\tilde{g}$. In addition those charts can be extended to $B(o_{\alpha},\delta r_{2})$
satisfying the properties (\ref{delhrad1}) and (\ref{delhrad2}). 

\begin{Prop}\label{lem3}
A flow solution $\phi_{*}{\bf g}$ in $\Sigma\times I$ with BR-norm bounded by $\Lambda$ has the following properties 
(at least on a smaller interval $I(\Lambda)$)

\begin{enumerate}
\item $\Lambda$ controls the $C(I,4,4)(H_{x_{\beta}})$ norm of the transition functions 
$x_{\alpha}(x_{\beta})$.

\item $\Lambda$ controls the $C(I,3,3)(H_{x_{\beta}})$ norm of ${\bf g}=(g,N,X_{{\bf x}_{\beta}})$ where 
$X_{{\bf x}_{\beta}}$ is the shift vector of the coordinates ${\bf x}_{\beta}$.

\item Let ${\bf g}(\lambda)$ be a path of metrics parameterized by $\lambda$. Then the BR-norm of 
${\bf g}'=\frac{d{\bf g}}{d\lambda}$ i.e.
\ben
\|(\phi_{*}{\bf g})'\|=\|g'\|_{C^{0}(I,2)(H_{{\mathcal{A}}(0)})}+\|K'\|_{C^{0}(I,1)(H_{{\mathcal{A}}(0)})}+
\sum_{j=0}^{j=1}\|(E'_{j},B'_{j})\|_{C^{0}(I,0)(H_{{\mathcal{A}}(0)})},
\een

controls the norms of $x_{\alpha}'=d x_{\alpha}/d\lambda$ and ${\bf g}'$ in $C(I,4,4)(H_{x_{\beta}})$ and 
$C(I,3,3)(H_{x_{\beta}})$ respectively. We will assume without lost of generality that the BR-norm of $(\phi_{*}{\bf g})'$
is bounded by $\Lambda$ too (i.e. the same as the bound for the BR-norm of the solution).

\end{enumerate}
\end{Prop} 

\begin{Remark}\label{RLem3} {\rm i. In {\it item 1} the time derivatives
of $x_{\alpha}$ (needed in the norm $C(I,4,4)(H_{x_{\beta}})$) are with respect to the time vector field defined by 
the chart $\{x_{\beta}(t),t\}$.

ii. Note that {\it item 1} in Proposition \ref{lem3} implies that when $\Sigma\times I$ is provided with the set of charts 
$\{\{(x_{\alpha}(t),t)\},\alpha=1,\ldots,m\}$ it makes it a $H^{4}$-four-manifold. 

iii. {\it Item 2} in Proposition \ref{lem3} 
shows that ${\bf g}$ is a $H^{3}$-metric in $\Sigma\times I$ with the $\{\{(x_{\alpha}(t),t)\}\}$ atlas.

iv. In {\it item 3} we use the convention that (as we have been assuming) the $C(I,3,3)(H_{x_{\beta}})$-norm of ${\bf g}'$ is the
$C(I,3,3)(H_{x_{\beta}})$-norm of each of its horizontal components. To find which the components are, pick two $\lambda$-independent
horizontal vectors $V^{a}$ and $W^{a}$ and compute
\ben
{\bf g}'_{ab}V^{a}W^{b}=g'_{ab}V^{a}W^{b},
\een

\n and
\ben
{\bf g}'_{ab}V^{a}T^{b}=-{\bf g}_{ab}V^{a}T'^{b}.
\een

\n Observing that
\ben
0=(\frac{\partial}{\partial t})'=N'T+NT' +X',
\een

\n we get
\ben
{\bf g}'_{ab}V^{a}T^{b}=\frac{X'^{b}}{N}{\bf g}_{ab}V^{a}.
\een

\n Finally
\ben
{\bf g'}_{ab}T^{a}T^{b}=-2{\bf g}_{ab}T^{a}T'^{b}=2\frac{N'}{N}.
\een

\n It follows that the components of ${\bf g}'$ form a set equivalent to $(g',N',X')$. If we prove that $(g',N',X')$ are in 
$C(I,3,3)(H_{x_{\beta}})$ with norm controlled by $\La$ the same will be true for ${\bf g}'$. We will do that when proving the
proposition.

v. It is straightforward to prove from {\it item 3} in Proposition \ref{lem3} that $(g',N',x')$ is in $C(I,3,3)(H_{\atlas})$ and
$(g'\dod,N'\dod,X'\dod)$ is in $C(I,2,2)(H_{\atlas})$ with norm controlled by $\La$. We will use this fact when discussing the initial
value formulation in the CMC gauge.}

\end{Remark}

\vspace{0.2cm} 
\noindent {\bf Proof:} 

{\it Item 1}. {\it Step 1}. We show first that the transition functions $x_{\alpha}(x_{\beta})$ have
$H^{4}_{x_{\beta}}$ norm  controlled by $\Lambda$. By Proposition \ref{AB} we know $\|Ric(t)\|_{H^{1}_{\atlas}}$ is controlled by $\Lambda$. 
For the same argument as in Proposition \ref{RICC} this implies that the norm $\|g\|_{H^{3}_{x_{\beta}}(B(o_{\beta},\delta r_{2}))}$ is controlled
by $\La$. Now, express the harmonic condition of the coordinate $x_{\alpha}^{k}$ over $\{x_{\beta}\}$. We have
\ben
g^{ij}\partial_{x^{i}_{\beta}}\partial_{x^{j}_{\beta}}x_{\alpha}^{k}+(\frac{1}{\sqrt{g}}\partial_{x^{i}_{\beta}}\sqrt{g}g^{ij})\partial_{x^{j}_{\beta}}x^{k}_{\alpha}=0.
\een

\n We can apply the elliptic estimates of Proposition \ref{Pe1} (I) to get that 

\n $\|x_{\alpha}\|_{H^{4}_{x_{\beta}}(B(o_{\beta},r_{2}/2)\cap B(o_{\alpha},r_{2}/2))}$
is controlled by $\La$.

{\it Step 2.} Next we prove that the norms $\|x_{\alpha}\dod\|_{H^{3}_{x_{\alpha}}(B(o_{\alpha},r_{2}/2))}$, 
$\|x_{\alpha}\doo\|_{H^{2}_{x_{\alpha}}(B(o_{\alpha},r_{2}/2))}$, $\|x_{\alpha}\dooo\|_{H^{1}_{x_{\alpha}}(B(o_{\alpha},r_{2}/2))}$ 
and $\|x_{\alpha}\doooo\|_{H^{0}_{x_{\alpha}}(B(o_{\alpha},r_{2}/2))}$ are controlled by $\Lambda$ and where the dot means
derivative with respect to $\partial_{t}=NT+X$ (and not the time vector field of the chart $\{(x_{\beta}(t),t)\}$ as we will do in {\it Step 3}). 
We will prove that studying the defining equation for $x_{\alpha}^{k}$ on the torus. Control will be obtained in the order
\ben
x_{\alpha}\Rightarrow x_{\alpha}\dod\Rightarrow x_{\alpha}\doo\Rightarrow x_{\alpha}\dooo\Rightarrow x_{\alpha}\doooo.
\een

\n We have defined $x^{k}_{\alpha}$ by solving
\beq\label{mm1}
\Delta_{\bar{g}(t)} x^{k}_{\alpha}=h^{k},
\eeq

\noindent subject to the condition
\beq\label{mm2}
\int_{T^{3}}x^{k}_{\alpha}dv_{\bar{g}(t)}=0.
\eeq

\n The idea is to differentiate consecutively these equations with respect to time an in each step use the control gotten in
the previous step. Differentiating both equations with respect to time we get
\begin{equation}\label{5}
\Delta (x^{k}_{\alpha})\dod=-\Delta\dod x^{k}_{\alpha},
\end{equation}

\noindent and
\begin{equation}\label{6}
\int_{T^{3}}(x^{k}_{\alpha})\dod dv_{\bar{g}(t)}=-\int_{T^{3}} x^{k}_{\alpha}\xi(-Nk+\nabla^{a}X_{a})dv_{\bar{g}}.
\end{equation}

\noindent Using formula (\ref{lapder}) for the derivative of the Laplacian we see that the RHS of equation (\ref{5}) 
is controlled in $H^{1}_{T^{3}}$
by $\Lambda$. We split $(x^{k}_{\alpha})\dod=(x^{k}_{\alpha})\dod_{0}+c$ where $c$ is a constant and $(x^{k}_{\alpha})\dod_{0}$ 
has average zero. From Proposition \ref{EEII} we conclude that $\|(x^{k}_{\alpha})\dod_{0}\|_{H^{3}_{T^{3}}}$ is 
controlled by $\Lambda$ and using equation (\ref{6}) we conclude that $c$ is controlled by $\Lambda$ too. As a result 
$\|x_{\alpha}\dod\|_{H^{3}_{x_{\alpha}}(B(o_{\alpha},r_{2}/2))}$ is controlled by $\La$ too. 

We show next that the norms $\|x_{\alpha}\doo\|_{H^{2}_{x_{\alpha}}(B(o_{\alpha},r_{2}/2))}$, 
$\|x_{\alpha}\dooo\|_{H^{1}_{x_{\alpha}}(B(o_{\alpha},r_{2}/2))}$ and
$\|x_{\alpha}\doooo\|_{H^{0}_{x_{\alpha}}(B(o_{\alpha},r_{2}/2))}$ are controlled by $\La$. We proceed exactly as 
we did before. Differentiate equations (\ref{5}) and (\ref{6}) and use Proposition \ref{EEII}. A lengthy but straightforward calculation 
shows that every time equation (\ref{5}) is differentiated we can apply Proposition \ref{EEII}. In other words the RHS of the
equation
\ben
\Delta x^{k (i)}_{\alpha}=\sum_{n=0}^{n=i}c_{n}\Delta^{(n)} x^{k (i-n)}_{\alpha},
\een

\n ($(\star)$ denotes the $\star$-derivative) for each $i=1,2,3,4$ has the structure of $h$ in Proposition \ref{EEII}. The claim the
follows.  

{\it Step 3}. We are now in condition to finish the proof of {\it item 1}. We will use the notation 
$H^{*}_{x_{\beta}}$ instead of $H^{*}_{x_{\beta}}(B(o_{\beta},r_{2}/2)\cap B(o_{alpha},r_{2}/2))$. 
Let us denote ${\bf x_{\beta}}=(x_{\beta}(t_{\beta}),t_{\beta})$
and ${\bf x_{0}}=(x_{\beta}(0),t_{0})$ where $\{x_{0}=x_{\beta}(0)\}$ is a time independent chart. The functions $t_{\beta}$ and $t_{0}$ are 
the same and just $t$ (the subindex is to differentiate $\partial_{t_{0}}$ from $\partial_{t_{\beta}}$). For any function $f$ the chain 
rule gives
\ben
\frac{\partial f}{\partial t_{0}}=\frac{\partial f}{\partial t_{\beta}}+\frac{\partial 
f}{\partial x_{\beta}^{i}}\frac{\partial x_{\beta}^{i}}{\partial t_{0}}.
\een 
   
\n From this we conclude that 
\beq\label{partX}
\frac{\partial}{\partial_{t_{\beta}}}=NT+X-\frac{\partial x_{\beta}^{i}}{\partial t_{0}}\frac{\partial}{\partial x_{\beta}^{i}}.
\eeq

\n Now suppose we want to show that $\|\partial_{t_{\beta}}x_{\alpha}^{k}\|_{H^{3}_{x_{\beta}}}$ is controlled by $\La$. Apply 
$\partial t_{\beta}$
using formula (\ref{partX}) over $x_{\alpha}^{k}$ (for $k=1,2,3$). We get
\ben
\frac{\partial x_{\alpha}^{k}}{\partial t_{\beta}}=x_{\alpha}^{k}\dod-x_{\beta}^{i}\dod\ \frac{\partial x_{\alpha}^{k}}{\partial x_{\beta}^{i}}.
\een

\n {\it Steps 1} and {\it 2} show that both terms on the RHS of the previous equation are controlled in $H^{3}_{x_{\beta}}$. 
Showing control over the other time derivatives follows the same strategy. To illustrate the procedure let us prove that 
$\|\partial_{t_{\beta}}^{2}x^{k}_{\alpha}\|_{H^{2}_{x_{\beta}}}$ is controlled by $\La$ too. 
The remaining two cases are done in the same fashion. Apply the expression
(\ref{partX}) twice over $x_{\alpha}^{k}$. We get
\ben
\frac{\partial^{2} x_{\alpha}^{k}}{\partial t_{\beta}^{2}}=\frac{\partial^{2} x_{\alpha}^{k}}{\partial t_{0}^{2}}-
\frac{\partial x^{i}_{\beta}}{\partial t_{0}}\frac{\partial }{\partial x^{i}_
{\beta}}(\frac{\partial x_{\alpha}^{k}}{\partial t_{0}})-\frac{\partial}{\partial t_{0}}(\frac{\partial x_{\beta}^{l}}
{\partial t_{0}}\frac{\partial x_{\alpha}^{k}}{\partial x^{l}_{\beta}})+
\frac{\partial x^{i}_{\beta}}{\partial t_{0}}\frac{\partial }{\partial x_{\beta}^{i}}(\frac{\partial x_{\beta}^{l}}{\partial t_{0}}\frac{\partial x_{\alpha}^{k}}{\partial x_{\beta}^{l}}).
\een

\n From {\it Steps 1} and {\it 2} we see that all the terms in the RHS of the previous equation except perhaps the third are controlled in $H^{2}_{x_{\beta}}$.
To treat the third we compute
\beq\label{ppp}
\frac{\partial}{\partial t_{0}}(\frac{\partial x_{\beta}^{l}}{\partial t_{0}}\frac{x_{\alpha}^{k}}{\partial x_{\beta}^{l}})=
\frac{\partial^{2}x_{\beta}^{l}}{\partial t_{0}^{2}}\frac{\partial x_{\alpha}^{k}}{\partial x_{\beta}^{l}}+
\frac{\partial x_{\beta}^{l}}{\partial t_{0}}\frac{\partial^{2}x_{\alpha}^{k}}{\partial t_{0}\partial x_{\beta}^{l}}.
\eeq

\n The first term on the RHS of this last equation is controlled in $H^{2}_{x_{\beta}}$ by {\it Steps 1} and {\it 2}. Let us consider the
factor 
\beq\label{pppx}
\frac{\partial}{\partial t_{0}}\frac{\partial x_{\alpha}^{k}}{\partial x_{\beta}^{l}}, 
\eeq

\n on the second term of the RHS of equation (\ref{ppp}). As we know that the factor $\frac{\partial x_{\beta}^{l}}{\partial t_{0}}$
is controlled in $H^{2}_{x_{\beta}}$ we would like to prove that the term (\ref{pppx}) is also controlled in $H^{2}_{x_{\beta}}$. Rewrite
(\ref{pppx}) as
\ben
\frac{\partial }{\partial t_{0}}(\frac{\partial x_{\alpha}^{k}}{\partial x_{0}^{m}}\frac{\partial x_{0}^{m}}{\partial x_{\beta}^{l}})=
\frac{\partial^{2} x_{\alpha}^{m}}{\partial x_{0}^{m}\partial t_{0}}\frac{\partial x_{0}^{m}}{\partial x_{\beta}^{l}}+
\frac{\partial x_{\alpha}^{k}}{\partial x_{0}^{m}}\frac{\partial }{\partial t_{0}}(\frac{\partial x_{0}^{m}}{\partial x_{\beta}^{l}}).   
\een

\n The first term in the previous equation is controlled by $\La$ in $H^{2}_{x_{\beta}}$ by {\it Steps 1} and {\it 2}. The second term is 
treated using
\ben
\frac{\partial x_{0}^{l}}{\partial x_{\beta}^{l}}\frac{\partial x_{\beta}^{l}}{\partial x_{0}^{j}}=\delta^{i}_{\ j}.
\een 

\n From this we compute
\ben
\frac{\partial^{2} x_{0}^{m}}{\partial t_{0}\partial x_{\beta}^{l}}=-\frac{\partial x_{0}^{j}}{\partial x_{\beta}^{l}}
\frac{\partial x_{0}^{m}}{\partial x_{\beta}^{s}}\frac{\partial^{2} x_{\beta}^{s}}{\partial t_{0}\partial x_{0}^{j}}.
\een 

\n It follows from {\it Steps 1} and {\it 2} that this terms is controlled in $H^{2}_{x_{\beta}}$ too.

{\it Item 2}. Checking this property is straightforward. We will use the notation $H^{*}_{x_{\beta}}$ instead of 
$H^{*}_{x_{\beta}}(B(o_{\beta},r_{2}/2))$. We will only check that $\|X_{{\bf x}_{\beta}}\|_{H^{3}_{x_{\beta}}}$ and 
$\|{\mathcal{L}}_{\partial t_{\beta}}X_{{\bf x}_{\beta}}\|_{H^{2}_{x_{\beta}}}$ are
controlled by $\La$. The remaining two cases are carried out in the same fashion. From equation \ref{partX} we get 
$X_{{\bf x_{\beta}}}=X-\frac{\partial x_{\beta}^{i}}{\partial t_{0}}\frac{\partial}{\partial x_{\beta}^{i}}$. We know 
$\|X\|_{H^{3}_{x_{\beta}}}$ is controlled by $\La$ and from {\it item 1} we know 
$\|\partial_{t_{0}}x_{\beta}^{k}\|_{H^{3}_{x_{\beta}}}$ is controlled by $\La$. It follows that $\|X_{{\bf x}_{\beta}}\|_{H^{3}_{\beta}}$
is controlled by $\La$ too. Let us consider now the Lie derivative of $X_{{\bf x}_{\beta}}$ in the direction of $\partial t_{\beta}$.
We apply ${\mathcal{L}}_{\partial t_{\beta}}$ first over $\frac{\partial x_{\beta}^{i}}{\partial t_{0}}\frac{\partial }{\partial x^{i}_{\beta}}$ 
and then over $X$. We compute 
\ben
{\mathcal{L}}_{\partial t_{\beta}}\frac{\partial x_{\beta}^{i}}{\partial t_{0}}\frac{\partial }{\partial x_{\beta}^{i}}=
(\frac{\partial}{\partial t_{\beta}}\frac{\partial x_{\beta}^{i}}{\partial t_{0}})\frac{\partial }{\partial x_{\beta}^{i}},
\een

\n and
\ben
\frac{\partial }{\partial t_{\beta}}\frac{\partial x_{\beta}^{i}}{\partial t_{0}}=\frac{\partial^{2} x_{\beta}^{i}}{\partial t_{0}^{2}}
-\frac{\partial x_{\beta}^{i}}{\partial t_{0}}\frac{\partial}{\partial x_{\beta}^{l}}(\frac{\partial x_{\beta}^{l}}{\partial t_{0}}).
\een

\n It follows from the equations above and {\it item 1} that the $H^{2}_{x_{\beta}}$-norm of ${\mathcal{L}}_{\partial t_{\beta}}
\frac{\partial x_{\beta}^{i}}{\partial t_{0}}\frac{\partial }{\partial x^{i}_{\beta}}$ is controlled by $\La$. Consider now 
${\mathcal{L}}_{\partial t_{\beta}}X$. We compute
\ben
{\mathcal{L}}_{\partial_{t_{\beta}}}X={\mathcal{L}}_{\partial_{t_{0}}}X-{\mathcal{L}}_{\frac{\partial x_{\beta}^{i}}{\partial t_{0}}
\frac{\partial }{\partial x_{\beta}^{i}}}X,
\een

\n and

\ben
{\mathcal{L}}_{\frac{\partial x_{\beta}^{i}}{\partial t_{0}}
\frac{\partial }{\partial x_{\beta}^{i}}}X=\frac{\partial x_{\beta}^{i}}{\partial t_{0}}\nabla_{\frac{\partial }{\partial x_{\beta}^{i}}}X
+X^{c}\nabla_{c}(\frac{\partial x_{\beta}^{i}}{\partial t_{0}}\frac{\partial }{\partial x_{\beta}^{i}}).
\een

\n It follows from these two equations, {\it item 1} and Proposition \ref{lem1} that $\|{\mathcal{L}}_{\partial_{t_{\beta}}}X\|_{H^{2}_{x_{\beta}}}$ is controlled by 
$\La$.

{\it Item 3}. We will show that $x_{\beta}'$ and $g',N',X_{x_{\beta}}'$ are controlled in $H^{4}_{x_{\beta}}$ and 
$H^{3}_{x_{\beta}}$ respectively. Here, prime means derivation with respect to $\lambda$. The control of 
their time derivatives (as $\frac{dg'}{dt}$ or $N'\dooo$) in their respective space, is carried out along similar lines and will not be
included. We start showing that $\|x_{\beta}'\|_{H^{3}_{x_{\beta}}}$ 
is controlled by $\Lambda$. We look at the equation defining $\{x_{\beta}\}$ in the torus. 
After differentiating in $\lambda$ we get
\begin{equation}\label{8}
\Delta (x^{k}_{\beta})'=-\Delta' (x^{k}_{\beta}),
\end{equation}

\noindent and
\begin{equation}\label{7}
\int_{T^{3}}(x^{k}_{\beta})'dv_{\bar{g}}=-\int_{T^{3}}x^{k}_{\beta}\xi\frac{tr g'}{2}dv_{\bar{g}}.
\end{equation}

\noindent As $g'$ is controlled in $H^{2}_{{\mathcal{A}}}$ by $\Lambda$ so is controlled the term on the
right of equation (\ref{7}). Using formula (\ref{lapder}) it is seen that the RHS of equation (\ref{8}) 
is controlled in $H^{1}_{T^{3}}$ by $\Lambda$. The elliptic estimates of Proposition \ref{EEII} show that 
$(x^{k}_{\beta})'$ is controlled in $H^{3}_{x_{\beta}}$ by $\Lambda$. Still we want to show that $\|(x_{\beta}^{k})'\|_{H^{4}_{x_{\beta}}}$
is controlled by $\Lambda$. To do this step further we prove next that 
$\|(g_{ij})'\|_{H^{3}_{x_{\beta}}}$ is controlled by $\Lambda$. As
a result if we write equation (\ref{8}) over the $\{x_{\beta}\}$ coordinates 
\begin{equation}
\frac{\partial}{\partial x_{\beta}^{i}}(\sqrt{g}g^{ij}\frac{\partial }{\partial 
x_{\beta}^{j}}(x^{k}_{\beta})')=-
\frac{\partial}{\partial x_{\beta}^{i}}((\sqrt{g}g^{ij})'\frac{\partial }{\partial 
x^{j}_{\beta}}(x^{k}_{\beta})),
\end{equation}

\n we can apply the elliptic estimates of Proposition \ref{Pe1} (I) to see
that $\|(x^{k}_{\beta})'\|_{H^{4}_{x_{\beta}}}$ is controlled by $\Lambda$ too. To show that $\|g'\|_{H^{3}_{x_{\beta}}}$ 
is controlled by $\La$ we proceed as follows. We look at the system 
of equations
\begin{equation}\label{9}
\frac{1}{2}g^{lm}\partial_{l}\partial_{m}g_{ij}+Q(\partial_{*}g_{*},g^{*})_{ij}=
E_{0,ij}+K_{i}^{s}K_{sj}-kK_{ij},
\end{equation}
\begin{equation}\label{10}
\nabla^{a}E_{0,ab}=(K\wedge B_{0})_{b},
\end{equation}
\begin{equation}\label{11}
(Curl\ E_{0})_{ab}=B_{1,ab}+\frac{3}{2}(B_{0}\times K)_{ab}-\frac{1}{2}kB_{0,ab}.
\end{equation}

\noindent The first equation is on the $\{x^{k}_{\beta}(\lambda)\}$ coordinate system (the other two are tensorial) and
for this reason we have to be specially careful when we differentiate with respect to $\lambda$. We utilize the following. 
For any two tensor $T$, it is 
\begin{equation}
T_{ij}=T(\frac{\partial}{\partial x^{i}(\lambda)},\frac{\partial}{\partial x^{j}(\lambda)})=
T(\frac{\partial}{\partial x^{m}(0)},\frac{\partial}{\partial x^{l}(0)})\frac{\partial x^{m}(0)}{\partial 
x^{i}(\lambda)}\frac{\partial x^{l}(0)}{\partial x^{j}(\lambda)}.
\end{equation}

\noindent Thus, when we differentiate with respect to $\lambda$ we get
\begin{equation}\label{12}
(T_{ij})'=(T')_{ij}+T_{il}(\frac{\partial x^{l}(0)}{\partial x^{j}(\lambda)})'+
T_{mj}(\frac{\partial x^{m}(0)}{\partial x^{i}(\lambda)})'.
\end{equation}

\noindent Observe that $(\frac{\partial x^{m}(0)}{\partial x^{i}(\lambda)})'$ is controlled in $H^{2}_{x_{\beta}}$. 
Differentiating the system of equations (\ref{9})-(\ref{11}) with respect to 
$\lambda$ and evaluate it in the $\{x_{\beta}\}$ coordinates. After a straightforward calculation we get
an elliptic system of the form
\begin{equation}
g^{lm}\partial_{l}\partial_{m}(g_{ij})'=((E_{0})'_{ij}+H^{1}_{ij},
\end{equation}
\begin{equation}
\nabla^{i}(E_{0})'_{ij}=H^{0}_{j},
\end{equation}
\begin{equation}
(Curl E_{0})'_{lm}=H^{0}_{lm},
\end{equation}

\noindent where $H^{1}_{ij}$ is a term controlled in $H^{1}_{x_{\beta}}$ by $\Lambda$ and $H^{0}_{j}$ and 
$\ H^{0}_{lm}$ are terms controlled in $H^{0}_{x_{\beta}}$ by $\Lambda$. Standard elliptic estimates show that
$\|(g_{ij})'\|_{H^{3}_{x_{\beta}}}$ is controlled by $\Lambda$. 

Now that we know $x'_{\beta}$ is controlled in
$H^{4}_{x_{\beta}}$ we conclude after inspecting equation (\ref{12}) that $g'$ is controlled in $H^{3}_{x_{\beta}}$
by $\Lambda$. It is straightforward to deduce, after differentiating equation (\ref{K-hat-equation}) in $\lambda$, that
$\|K'\|_{H^{2}_{x_{\beta}}}$ is controlled by $\Lambda$. Differentiating the lapse equation and 
using standard elliptic estimates we get that $N'$ is controlled in $H^{3}_{x_{\beta}}$. Using the definition of admissible gauge
we get that $X'$ is controlled in $H^{3}_{x_{\beta}}$ too.\ep

Before going into the inital value formulation let us mention that the space of flow solutions on an interval $I$ is complete 
under the BR-norm. The proof is straightforward and will not be included. 

\begin{Prop}\label{compl}
Let $(\Sigma,\atlas)$ be a $H^{4}$-manifold. The space of $H^{3}$-flow solutions over an interval $I$ is complete under the   
BR-norm.
\end{Prop}

{\center \section{Applications.}\label{4}}

{\center \subsection{The inital value formulation in the CMC gauge.}\label{3.5}}

\vspace{0.2cm}
\begin{T}\label{IVF} Let $(\Sigma,\atlas)$ be a $H^{4}$-manifold and say $(g_{0},K_{0})$ 
is an initial state in $H^{3}\times H^{2}$ with $k_{0}<0$. Then

\begin{enumerate}

\item There is a unique $H^{3}$-flow solution over an interval $I=(k_{-1},k_{1})$ with $-\infty\leq k_{-1}<k_{0}<k_{1}\leq 0$. 
Moreover the size $inf\{|k_{-1}-k_{0}|,|k_{0}-k_{1}|\}$ of the time interval on which the solution is guaranteed to exist is 
controlled from below by 
$1/\nu(k_{0})$, $\ln |k_{0}|$, ${\mathcal{V}}(g_{0},K_{0})$ and ${\mathcal{E}}_{1}(k_{0})$.

\item There is continuity with respect to the initial conditions if we measure the space of initial conditions with
the $H^{3}_{\atlas}\times H^{2}_{\atlas}$ norm and the space of flow solutions with the BR-norm.

\item Because of {\it item 1} above, we have the following continuity principle: a flow solution 
$((g,K),(N,X))(k)$ is defined 
until past of $k^{*}<0$ (or before $k_{*}<0$ if the flow is running in the past direction) iff 
$lim sup_{k\rightarrow k^{*}} 1/\nu+{\mathcal{V}}(k)+{\mathcal{E}}_{1}(k)<\infty$.

\end{enumerate} 

\end{T}

\begin{Remark} {\rm i. As an outcome of the proof it is seen that if an initial state is $C^{\infty}$ the
flow solution is $C^{\infty}$. Moreover the growth of high-order Bel-Robinson energies ${\mathcal{E}}_{i}$ ($i\geq 2$) 
is at most exponential (i.e. ${\mathcal{E}}_{i}\leq Ce^{rt}$, $i\geq 2$) on any compact subinterval of $I$ with constants $C$ and $r$
depending on the BR-norm of the solution on the subinterval. 

ii. The Theorem \ref{IVF} does not make any a priori hypothesis on the topology of
$\Sigma$. On manifolds with $Y(\Sigma)\leq 0$ we know the range of $k$ is always a subset of $(-\infty,0)$ but on manifolds with 
$Y(\Sigma)>0$ the flow may be defined until $k_{1}=0$ and beyond. The Theorem \ref{IVF} does not discuss a continuity criteria in this case (at 
$k=k_{1}=0$). One should be able to prove however that if $\|\hK\|_{L^{2}_{g}}\geq M>0$ all along the flow and $\lim_{k\rightarrow k_{1}=0}
1/\nu(k)+{\mathcal{V}}(k)+{\mathcal{E}}_{1}(k)<\infty$ then the flow is extendible behind $k_{1}=0$. This situation corresponds to
CMC solutions which are not time symmetric at the slice $\{k=k_{1}=0\}$ (i.e. $K\neq 0$ at $\{k=0\}$). The essential point in this case
is that the lapse equation is solvable and the lapse $N$ corresponding to the choice of time $t=k$ is not degenerating under the assumptions. 
If the condition above on $\|\hK\|_{L^{2}_{g}}$ is not imposed one may still be able to elaborate a criteria to guarantee that the flow is extended 
beyond $k_{1}=0$ in which case the CMC solution is time-symmetric. In this case one should work with the choice of time 
$t=k^{\frac{1}{3}}$. Non of these issues we will treat in this article. 

iii. The Theorem \ref{IVF} above should hold for initial states 
with $H^{i}\times H^{i-1}$ regularity ($i\geq 4$) as well, namely they should give rise to a unique $H^{i}$-flow solution and the continuity criteria being the same
as for $H^{3}\times H^{2}$ regularity. 

iv. A comment on the uniqueness of the solutions is in order. Call $S$ the map that returns 
$C^{\infty}$-solutions from $C^{\infty}$ initial states. Supply the $C^{\infty}$ initial states with the $H^{3}_{\atlas}\times H^{2}_{\atlas}$-norm and 
the $C^{\infty}$ solutions
with the BR-norm (for an appropriate time interval). As will be shown, the map $S$ is Lipschitz and therefore admitting a unique extension
to $H^{3}_{\atlas}\times H^{2}_{\atlas}$. In this sense the solutions are uniquely determined from the initial data. But also, 
as will be proved (and as is required in the definition of {\it flow solutions}), the flow solutions thus constructed are also 
space-time solutions. 
It follows from a well known result (see for instance \cite{Fma} Theorem 4.27 and references therein) that any space-time solution with the
same initial data is diffeomorphic to the space-time solution arising from the flow solution.}

\end{Remark}

\vspace{0.2cm}
\n {\bf Proof:} 

We will prove Theorem \ref{IVF} towards the future, the proof towards the past is naturally the same. We choose $t=k$. 

{\it Item 1}. {\it Step 1.} Smooth the atlas $\atlas$ to make it $C^{\infty}$ but compatible with $\atlas$ at the $4$-th level of regularity. 
We prove that there are $\epsilon$, $\Lambda$ and $I$ such that for any $C^{\infty}$ initial state $(g,K)$ 
with $\|g-g_{0}\|_{H^{3}_{\mathcal{A}}}\leq 
\epsilon$ and $\|K-K_{0}\|_{H^{2}_{{\mathcal{A}}}}\leq \epsilon$ the $C^{\infty}$-solution $\phi_{*}{\bf g}$
is defined on $I$ and has BR-norm bounded by $\Lambda$. For that we start proving that there are $\epsilon$, $\Lambda$ 
and $I$ such that while the $C^{\infty}$-flow solution is defined on $I$, the BR-norm over the interval where the solution is defined is bounded by $\Lambda$. Take $\Lambda$
equal to $C(\|g_{0}\|_{H^{2}_{{\mathcal{A}}}}+\|K_{0}\|_{H^{1}_{\mathcal{A}}}+\|(E_{0}(t_{0}),B_{0}(t_{0}))\|_{H^{0}_{\atlas}}+
\|(E_{1}(t_{0}),B_{1}(t_{0}))\|_{H^{0}_{\atlas}})$ where $C$ is a constant greater than one to be fixed later. 
Suppose there exists $(g_{i},K_{i})$ $C^{\infty}$ initial states converging in 
$H^{3}_{{\mathcal{A}}}\times H^{2}_{{\mathcal{A}}}$ to $(g_{0},K_{0})$
such that the BR-norm $\|(g,K)\|_{BR}$ over a time interval $I=[t_{0},t_{i}]$ is equal to $2\La$, with $t_{i}\rightarrow t_{0}$. 
We can write 

\ben
\|g(t)-g_{0}\|_{H^{2}_{{\mathcal{A}}}}\leq \int_{t_{0}}^{t} \|g\dod\|_{H^{2}_{{\mathcal{A}}}}dt,
\een
\ben
\|K(t)-K_{0}\|_{H^{1}_{{\mathcal{A}}}}\leq \int_{t_{0}}^{t} \|K\dod\|_{H^{1}_{{\mathcal{A}}}}dt.
\een

\noindent Observe that the integrands are controlled by $\Lambda$ by Propositions \ref{lem1}. Recall too that 
$\|{\bf J}({\bf W}_{1})\|_{L^{2}_{g}}$ and $\|\dt\|_{H^{2}_{\atlas}}$ are controlled by $\La$. It follows
from the Gauss equation that $sup_{\{t\in [t_{0},t_{i}]\}} \{{\mathcal{E}}_{1}(t)-{\mathcal{E}}_{1}(t_{0})\}\rightarrow 0$ as 
$t_{i}\rightarrow t_{0}$. We see therefore that when $t_{i}-t_{0}$ is sufficiently small, for a constant $C$ that was fixed big enough\footnote{$C$ is chosen big enough to have
${\mathcal{E}}_{1}(t_{0}\leq  C(\|(E_{0}(t_{0}),B_{0}(t_{0}))\|_{H^{0}_{\atlas}}+
\|(E_{1}(t_{0}),B_{1}(t_{0}))\|_{H^{0}_{\atlas}})$.}, the BR-norm of the solution on the interval $[t_{0},t_{i}]$ is less than $\La$ which is a contradiction. 

Next we prove that for such $\epsilon$, $\La$ and $I$ the $C^{\infty}$-solution $\phi_{*}{\bf g}$ is defined in all of $I$. To show that 
it is enough to prove that $Q_{i}$, $i\geq 2$ remains bounded (while defined) on $I$. In fact by Theorem \ref{teo:cc1} 
and Proposition \ref{SvsBR} we know that: i. if the flow is
defined until $t_{*}$ in $I$, ii. its BR-norm is bounded by $\La$ and all $Q_{i}$ for $i\geq 0$ remain controlled by $\La$, 
then the flow $(g,K)(t)$ must converge in $C^{\infty}$
to a $C^{\infty}$ state. Therefore the flow can be continued beyond $t_{*}$ which is a contradiction. 
Using Proposition \ref{inductionE}, bound the integrand involving ${\bf J}(\W_{i})$ in the Gauss equation for ${\bf W}_{i}$ as 
\ben
\begin{split}
\int_{\Sigma}|N(E_{i}^{ab}{\bf J}({\bf W}_{i})_{aTb}+B_{i}^{ab}{\bf J}(\W_{i})^{*}_{aTb})|dv_{g}\leq &C_{1}({\mathcal{E}}_{i-1},\nu,
\|\hat{K}\|_{L^{2}})Q_{i}^{\frac{1}{2}}(Q_{i}^{\frac{1}{2}}\\
&+C_{2}({\mathcal{E}}_{i-1},\nu,\|\hat{K}\|_{L^{2}})).
\end{split}
\een

\n Thus, from the Gauss equation we get the inequality
\ben
|Q_{i}\dod|\leq C_{1}({\mathcal{E}}_{i-1},\La)(Q_{i}+C_{2}({\mathcal{E}}_{i-1},\La)Q_{i}^{\frac{1}{2}}).
\een

\n Proceeding inductively in $i\geq 2$ using this inequality we get that $Q_{i}$ for $i\geq 2$ has at most exponential growth 
(with constants controlled by $\Lambda$) as desired. 

{\it Step 2:} Until now we have a well defined map $S$ from $C^{\infty}$ initial states $\epsilon$-close to $(g_{0},K_{0})$ in
$H^{3}_{{\mathcal{A}}}\times H^{2}_{{\mathcal{A}}}$ into $C^{\infty}$ solutions defined on $I$ with uniformly 
bounded BR-norm. We will prove now that the map $S$ is Lipschitz if we take $\epsilon$ sufficiently small. It will follow from the completeness of the space of flow solutions that
$S$ can be uniquely extended to a continuous map from $H^{3}\times H^{2}$ initial states $\epsilon$-close to $(g_{0},K_{0})$ 
in $H^{3}_{{\mathcal{A}}}\times H^{2}_{{\mathcal{A}}}$ into $H^{3}$-flow solutions in the sense of Definition \ref{g}. 
To show that $S$ is Lipschitz we proceed as follows. First we prove that if $\epsilon$ is sufficiently small, any two states $(g_{1},K_{1})$
and $(g_{2},K_{2})$ in the ball of center $(g_{0},K_{0})$ and radius $\epsilon$ in $H^{3}_{\atlas}\times H^{2}_{\atlas}$ can be joined 
by a path $(g,K)(\lambda)$, with $\lambda\in [0,1]$, of $C^{\infty}$ initial states and with $\lambda$-derivative controlled in $H^{3}_{\atlas}\times H^{2}_{\atlas}$ by
$\|(g_{0},K_{0})\|_{H^{3}_{\atlas}\times H^{2}_{\atlas}}$ (for any two $C^{\infty}$ states $(g_{1},K_{1})$ and $(g_{2},K_{2})$ as
before, denote by ${\bf g}(\lambda)$ the solutions with initial states
$(g,K)(\lambda)$, $\lambda\in [0,1]$). Then we prove that we can choose $\epsilon$ sufficiently small in such a way that there is a 
subinterval $I'$ of $I$ such that the BR-norm of ${\bf g}'=\frac{d{\bf g}}{d\lambda}$ on $I'$ (see Proposition \ref{lem3} {\it item 3})
\ben
\|(\phi_{*}{\bf g})'\|_{BR}=\|g'\|_{C^{0}(I',2)(H_{{\mathcal{A}}})}+\|K'\|_{C^{0}(I',1)(H_{{\mathcal{A}}})}+
\sum_{j=0}^{j=1}\|(E'_{j},B'_{j})\|_{C^{0}(I',0)(H_{{\mathcal{A}}})},
\een

\n is as small as we want. By Proposition \ref{lem3} this is enough to 
guarantee that the map $S$ is Lipschitz\footnote{By Proposition \ref{lem3} 
when we differentiate the components of ${\bf g}(\lambda)$ with respect to $\lambda$ we do not loose regularity. 
The claim that the map is Lipschitz follows essentially by integrating the fields along the path.} 

Let us start with the construction of the paths $(g,K)(\lambda)$. Define $(\tilde{g},\tilde{K})(\lambda)=\lambda(g_{1},K_{1})+(1-\lambda)
(g_{2},K_{2})$. Then, for every $\lambda$ find the transverse traceless part of $\tilde{K}$ with respect to $\tilde{g}$ and call
it $\tilde{K}_{TT}$ (see \cite{I}). Now note that if $\hat{K_{0}}\neq 0$ then for $\epsilon>0$ sufficiently small
it is $\tilde{K}_{TT}\neq 0$ (there is continuity in the York decomposition) and if $\hat{K}_{0}=0$ then
for $\epsilon$ sufficiently small $\tilde{g}$ is conformally deformable to a metric of constant negative scalar
curvature. Thus, according to \cite{I}, the energy constraint is in all cases conformally solvable. It is direct to see that 
the path of initial states $(g,K)(\lambda)$ constructed in this way has 
$\|(g',K')\|_{H^{3}_{{\mathcal{A}}}\times H^{2}_{{\mathcal{A}}}}$-norm controlled by the $H^{3}_{\atlas}\times H^{2}_{\atlas}$-norm of 
$(g_{0},K_{0})$. 

We next show that $\epsilon$ can be chosen small enough in such a way that at every $\lambda \in [0,1]$ and 
for every path of $C^{\infty}$ initial states as described above, the BR-norm of ${\bf g}'$ on a uniform interval $I'\subset I$ 
is as small as we want. We will get control on $\|{\bf g'}\|_{BR}$ from a contradiction argument. 
We need some preliminary discussion on the $\lambda$-derivative of the Weyl fields ${\bf W}={\bf W}_{0}$ 
and ${\bf W}={\bf W}_{1}$.
 
We first note that the $\lambda$-derivative of a Weyl field is not necessarily a Weyl field as we have
\ben
({\bf W}_{abcd})'{\bf g}^{bd}=-{\bf W}_{abcd}({\bf g}^{bd})'.
\een

\noindent Note that $({\bf W}_{abcd})'{\bf g}^{ac}{\bf g}^{bd}=0$. We can construct from ${\bf W}'$ a Weyl field 
by making it traceless according to the formula
\begin{equation}\label{FFFF1}
\tilde{{\bf W}'}_{abcd}={\bf W}'_{abcd}-\frac{1}{2}({\bf g}_{ac}{\bf W}_{bd}'+{\bf g}_{bd}{\bf W}'_{ac}-
{\bf g}_{bc}{\bf W}_{ad}'-{\bf g}_{ad}{\bf W}_{bc}'),
\end{equation}

\noindent where we have defined ${\bf W}'_{ac}={\bf W}_{abcd}'{\bf g}^{bd}$, which makes it a $(2,0)$ symmetric and traceless tensor. 
We conclude from this that if we control ${\bf g}'$ in $H^{2}_{{\mathcal{A}}}$ (i.e. all of its components at every time slice) 
and $Q(\tilde{{\bf W}}')$ then we
control $\|{\bf W}'\|_{H^{0}_{\atlas}}$ (at every time slice). We will get control on $Q(\tilde{{\bf W}}')$ using the Gauss equation and to use the Gauss equations we need 
an expression for the divergence of $\tilde{\bf W}'$. We compute
\ben
\begin{split}
\tilde{{\bf J}}_{bcd}=\bn^{a}\tilde{\bf W}'_{abcd}=&\bn^{a}{\bf W}'_{abcd}-\frac{1}{2}({\bf g}_{ac}\bn^{a}{\bf W}'_{bd}+{\bf g}_{bd}\bn^{a}{\bf W}'_{ac}
-{\bf g}_{bc}\bn^{a}{\bf W}'_{ad}\\
&-{\bf g}_{ad}\bn^{a}{\bf W}'_{bc}).
\end{split}
\een

\n that we arrange in the form
\begin{equation}\label{rear}
\tilde{{\bf J}}_{bcd}=\bn^{a}{\bf W}'_{abcd}-
\frac{1}{2}(\bn_{c}{\bf W}'_{bd}-\bn_{d}{\bf W}'_{bc})-\frac{1}{2}({\bf g}_{bd}\bn^{a}{\bf W}'_{ac}
-{\bf g}_{bc}\bn^{a}{\bf W}'_{ad}).
\end{equation}

\n We have therefore three different terms that we have to estimate on the RHS of the last equation.
The first term can be computed by
\begin{equation}\label{125Fin}
\bn^{a} {\bf W}'_{abcd}=-\bn^{'a}{\bf W}_{abcd} + {\bf J}'_{bcd}.
\end{equation}

\noindent if we compute ${\bf J}'_{bcd}$. Contracting the last equation we get 
\beq\label{126Fin}
\bn^{a}{\bf W}'_{ac}={\bf J}'_{bcd}{\bf g}^{bd}.
\eeq

\n which can be used instead of each summand in the third term 
of equation (\ref{rear}) if we compute ${\bf J}'_{bcd}$. The second term $\bn_{c}{\bf W}'_{bd}-\bn_{d}{\bf W}'_{bc}$ in equation (\ref{rear}) is calculated by differentiating and 
contracting the identity
\ben
\bn_{a}{\bf W}_{bcde}+\bn_{b}{\bf W}_{cade}+\bn_{c}{\bf W}_{abde}=\frac{1}{3}{\bf\epsilon}_{fabc}{\bf J}^{*f}_{\ \ de}.
\een

\noindent Differentiating we get
\ben
(\bn_{a}'{\bf W}_{bcde}+\bn_{b}'{\bf W}_{cade}+\bn_{c}'{\bf W}_{abde}){\bf g}^{bd}+
\bn_{a}{\bf W}'_{ce}+\bn^{b}{\bf W}'_{beca}-\bn_{c}{\bf W}'_{ae}=
\een
\ben
\ \ \ \ =\frac{1}{3}{\bf\epsilon}'_{fabc}
{\bf J}^{*f}_{\ \ de}+\frac{1}{3}{\bf\epsilon}_{fabc}{\bf J}^{*f}_{\ \ de}{'},
\een

\n and rearranging terms
\beq\label{F122}
\begin{split}
\bn_{a}{\bf W}'_{ce}-\bn_{c}{\bf W}'_{ae}=&-(\bn_{a}'{\bf W}_{bcde}+\bn_{b}'{\bf W}_{cade}+\bn_{c}'{\bf W}_{abde}){\bf g}^{bd}\\
&-\bn^{b}{\bf W}'_{beca}+\frac{1}{3}{\bf\epsilon}'_{fabc}{\bf J}^{*f}_{\ \ de}+\frac{1}{3}{\bf\epsilon}_{fabc}{\bf J}^{*f}_{\ \ de}{'}.
\end{split}
\eeq

\n Now, ${\bf J}({\bf W}_{0})'=0$ and ${\bf J}({\bf W}_{1})'$ has an expression of the form
\begin{equation}\label{F124}
{\bf J}({\bf W}_{1})'=\dt'*\bn {\bf W}_{0} + \dt*\bn' {\bf W}_{0}+\dt*\bn {\bf W}_{0}'+T'*{\bf Rm}*{\bf W}_{0}+
T*{\bf Rm}*{\bf W}_{0}'.
\end{equation}

\n Recall finally that ${\bf J}^{*}_{abc}=\frac{1}{2}{\bf J}_{alm}\epsilon^{lm}_{\ \ bc}$. Using the discussion above, let us give a schematic representation of the currents $\tilde{{\bf J}}(\tilde{\W}'_{0})$ and $\tilde{{\bf J}}(\tilde{\W}'_{1})$. The expressions will be
used in the Gauss equations for $\tilde{\W}'_{0}$ and $\tilde{\W}'_{1}$. The expression for $\tilde{\bf J}(\tilde{\W}'_{0})$ comes from 
using equations (\ref{125Fin}), (\ref{126Fin}) and (\ref{F122}) together with ${\bf J}(\W_{0})'=0$ in equation (\ref{rear}). We can write 
(schematically)
\beq\label{schJJ1}
\tilde{{\bf J}}(\tilde{\W}'_{0})=\bn'*\W_{0}.
\eeq

\n To get a schematic expression for $\tilde{{\bf J}}(\tilde{\W}'_{1})$ use again equations (\ref{125Fin}), (\ref{126Fin}) and (\ref{F122}) 
in equation (\ref{rear}). We can write (schematically)
\ben
\tilde{\bf J}(\tilde{\W}'_{1})=\bn'*\W_{1}+{\bf J}(\W_{1})'*{\bf g}+\epsilon'*\epsilon*{\bf J}(\W_{1})+\epsilon*\epsilon*{\bf J}(\W_{1})'.
\een

\n with the expression for ${\bf J}(\W_{1})'$ borrowed from equation (\ref{F124}).  

From Proposition \ref{lem3} ({\it item 3}) it is direct to show (see also iv. and v. in Remark \ref{RLem3}) that the BR-norms of 
${\bf g}$ and ${\bf g}'$ over an interval $I'$ 
control the norms
\ben
\|\bn '(t)\|_{H^{2}_{\atlas}},\|\dt'(t)\|_{H^{2}_{\atlas}},\|{\bf W}_{0}'(t)\|_{H^{1}_{\atlas}},\|\epsilon'(t)\|_{H^{3}_{\atlas}},
\|T'(t)\|_{H^{2}_{\atlas}}. 
\een

\n for every $t$ in $I'$. Moreover, as $\|{\bf g}'\dod\|_{H^{2}_{\atlas}}$ is controlled by the BR-norms of ${\bf g}$ and ${\bf g}'$ 
(see v. in Remark \ref{RLem3}), we have
\beq\label{A}
\|{\bf g}'(t)-{\bf g}'(t_{0})\|_{H^{2}_{\atlas}}\leq C(t-t_{0}),
\eeq

\n for every $t$ in $I'$, where $C$ depends on the BR-norm of ${\bf g}$ and ${\bf g}'$ on $I'$. The Gauss equation applied to $\tilde{\bf W}_{0}$ gives  
the inequality
\beq\label{GF1}
|Q(\tilde{\bf W}'_{0})\dod|\leq C Q(\tilde{{\bf W}}'_{0}),
\eeq

\n where $C$, again, depends on the BR-norms of ${\bf g}$ and ${\bf g}'$ on $I'$. To see that, use the schematic expression (\ref{schJJ1}) 
for the current $\tilde{{\bf J}}(\tilde{\W}'_{0})$ in the Gauss equation (equation (\ref{Gausseq}))
\ben
\dot{Q}(\tilde{\W}'_{0})=-\int_{\Sigma}2NE^{ij}(\tilde{\bf W}'_{0})\tilde{\bf J}(\tilde{{\bf W}}'_{0})_{iTj}+2NB^{ij}(\tilde{\W}'_{0})
\tilde{{\bf J}}^{*}(\tilde{{\bf W}}'_{0})_{iTj}
+3NQ_{abTT}\dt^{ab}dv_{g},
\een

\n together with the control on the norms $\|\dt\|_{H^{2}_{\atlas}}$ and $\|\bn'\|_{H^{2}_{\atlas}}$ mentioned before. 
Similarly one obtains an inequality for the time derivative of the Bel-Robinson energy associated to $\tilde{\W}'_{1}$
\beq\label{GF2}
|Q(\tilde{{\bf W}}'_{1})\dod|\leq C(Q(\tilde{\bf W}'_{1})+Q(\tilde{\bf W}'_{1})^{\frac{1}{2}}).
\eeq

\n $C$, as above, depends on the BR-norms of ${\bf g}$ and ${\bf g}'$ over $I'$. 
Now, from the inequalities (\ref{GF1}) and (\ref{GF2}) we have 
\beq\label{B}
|Q(\tilde{\bf W}'_{0})(t)-Q(\tilde{\bf W}'_{0})(t_{0})|\leq C(t-t_{0}),
\eeq
\beq\label{C}
|Q(\tilde{\bf W}_{1}')(t)-Q(\tilde{\bf W}_{1}')(t_{0})|\leq C(t-t_{0}).
\eeq

\n where, again, $C$ depends on the BR-norms of ${\bf g}$ and ${\bf g}'$ on $I'$. Fix two $\lambda$-independent horizontal vector fields 
$V^{a}$ and $W^{1}$. For any Weyl field $\W=\W_{0}$ or $\W=\W_{1}$
we have
\beq\label{FF1}
E'_{ab}V^{a}W^{b}=\W'_{acbd}T^{c}T^{d}V^{a}W^{b}+\W_{acbd}T'^{c}T^{d}V^{a}W^{b}+\W_{acbd}T^{c}T'^{d}V^{a}W^{b},
\eeq

\n and recall (see iv. in Remark \ref{RLem3})
\beq\label{FF2}
T'^{a}=-\frac{X'^{a}}{N}-\frac{N'}{N}T^{a}.
\eeq

\n On the other hand
\beq\label{FF3}
{\bf W}'_{abcd}=\tilde{{\bf W}'}_{abcd}+\frac{1}{2}({\bf g}_{ac}{\bf W}_{bd}'+{\bf g}_{bd}{\bf W}'_{ac}-
{\bf g}_{bc}{\bf W}_{ad}'-{\bf g}_{ad}{\bf W}_{bc}'),
\eeq

\n and 
\beq\label{FF4}
{\bf W}'_{ac}=-\W_{abcd}{\bf g}'^{bd}.
\eeq

\n From (\ref{FF1}), (\ref{FF2}), (\ref{FF3}) and (\ref{FF4}) we can write \footnote{Note that inside expression (\ref{ESE}) there are norms
with respect to $\atlas$ and norms with respect to $g$. We use implicitly that for any space-time tensor ${\bf U}$ we have
$A(\La)\|{\bf U}\|_{H^{2}_{g}}\leq \|{\bf U}\|_{H^{2}_{\atlas}}\leq B(\La)\|{\bf U}\|_{H^{2}_{g}}$.} 
for $E=E(\W_{0})$ or $E=E(\W_{1})$
\beq\label{ESE}
\begin{split}
\|E'\|_{H^{0}_{\atlas}}^{2}&\leq C(\La)(\|\tilde{\W}'\|^{2}_{H^{0}_{g}}+\|\W\|_{H^{0}_{g}}^{2}\|{\bf g}'\|^{2}_{H^{2}_{\atlas}})\\
&\leq C(\La)(Q(\tilde{\W}')+Q(\W)\|{\bf g}'\|_{H_{\atlas}^{2}}).
\end{split}
\eeq

\n Doing the same for $B(\W_{0})$ or $B(\W_{1})$ and putting altogether gives
\beq\label{DF}
\begin{split}
\|(E'_{0},B'_{0})\|_{H^{0}_{\atlas}}+\|(E'_{1},B'_{1})\|_{H^{0}_{\atlas}}\leq &C(\La)(Q^{\frac{1}{2}}(\tilde{\W}'_{0})+
Q^{\frac{1}{2}}(\tilde{\W}'_{1})+(Q^{\frac{1}{2}}(\W_{0})\\
&+Q^{\frac{1}{2}}(\W_{1}))\|{\bf g}'\|_{H^{2}_{\atlas}}).
\end{split}
\eeq

\n Note finally that by Proposition \ref{lem3} $\|\frac{d}{dt}g'\|_{H^{2}_{\atlas}}$ and $\|\frac{d}{dt}K'\|_{H^{1}_{\atlas}}$ are
controlled by the BR-norm of ${\bf g}$ and ${\bf g}'$ over an interval $I'$. This implies
\beq\label{mier1}
\|g'(t)-g'(t_{0})\|_{H^{2}_{\atlas}}\leq C(t-t_{0}),
\eeq
\beq\label{mier4}
\|K'(t)-K'(t_{0})\|_{H^{1}_{\atlas}}\leq C(t-t_{0}),
\eeq

\n where $C$, as above, depends on the BR-norm of ${\bf g}$ and ${\bf g}'$ on $I'$. 

We start the argument by contradiction. We will end up showing that the BR-norm of ${\bf g}'$ over a time interval $I'\subset I$ can
be made as small as we want if $\epsilon$ is chosen small enough. Suppose that there is $(g_{1,i},K_{1,i})$ and $(g_{2,i},K_{2,i})$ converging
to $(g_{0},K_{0})$ in $H^{3}_{\atlas}\times H^{2}_{\atlas}$ and such that for some $\lambda_{i}\in[0,1]$ the BR-norm of ${\bf g}_{i}'(\lambda_{i})$ 
over an interval $I'=[t_{0},t_{i}]$, with $t_{i}\rightarrow t_{0}$, is equal to a positive constant $\tilde{C}$. We note that as 
$(g_{j,i},K_{j,i})$, $j=1,2$ converge to $(g_{0},K_{0})$ in $H^{3}_{\atlas}\times H^{2}_{\atlas}$
it is $(g_{i}',K_{i}')(\lambda)\rightarrow 0$ in $H^{3}_{\atlas}\times H^{2}_{\atlas}$ and uniformly in $\lambda\in [0,1]$. This fact has several
consequences on the $\lambda$-derivative of the fields at time $t=t_{0}$. Namely, for the path $(g,K)(\lambda)$ it is 
$\|N'(t_{0})\|_{H^{3}_{\atlas}}\rightarrow 0$ and $\|X'(t_{0})\|_{H^{3}_{\atlas}}\rightarrow 0$. From (\ref{FF2}) we conclude that 
$\|T'(t_{0})\|_{H^{3}_{\atlas}}\rightarrow 0$
and therefore $\|{\bf g}'(t_{0})\|_{H^{3}_{\atlas}}\rightarrow 0$. Now, 
decompose $\W_{0}(t_{0})$ and $\W_{1}(t_{0})$ into vertical and horizontal components using formulas (\ref{EB1})-(\ref{EB4}). 
Differentiating with respect to $\lambda$ we get that $\|\W'_{0}(t_{0})\|_{H^{0}_{\atlas}}\rightarrow 0$ and 
$\|\W'_{1}(t_{0})\|_{H^{0}_{\atlas}}\rightarrow 0$.
It follows from (\ref{FFFF1}) that $\|\tilde{\W}_{0}'(t_{0})\|_{H^{0}_{\atlas}}\rightarrow 0$ and $\|\tilde{\W}_{1}'(t_{0})\|_{H^{0}_{\atlas}}\rightarrow 0$. 
In particular we have $Q(\tilde{\W}_{0}'(t_{0}))\rightarrow 0$ and $Q(\tilde{\W}_{1}'(t_{0}))\rightarrow 0$. Taking this into account, 
equations  (\ref{A}), (\ref{B}), (\ref{C}), (\ref{DF}) and (\ref{mier1}), (\ref{mier4}), show that if $t_{i}-t_{0}$ is taken small enough 
the BR-norm of ${\bf g}'$ over $I'=[t_{0},t_{i}]$ is less than $\tilde{C}$ which is a contradiction. 

{\it Item 2}. This {\it item} follows from the fact that the map $S$ from initial states into flow solutions is Lipschitz.

{\it Item 3}. This {\it item} follows from {\it item 1}.\ep

{\center \subsection{Long time flows.}\label{3.6}}

We give here an application showing that a ($C^{\infty}$) CMC-flow over a three-manifold with non-positive Yamabe invariant is a long-time
flow if the corresponding globally hyperbolic space-time is future geodesically complete and the first order Bel-Robinson energy of the CMC flow 
remains bounded above. 

\vspace{0.2cm}
\begin{T} Say $Y(\Sigma)\leq 0$ and $({\bf M}\sim \Sigma\times \field{R},{\bf g})$ a smooth $C^{\infty}$ maximally globally hyperbolic 
space-time. Say $(g,K)(k)$ with $k\in [a,b)$
is a CMC flow where $b$ is the lim sup of the range of $k$ and say ${\mathcal{E}}_{1}\leq \Lambda$. Then
if $({\bf M},{\bf g})$ is future geodesically complete the CMC flow is a long time flow i.e. $b=0$.
\end{T}

A CMC slice with mean curvature $k_{0}$ will be denoted by $\Sigma_{k_{0}}$. Let $t$ be a global
smooth time function on ${\bf M}$. Assume the range of $t$ is $(-\infty,\infty)$. Let 
$T=-\bn t/|{\bf g}(\bn t,\bn t)|^{\frac{1}{2}}$ be the unit future pointing normal to 
the foliation induced by $t$. Observe that if $\xi:[c,d]\rightarrow {\bf M}$ is a curve then 
$t(\xi)\dot{}=dt(\dot{\xi})={\bf g}(\bn t,\dot{\xi})$, therefore if ${\bf g}(T,\dot{\xi})\neq 0$ everywhere 
the curve cannot be closed (i.e. $\xi(c)=\xi(d)$).

\vspace{0.2cm}
{\bf Proof:} We will proceed by contradiction and assume that $b<0$. Under that assumption we first prove that 
the CMC foliation lies entirely between $\{t=t_{0}\}$ and $\{t=t_{1}\}$, for some $t_{0}$ and $t_{1}$. Suppose this
is not the case, then there exists a sequence of points $p_{i}$ in $\Sigma_{k_{i}}$ with $k_{i}\rightarrow b$ and
$t(p_{i})\rightarrow \infty$. For a given $i$ denote by $\gamma_{i}$ a past directed geodesic starting at
$p_{i}$ and maximizing the Lorentzian distance to the initial CMC slice $\Sigma_{a}$. Clearly $\gamma_{i}$ are 
perpendicular to
$\Sigma_{a}$. Also because the lapse for the CMC time $k$ is bounded above by $3/b^{2}$ the length of the geodesics
$\gamma_{i}$ is controlled above by $b$. There is a subsequence of geodesics $\gamma_{i_{j}}$ converging into an
inextendible geodesic $\gamma_{\infty}$ of bounded length and perpendicular to $\Sigma_{a}$. This is a contradiction
as we have assumed the space-time was future geodesically complete.

Now ${\mathcal{E}}_{1}$ is assumed less than $\Lambda$ and if $b<0$ then $Vol_{g(k)}(\Sigma)$ is bounded above for
all $k$ in $[a,b)$, therefore for the CMC flow not to be extendable beyond $b$ there must be a sequence of points $p_{i}$ in $\Sigma_{k_{i}}$ with $k_{i}\rightarrow b$ such that the volume
radius $\nu_{g(k_{i})}(p_{i})\rightarrow 0$. Choose it in such a way
that $\nu_{g(k_{i})}(p)\geq \nu_{g(k_{i})}(p_{i})$ for all $p\in \Sigma_{k_{i}}$. We will be arguing with
the rescaled space-time $\lambda_{i}^{2}{\bf g}$ with $\lambda_{i}=1/\nu_{g(k_{i})}(p_{i})$. Denote with a subindex
$\lambda_{i}$ any new rescaled quantity, for example $k_{\lambda_{i}}=k_{i}/\lambda_{i}$, $g_{\lambda_{i}}=\lambda^{2}_{i}g(k_{i})$
and so on. Noting that $\int_{\Sigma}|K|^{4}dv_{g}$, $\int_{\Sigma}|Ric|^{2}dv_{g}$ and $Q_{0}$ scale as a $distance^{-1}$ and $Q_{1}$ 
as a $distance^{-3}$, 
the sequence $(\Sigma,p_{i},g_{\lambda_{i}},K_{\lambda_{i}})$ is converging (strongly) in $H^{3}\times H^{2}$ into a flat 
(and complete)
initial state $(\bar{\Sigma},p_{\infty},g_{\infty},K_{\infty})$ where $\bar{\Sigma}$ is some three-manifold, $g_{\infty}$ is a flat
metric and $K_{\infty}=0$. In particular there is sequence of geodesic loops $l_{i}$ in $\Sigma_{g_{\lambda_{i}}}$ (say of length one)
and based at $p_{i}$. Parameterized with the arc-length $s$ we write 
$l_{i}:[0,1]\rightarrow \Sigma_{g_{\lambda_{i}}}\subset {\bf M}$. 
As noted above there must be for any $i$ a $s_{i}$ with ${\bf g}_{\lambda_{i}}(\dot{l}_{i}(s_{i}),T_{\lambda_{i}})=0$.
At $l_{i}(s_{i})$ consider the orthonormal frame $(\dot{l}_{i},e_{2},e_{3},T_{\lambda_{i}})$. Parallel transport it
along $l_{i}$ and call the resulting frames as $e_{\delta}$, $\delta=1,2,3,4$. As observed above the states are becoming flatter
\footnote{It is important here that $K\rightarrow 0$ in $H^{2}$. For this we need the assumption that ${\mathcal{E}}_{1}$ and not 
just ${\mathcal{E}}_{0}=Q_{0}$ are bounded along evolution.}, therefore 
${\bf g}_{\lambda_{i}}(\bn_{\dot{l}_{i}} \dot{l}_{i},e_{\delta})\rightarrow 0$. Now think the loops $l_{i}$
as inside the scaled space-time $({\bf M},\lambda^{2}_{i}{\bf g})$. Call $l_{\infty}(s_{\infty})$ a limit point of 
$l_{i}(s_{i})$ in ${\bf M}$. As the scaled space-time is becoming closer and closer to Minkowski around $l_{\infty}(s_{\infty})$
the sequence of loops and frames converge to a solution of the system, i. ${\bf g}_{Min}(\bn_{\dot{l}_{\infty}}\dot{l}_{\infty},e_{\delta})=0$,
ii. $\bn_{\dot{l}_{\infty}} e_{\delta}=0$, where ${\bf g}_{Min}$ is the Minkowski metric on $\field{R}^{4}$ and $
l_{\infty}:[0,1]\rightarrow \field{R}^{4},\ {\rm and \ } {\bf g}_{Min}(\dot{l}(s_{\infty}),\dot{l}(s_{\infty}))=1$. Necessarily, such 
solution must be a piece of segment of length one and therefore not closed. \ep

\begin{Remark} {\rm The same proof works to prove that a $C^{\infty}$ CMC flow solution over a three-manifold with non-positive Yamabe invariant is either
a long time flow or the CMC foliation reaches the end of the maximally globally hyperbolic solution, in the sense that for any compact subset
of the maximally globally hyperbolic space-time, there is a CMC slice which is disjoint from it (and to the future of a fixed CMC slice).}
\end{Remark}

\end{document}